\documentclass[11pt,preprint]{aastex}
\usepackage{graphicx}
\usepackage{amsmath,amssymb} 
\usepackage{subfigure}

\begin{document}

\newcommand{\rhat}{\hat{\boldsymbol r}}
\newcommand{\that}{\hat{\boldsymbol \theta}}
\newcommand{\phat}{\hat{\boldsymbol \phi}}
\newcommand{\bfnabla}{\boldsymbol \nabla}
\newcommand{\bfv}{\boldsymbol v}
\newcommand{\bfomega}{\boldsymbol \omega}
\newcommand{\ba}{\begin{eqnarray}}
\newcommand{\ea}{\end{eqnarray}}
\newcommand{\eh}{\hat{\boldsymbol e}}
\newcommand{\cp}{\varpi}
\newcommand{\ol}{\overline}
\newcommand{\sgn}{\mathop{\rm sgn}}
\newcommand{\ephi}{{\bf e}_\varphi}
\newcommand{\Fdis}{F_{\rm w,dis}}
\newcommand{\eps}{\epsilon}
\newcommand{\La}{\langle}
\newcommand{\Ra}{\rangle}


\title{Angular Momentum Transport by Acoustic Modes
  Generated in the Boundary Layer I: Hydrodynamical Theory and Simulations.}

\author{Mikhail A. Belyaev\altaffilmark{1} \& Roman
  R. Rafikov\altaffilmark{1} \& James M. Stone\altaffilmark{1}}
\altaffiltext{1}{Department of Astrophysical Sciences, 
Princeton University, Ivy Lane, Princeton, NJ 08540; 
rrr@astro.princeton.edu}


\begin{abstract}
The nature of angular momentum transport in the boundary layers
of accretion disks has been one of the central and long-standing 
issues of accretion disk theory. In this work we
demonstrate that acoustic waves excited by supersonic shear in the boundary 
layer serve as an efficient mechanism of mass, momentum and energy 
transport at the interface between the disk and the accreting 
object. We develop the theory of angular momentum transport by 
acoustic modes in the boundary layer, and support our findings 
with 3D hydrodynamical simulations, using an isothermal equation 
of state. Our first major result is the identification of three 
types of global modes in the boundary layer. We derive dispersion
relations for each of these modes that accurately capture the pattern
speeds observed in simulations to within a few percent. Second, we
show that angular momentum transport in the boundary layer is 
intrinsically nonlocal, and is driven by radiation
of angular momentum away from the boundary layer into both 
the star and the disk. The picture of angular momentum transport 
in the boundary layer by waves that can travel large distances 
before dissipating and redistributing angular momentum and energy 
to the disk and star is incompatible with the conventional notion of local
transport by turbulent stresses. Our results have important
implications for semianalytical models that describe the spectral
emission from boundary layers.
\end{abstract}

\keywords{accretion, accretion disks -- hydrodynamics --- waves -- instabilities}


\section{Introduction}


Understanding the structure and dynamics of the boundary layer (BL) is
one of the outstanding
theoretical problems in accretion disk theory. We define the BL as the
thin transitional region where the disk attaches to the star, gas
slows down from the orbital to the stellar rotation velocity, and
$d \Omega/ d\cp > 0$ ($\Omega$ is the angular velocity of the fluid
and $\cp$ is the cylindrical radius). In this work, we do not
differentiate between
the different types of accreting objects (a neutron star, a white 
dwarf, or a young star), as long as they have a
surface. This assumes that the compact object has a
weak enough magnetic field that the disk extends all the way down
to the surface of the star without being disrupted
and channeled to the magnetic poles. The criteria for this to
occur is that the magnetic pressure should be less than the ram pressure of
the gas at the surface of the star \citep{GhoshLamb}. 

Within the BL, material slows down from the Keplerian rotational
velocity to the rotational velocity inside the star. Since the radial
width of the BL is significantly less than that of the star,
this implies that the BL is a region of intense shear. In fact, the
rate of energy dissipation inside the BL is expected to be as high 
as for the rest of the disk combined \citep{Kluzniak}, resulting
in intense heating of the gas. 

An open question in BL theory is the mechanism of angular momentum
transport. Due to a rising rotation profile, the BL is linearly stable to
the magnetorotational instability (MRI) which is responsible for
accretion in the disk \citep{BalbusMRI}. Indeed, \citet{Pessah} has
shown that if $d \Omega/ d\cp
> 0$, then the stresses due to a sheared magnetic field are
inefficient at transporting angular momentum and oscillate about
zero. Thus, some other mechanism must be
responsible for angular momentum transport within the BL. As we have
previously shown in \citet{BRS}, a likely candidate for this mechanism is 
excitation of acoustic waves in the BL accompanied by radiation of 
angular momentum away from the layer. This acoustic
radiation arises due to sonic instabilities in the BL, which are a
type of shear instability akin to the Papaloizou-Pringle instability
\citep{PP, NGG} that operate in supersonic flows \citep{Glatzel,BR}.

The focus of our present work will be the detailed description of 
acoustic modes in the BL and their relation to angular momentum 
transport and mass. We will study these modes both theoretically by solving for
their dispersion relation and computationally with 2D, 3D
unstratified, and 3D stratified hydro simulations of the BL with an
isothermal equation of state. 

Understanding the physics of acoustic modes is an important step
towards understanding the temporal variability in the BL and
constructing semianalytic models of BL spectra. Currently, virtually
all semianalytic models of the BL
assume that angular momentum transport is mediated by local
turbulent stresses and adopt an effective $\alpha$-prescription based
on Navier-Stokes viscosity \citep{PN, InogamovSunyaev,
  PiroBildsten}. We argue, however, that angular momentum in the
BL is transported by global waves, rather than local turbulence. These waves
can potentially travel long distances before dissipating, meaning that
angular momentum and energy transport is inherently nonlocal and 
cannot be well-described
by any model based on a local anomalous viscosity. 

The paper is organized as follows. In \S \ref{numericalmodel}, we describe
the model setup and numerical methods that we use for studying the
BL, and in \S \ref{gensec} we provide a general overview of our
simulation results. We
find that sonic instabilities are a ubiquitous outcome in our
simulations and excite acoustic modes that radiate angular momentum
into both the star and the disk. In \S \ref{disrel}, we show that the
acoustic modes observed in simulations have a dispersion relation that is
well-described by a few simple modifications to the dispersion
relation of the supersonic vortex sheet \citep{Miles2,Gerwin,BR}. In \S
\ref{angmomsec}, we show that radiation from $k_z =
0$ acoustic modes entirely accounts
for the angular momentum transport in our simulations
and discuss the implications of this result for semi-analytic
models of the BL. Finally, in \S \ref{stratsec} we discuss the effects of
disk stratification in the $z$-direction on acoustic modes and angular
momentum transport, and we find
that stratification makes no fundamental difference to our results.


\section{Numerical Model}
\label{numericalmodel}



\subsection{Governing Equations}

For our simulations we use the Godunov code Athena \citep{Stone} to solve
the Euler equations of fluid dynamics in cylindrical geometry with an isothermal
equation of state. These equations are
\ba
\label{eq:cont}
\frac{\partial \rho}{\partial t} + \bfnabla \cdot (\rho \bfv) = 0,
\\
\label{eq:mom}
\frac{\partial (\rho \bfv)}{\partial t} + \bfnabla \cdot (\rho \bfv
\bfv) + \bfnabla P + \rho \bfnabla \Phi = 0, \\
\label{eq:state}
P = \rho s^2.
\ea
Here, $\bfv$ is the velocity, $\rho$ is either the density
(3D) or the surface density (2D), $P$ is the pressure, $\Phi$ is the
gravitational potential, and $s$ is the sound speed. We also use $\cp$
throughout this work to denote the cylindrical radius. 


\subsection{Nondimensional Units}

Our initial setup, described in more detail below, consists of a star
which attaches to a constant density disk via a thin interfacial region. 
Following the convention of \citet{BRS}, we nondimensionalize
quantities by setting the radius of the star to $\cp_\star=1$, which
defines a unit of length, the Keplerian angular velocity to
$\Omega_K(\cp_\star) = 1$, which defines a unit
of time, and $\rho = 1$ within the disk, which defines a
characteristic density. Expressed in these units, the orbital timescale
at the surface of the star is $P_K = 2 \pi$. 

We also define a characteristic Mach number as $M \equiv 1/s$, which
is the true Mach number of the flow in the disk at the surface of the
star, since $V_K(\cp_\star)=1$. For a
given set of initial conditions, the Mach number is the sole {\it
  physical} parameter
that determines the time-evolution of the system \citep{BRS}, assuming
an isothermal equation of state. Thus,
the solutions to equations (\ref{eq:cont})-(\ref{eq:state}) form a
one-parameter family in the Mach number. However, in practice there are
also {\it numerical} parameters, e.g. the size of the
simulation domain, that influence the outcome of a simulation.


\subsection{Gravitational Potential}
\label{gravsec}

We take the gravitational potential to be fixed throughout the course
of a simulation. This amounts to ignoring the self-gravity of the
outer layers of the star (Cowling's approximation,
\citet{Cowling}), which was discussed in the context of the BL in
\citet{BR}. 

We run both vertically stratified and unstratified
simulations. For vertically unstratified runs, we use the
cylindrically symmetric potential
\ba
\Phi = -\frac{1}{\cp},
\ea
whereas for vertically stratified runs, we use the point mass potential
\ba
\Phi = -\frac{1}{\sqrt{\cp^2+z^2}},
\ea
where $z$ is the height above the midplane.


\subsection{Initial Rotation Profile}
\label{rotsec}

Initially, the star attaches to the disk via a thin
interface, inside of which the
rotational velocity, $v_\phi$, rises very nearly linearly 
from zero (the velocity in the
star) to the Keplerian orbital velocity (the velocity in the
disk). The initial rotation profile is given by
\ba
\label{regions}
\Omega(\cp) = \left\{
     \begin{array}{lc}
       0 & \cp < 1-\frac{\delta_{BL,0}}{2} \ \ \ \ \ \text{``star''} \\
       \cp^{-3/2}\left(\frac{\cp-1}{\delta_{BL,0}} + \frac{1}{2}\right) & 1-\frac{\delta_{BL,0}}{2} \le \cp \le 1 +
       \frac{\delta_{BL,0}}{2} \ \ \ \ \ \text{``interface''}\\
       \cp^{-3/2} & \cp > 1 + \frac{\delta_{BL,0}}{2} \ \ \ \ \ \text{``disk''}
     \end{array}
   \right. ,
\ea 
and is the same as in \citet{BRS} (see their Figure 1). For all our
runs, we use $\delta_{BL,0} = .01$, so the interface
is as narrow as possible, while still being resolved with $\sim 10$
cells. Our results are independent of $\delta_{BL,0}$, since over the
course of a simulation, the interface thickens into a
self-consistent BL with width $\gg \delta_{BL,0}$ due to angular 
momentum transport by acoustic modes.


\subsection{Initial Density and Pressure Profiles}
\label{idpsec}

The initial pressure profile is specified everywhere throughout the 
domain through the equation of hydrostatic equilibrium
\ba
\frac{1}{\rho}\frac{\partial P}{\partial \cp} = -\frac{\partial
  \Phi}{\partial \cp} + \Omega^2 \cp,
\label{hydrostatr}
\\
\frac{1}{\rho}\frac{\partial P}{\partial z} = -\frac{\partial \Phi}{\partial z}.
\label{hydrostatz}
\ea
The density is directly proportional to the pressure and is related to
it through the equation of state (\ref{eq:state}).

For the 2D and 3D unstratified simulations, the initial density is
$\rho(\cp, t = 0) = 1$
within the disk, and $\rho(\cp, t = 0) = \exp[-\Phi(\cp)/s^2]$ within the
star, where $\Omega = 0$. For the
stratified simulations, these expressions are also accurate for the
midplane density, $\rho_0(\cp,t=0) \equiv \rho(\cp, z = 0, t=0)$, as
long as we use the midplane potential
$\Phi_0(\cp) \equiv \Phi(\cp, z=0)$. The initial density throughout
the simulation domain for a stratified simulation can be expressed in terms of
the initial midplane density as $\rho(\cp, z, t=0) =
\rho_0(\cp, t=0)\exp[-(\Phi(\cp,z) -
  \Phi_0(\cp))/s^2]$. This formula is valid everywhere, including the
interfacial region, since we have rotation on cylinders throughout the
simulation domain, i.e. $\Omega$
is a function only of $\cp$ (equation [\ref{regions}]).


\subsection{Boundary Conditions}
\label{bcsec}

In all of our simulations, we use periodic boundary conditions (BCs) in the
$\phi$ direction and ``do-nothing'' BCs in the $\cp$
direction. The ``do-nothing'' BC simply means that all
fluid quantities retain their initial, equilibrium values for the
duration of the simulation at the inner and outer
$\cp$-boundaries. A ``do-nothing'' BC in $\cp$ is advantageous
to an outflow BC, especially at the inner edge of the domain inside
the star, since it
prevents a systematic change in the density profile over the course of
the simulation due to small deviations from equilibrium. A
``do-nothing'' BC is also advantageous to a reflecting BC, since it
allows sound waves to leave the domain. However, a ``do-nothing'' BC
does not absorb radiation perfectly (neither does an outflow BC), and
there is some partial reflection of waves from the radial boundaries,
although the reflection coefficient is significantly less than unity
(see \S \ref{starcl}).

For the 3D unstratified and stratified simulations we use periodic BCs
in the $z$-direction. The reason for using periodic BCs in the
$z$-direction as opposed to outflow BCs, even in the
stratified case, is that it is hard to achieve a good numerical
equilibrium with outflow BCs. The application of outflow BCs is
potentially more straightforward in spherical geometry (in the
$\theta$-direction), since in that case the gravity vector is
parallel to the $\theta$-boundary.


\subsection{Initial Perturbations and Burn-In}
\label{burnpertsec}

Although we initialize our density and pressure profiles to satisfy
the equations of hydrostatic equilibrium (equations [\ref{hydrostatr}]
and [\ref{hydrostatz}]), we find that we need to burn-in our
simulations for $t \sim 100$ to let any initial transients damp out
and achieve a good numerical hydrostatic
equilibrium. This burn-in procedure was previously used by
\citet{Armitage} in studying the BL, and we find it easier than
setting up an exact numerical equilibrium \citep{Zingale}.

After the burn-in period, we introduce random perturbations to $v_\cp$
of amplitude $\delta v_\cp \sim .001$ in the region $\cp > 1$. These
random perturbations to the radial
velocity seed the sonic instability and facilitate the growth of
acoustic modes. The reason for only introducing perturbations
in the region $\cp > 1$ is that waves traveling from high density to low
density regions amplify by a factor of $\rho^{-1/2}$ to conserve
energy. Since the density in the star is up to $10^6$ times higher
than that in the disk, we want to avoid any transient waves that
amplify as they travel out of the star and into the disk.


\subsection{Simulation-Specific Parameters}

Table \ref{simtable} lists the simulation specific-parameters for each of our
runs. Typically, our simulations are run until a time of $t \approx
400-600$ ($70-100$ orbits at the inner edge of the disk), where $t=0$
corresponds to the time when the random perturbations are introduced
to the radial velocity. We find that this
is long enough for the boundary layer to widen to an approximately
steady-state width, after an initial period of rapid growth. Some of
our simulations are run for over $100$ orbits, and we observe little
change in the behavior of the fluid in the vicinity of the BL or in
the width of the boundary layer at the end of these runs as compared
to our typical ones.

The four most important parameters we vary from simulation to
simulation are dimensionality (2D or 3D), Mach number ($M=6$ or
$M=9$), stratification in the $z$-direction (unstratified or
stratified), and the $\phi$
extent of the simulation domain. Comparing the effects of
dimensionality and stratification across simulations with all other
parameters held constant allow us to probe how the thickness of the
disk affects the dynamics of the system. Varying the
Mach number gives us a handle on how the phenomena we discuss
translate over to astrophysical systems which tend to have
higher Mach numbers (e.g. $M \sim 100$ for white dwarfs). Finally,
changing the $\phi$ extent of the simulation domain gives us a measure
of control over the wavelength of the dominant modes present in the
box, allowing us to probe the dispersion relation for our model (\S
\ref{disrel}).

\begin{table}[!h]
\centering
\begin{tabular}{|c|c|c|c|c|c|c|c|}
\hline
label & dimension & $M$ & stratified & $\cp$-range &
$\phi$-range & $z$-range & $N_\cp \times N_\phi \times N_z$ \\
\hline
3D6a & 3D & 6 & no & (.7, 2.5) & (0, 2$\pi$) & (-1/12,1/12) & $2048 \times
2048 \times 32$ \\
3D6b & 3D & 6 & no & (.7, 2.5) & (0, 2$\pi$) & (-1/6,1/6) & $2048 \times
2048 \times 64$ \\
3D6c & 3D & 6 & yes & (.7, 2.5) & (0, 2$\pi$) & (-1/3,1/3) & $1536
\times 1536 \times 128$ \\
3D6d & 3D & 6 & no & (.7, 2.5) & (0, $\pi$/3) & (-1/12,1/12) & $2048 \times
384 \times 32$ \\
3D9a & 3D & 9 & no & (.85, 2.5) & (0, 2$\pi$) & (-1/18,1/18) & $2048 \times
2048 \times 32$ \\
3D9b & 3D & 9 & no & (.85, 2.5) & (0, 2$\pi$) & (-1/9,1/9) & $2048 \times
2048 \times 64$ \\
3D9c & 3D & 9 & yes & (.85, 2.5) & (0, 2$\pi$) & (-2/9, 2/9) & $1536 \times
1536 \times 128$ \\
3D9d & 3D & 9 & no & (.85, 2.5) & (0, $\pi$/3) & (-1/18,1/18) & $2048 \times
384 \times 32$ \\
3D9e & 3D & 9 & no & (.85, 2.5) & (0, 2$\pi$/7) & (-1/18,1/18) & $2048 \times
384 \times 32$ \\
3D9f & 3D & 9 & no & (.85, 2.5) & (0, 2$\pi$/7) & (-1/18,1/18) & $2048 \times
384 \times 64$ \\
2D6a & 2D & 6 & -- & (.7, 2.5) & (0, 2$\pi$) & -- & $2048 \times
2048 \times 1$ \\
2D6b & 2D & 6 & -- & (.7, 2.5) & (0, $\pi$/2) & -- & $2048 \times
512 \times 1$ \\
2D6c & 2D & 6 & -- & (.7, 2.5) & (0, $\pi$/4) & -- & $2048 \times
256 \times 1$ \\
2D6d & 2D & 6 & -- & (.7, 2.5) & (-.4, .4) & -- & $2048 \times
512 \times 1$ \\
2D9a & 2D & 9 & -- & (.85, 2.5) & (0, 2$\pi$) & -- & $4096 \times
4096 \times 1$ \\
2D9b & 2D & 9 & -- & (.85, 2.5) & (0, $\pi$/2) & -- & $4096 \times
1024 \times 1$ \\
2D9c & 2D & 9 & -- & (.85, 2.5) & (0, $\pi$/4) & -- & $4096 \times
512 \times 1$ \\
\hline
\end{tabular}
\caption{Summary of simulation parameters. The columns from left to
  right are: simulation label, dimensionality (2D or 3D), 
  Mach number, stratification (yes = stratified, no = unstratified),
  radial extent of the simulation domain, azimuthal extent, vertical
  extent (only in 3D), number of grid points in the radial $\times$
  azimuthal $\times$ vertical directions.}
\label{simtable}
\end{table}

 
\section{General Overview}
\label{gensec}


We now present a general overview of our simulation results that will
frame the discussion throughout the paper.


\subsection{Notation for Measured Quantities}

We begin by introducing
some notation, regarding how simulation quantities are measured. For
3D simulations, we define a $z$-averaged density as
\ba
\Sigma(\cp, \phi) \equiv \frac{1}{\Delta z} \int dz \rho,
\ea
where the integral is performed over the entire $z$-extent of the
domain, and $\Delta z$ is the thickness of the simulation domain in
the $z$ direction. The $\phi$-averaged density and perturbations to the
density are then defined as
\ba
\Sigma_0(\cp) &\equiv& \frac{1}{\Delta \phi} \int d \phi \Sigma \\
\delta \Sigma(\cp,\phi) &\equiv& \Sigma - \Sigma_0 \\
\label{deltarhoeq}
\delta \rho(\cp, \phi, z) &\equiv& \rho - \rho_0.
\ea
Next, we define the density-weighted averaging operator represented by a set of
triangle brackets as
\ba
\La f \Ra \equiv \frac{1}{\Delta z \Delta \phi \Sigma_0}\int dz d\phi
\rho f. 
\ea
In 2D, the density-weighted average is the same, except that there is
no integral over the $z$-dimension.
Sometimes, we perform density-weighted averages in only the
$z$-direction so we also define the operation
\ba
\La f \Ra_z \equiv \frac{1}{\Delta z \Sigma}\int dz \rho f. 
\ea

Density-weighted averages are the proper way of measuring simulation
quantities, since mathematically exact analogs to the fluid equations can be
formulated in terms of them \citep{BalbusPapaloizou}. For instance,
one-dimensional forms for the continuity and angular
momentum transport equations are
\ba
\label{1Dcont}
\frac{\partial \Sigma_0}{\partial t} +
\frac{1}{\cp}\frac{\partial}{\partial \cp} \left(\cp \Sigma_0 \La v_\cp
\Ra \right) &=& 0 \\
\label{1Dangmom}
\frac{\partial}{\partial t}\left(\cp^2 \Sigma_0 \Omega \right) +
\frac{1}{\cp} \frac{\partial}{\partial \cp} \left(\cp^2 \Sigma_0 \La
v_\phi v_\cp \Ra \right) &=& 0,
\ea
where we have assumed molecular viscosity is negligible in equation
(\ref{1Dangmom}), and we have defined the angular velocity as
\ba
\Omega \equiv \frac{\La v_\phi \Ra}{\cp}
\ea

Equation (\ref{1Dangmom}) expresses angular momentum conservation and
the second term is the divergence of the
angular momentum flux,
\ba
F_L \equiv \cp \Sigma_0 \La v_\phi v_\cp \Ra.
\ea
Throughout the paper we shall also refer to the angular momentum
current, which is defined as
\ba
\label{CLdef}
C_L \equiv 2 \pi \cp F_L.
\ea
The angular momentum current has the intuitive interpretation that the
rate of change of angular momentum in an annular region is
the difference in angular momentum current between the two edges
of the annulus.

The angular momentum flux, $F_L$, can be decomposed into two parts
\citep{BalbusPapaloizou}, which are the advective and stress terms
\ba
\label{FLdecompose}
F_L &=& F_A + F_S \\
F_A &\equiv& \cp^2 \Sigma_0 \Omega \La v_\cp \Ra \\
\label{FSgen}
F_S &\equiv& \cp \Sigma_0 \La \delta v_\phi v_\cp \Ra \\
\delta v_\phi &\equiv& v_\phi - \cp \Omega.
\ea
These definitions are trivially generalized to 3D case by replacing
$\Sigma_0$ with $\rho_0$.

The advective term corresponds to angular momentum transport by the
bulk motion of the fluid and can be written in terms of the mass
accretion rate, $\dot{M}$, as 
\ba
F_A = -\frac{\dot{M}}{2 \pi \cp} \cp^2 \Omega \\
\label{Mdotdef}
\dot{M} \equiv -2\pi\cp\Sigma_0 \La v_\cp \Ra
\ea 
The stress term, on the other hand,
corresponds to angular momentum transport by turbulence and/or
waves. In the standard $\alpha$-viscosity treatment, the stress
term is alternately written as \citep{SS, PN, BalbusPapaloizou}
\ba
\label{alphadef1}
F_S &=& \cp \alpha \Sigma_0 s^2 \\
\label{alphadef2}
F_S &=& \cp \alpha_\nu \Sigma_0 s^2 \left(-\frac{d \ln \Omega}{d \ln
  \cp}\right) \\
\label{alphadef3}
F_S &=& \cp \alpha_{\text{PN}} \Sigma_0 s^2
 \left(-\frac{d\ln \Omega}{d \ln \cp}\right) \left[1 + \left(\frac{s}{\Omega}\frac{d \ln P}{d \cp}\right)^2\right]^{-1/2}.
\ea
For any of these (or any other) $\alpha$ viscosity
model to provide a good description of the BL, $\alpha$ should
be approximately independent of $\cp$. However, all conventional $\alpha$
viscosity models are based on the
notion of {\it local} turbulent stresses. As we discuss in \S
\ref{angmomsec}, in our BL simulations angular
momentum transport is facilitated by waves not turbulence, 
and so is not well described by any local model.


\subsection{Simulation Results}
\label{simresult}

We now paint a picture of the typical evolution of one of our
simulations. Due to the large shear in the interfacial region
(\S\ref{rotsec}), our initial rotation profile
is unstable to the sonic instability \citep{Glatzel, BR,
  BRS}. Consequently, acoustic modes are excited starting from the initial seed
perturbations to
the radial velocity (\S \ref{burnpertsec}). The amplitude of these
acoustic modes increases with time until
$t \sim 60$ when the sonic instability saturates. 

Around $t \sim 60$, there is typically a single mode present with a
well-defined azimuthal pattern number, $m$, and pattern speed,
$\Omega_P$. This mode, which is discussed in detail in \S \ref{uppersec},
has $\Omega_P \sim 1 - s$ and radiates angular momentum in the form of
sound waves away from the
vicinity of the interfacial region into both the disk and the
star. This radiation of angular momentum causes the interfacial region
to widen and drives accretion of material from the innermost portion
of the disk onto the star.

\begin{figure}[!h]
\centering
\subfigure[]{\includegraphics[width=0.8\textwidth]{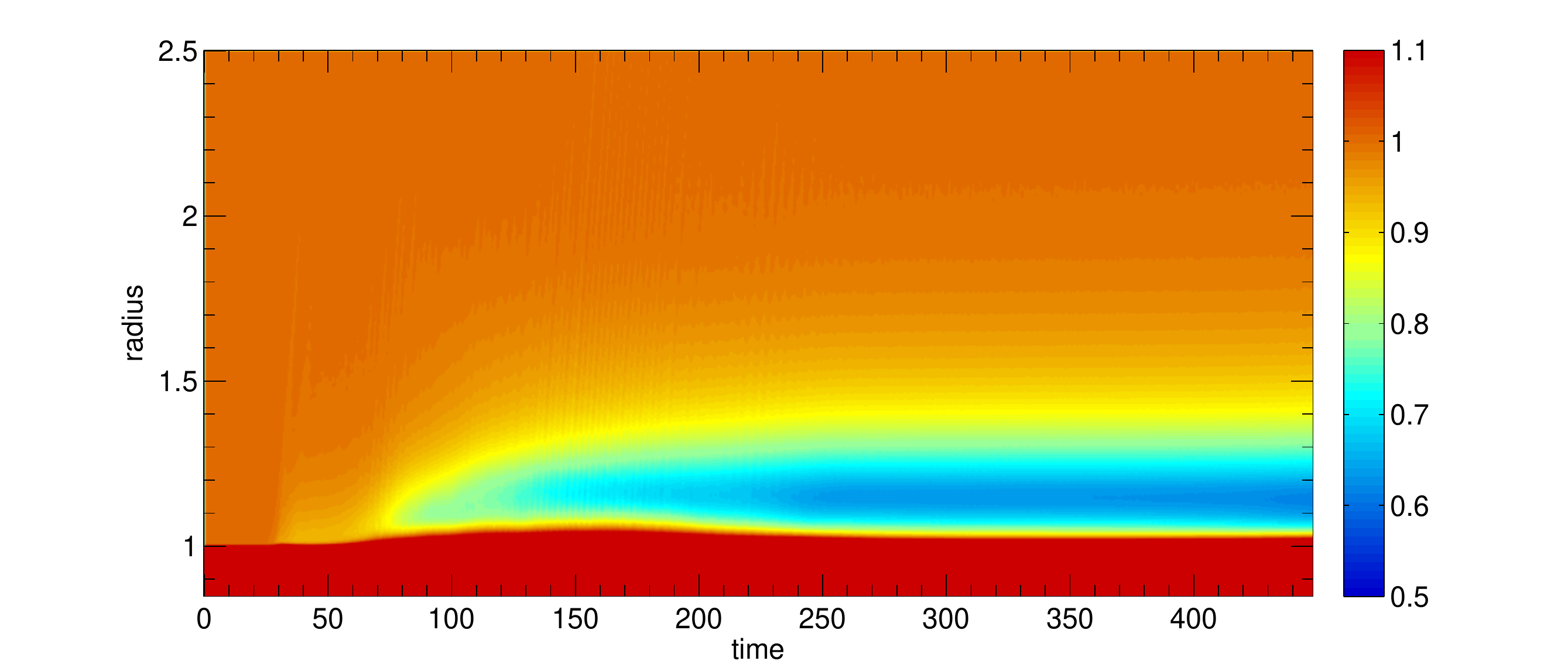}}
\subfigure[]{\includegraphics[width=0.8\textwidth]{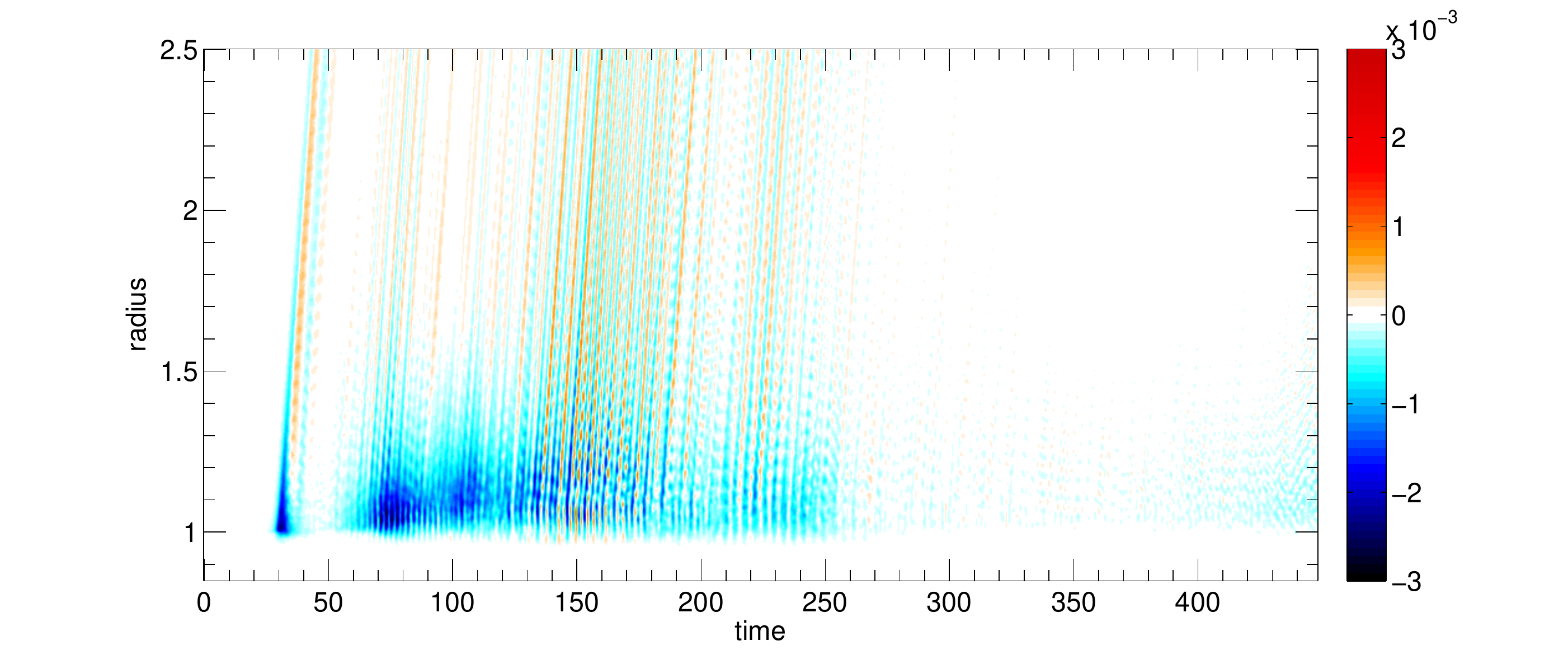}}
\caption{Panels a \& b show radius-time plots of $\Sigma_0$ and $\La
  v_\cp \Ra$, respectively for simulation 3D9a.}
\label{spacetimedensity}
\end{figure}

\begin{figure}[!h]
\centering
\subfigure[]{\includegraphics[width=0.49\textwidth]{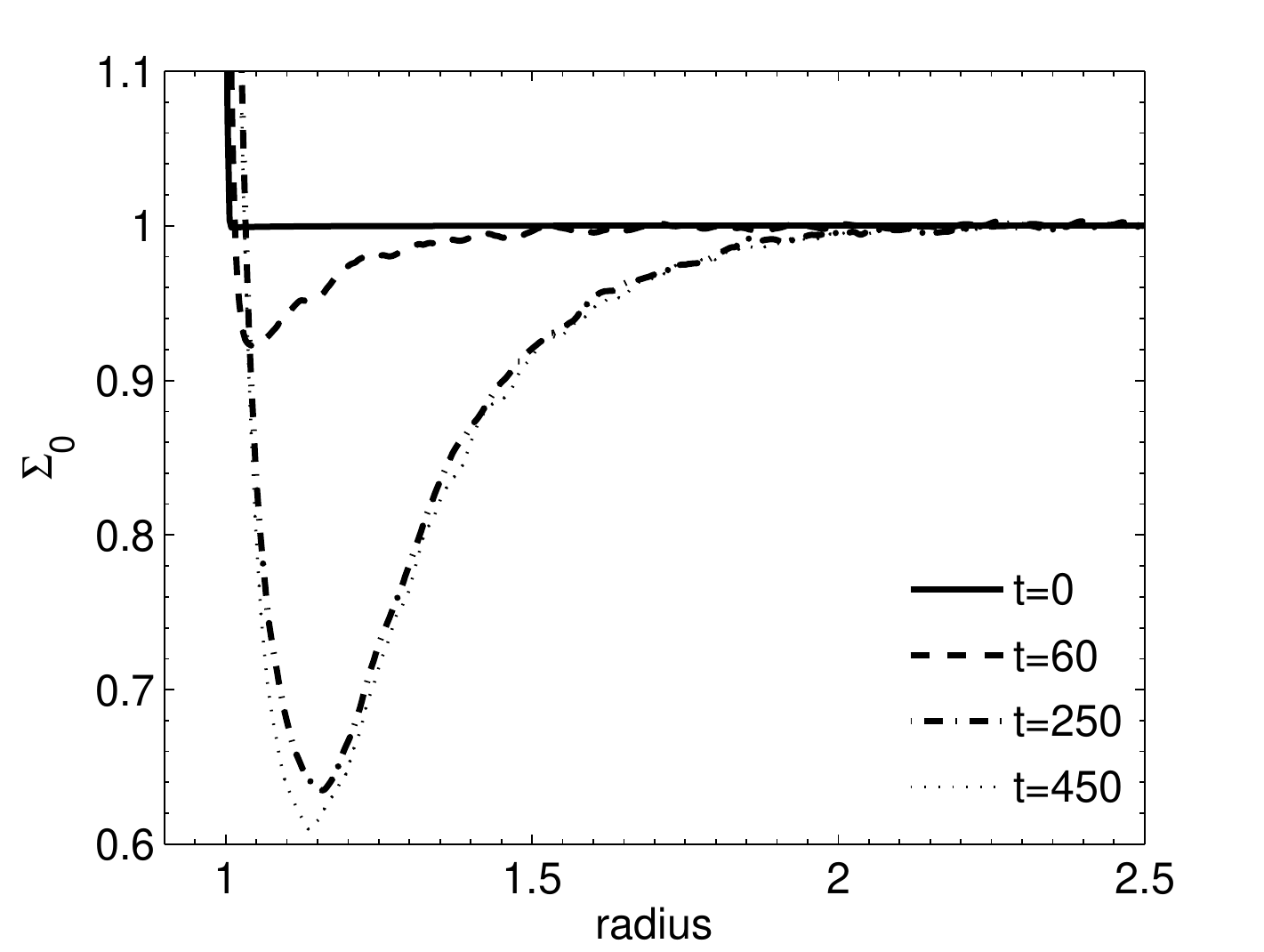}}
\subfigure[]{\includegraphics[width=0.49\textwidth]{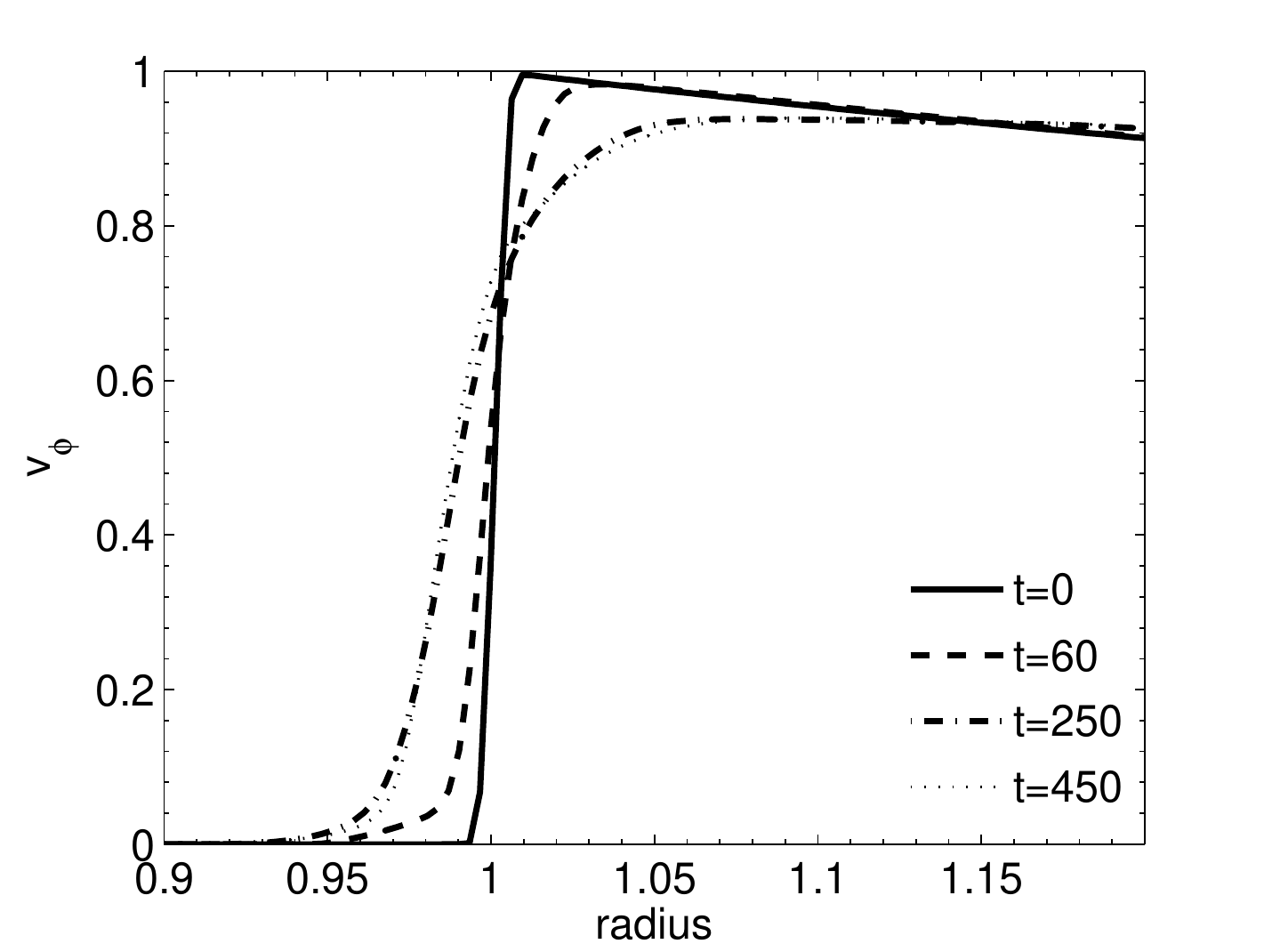}}
\caption{Panels a \& b show $\Sigma_0$ and $\cp \Omega$ vs. time,
  respectively, for simulation 3D9a.}
\label{densomega}
\end{figure}

Fig. \ref{spacetimedensity}a shows a radius-time
image of $\Sigma_0$ for
simulation 3D9a. From the image, it is clear that a gap in the density
is opened up in
the innermost portion of the disk starting at around $t \sim 60$,
which is around the same time when the sonic instability
saturates. The resulting radial density gradient gives rise to partial
pressure support for fluid in the inner disk and causes
the rotation profile to deviate from Keplerian, an issue that we 
return to in \S \ref{sect:Mdot}. Figs. 
\ref{densomega}a \& b show plots of $\Sigma_0$
and $\cp \Omega$, respectively,
 at various times for simulation 3D9a. It is evident
that as the inner region of the disk is depleted of mass, the velocity
profile becomes increasingly sub-Keplerian. Both effects can be
attributed to radiation of angular momentum away from the inner disk
which drives accretion onto the star.

Fig. \ref{spacetimedensity}b shows a radius-time image of the radial
velocity, and it is clear that there is radial infall in the inner
region of the disk, which occurs during the same time as depletion of
the inner disk in Fig. \ref{spacetimedensity}a. The striations in
Fig. \ref{spacetimedensity}b correspond to density waves that emanate
away from the boundary layer, and their ``slant'' indicates that by
causality they are indeed emitted from the vicinity of the BL.

It is evident from Fig. \ref{spacetimedensity}b that after a period of
vigorous infall, the radial velocity and hence the accretion rate in
the inner disk drops off around $t \sim 250$. This can be
attributed to a reduction in the
amplitude of waves emitted from the BL, so the rate of angular
momentum transport is diminished. Moreover, the pattern speed drops
significantly during this period, and it is possible for trapped shocks
to develop between the BL and an evanescent region in the disk
\citep{BRS}. The reason for this shift
in behavior seems to be related to the angular momentum budget of the
inner disk, and we postpone a more detailed discussion of this issue 
until \S \ref{angmomsec}, where we investigate angular momentum transport in
depth. 

Finally, we comment on a result of \citet{BRS}, where they found the
width of the BL to extend across $6-7$ scale heights in the 2D hydro
case. Defining the BL as the region in which
\ba
\label{BLwidth}
0.1 < \langle v_\phi(\cp)\rangle/v_K(\cp) < 0.9, 
\ea
we find general agreement with the 2D results of \citet{BRS} in our 3D
simulations. 

\begin{figure}[!h]
\centering
\includegraphics[width=0.7\textwidth]{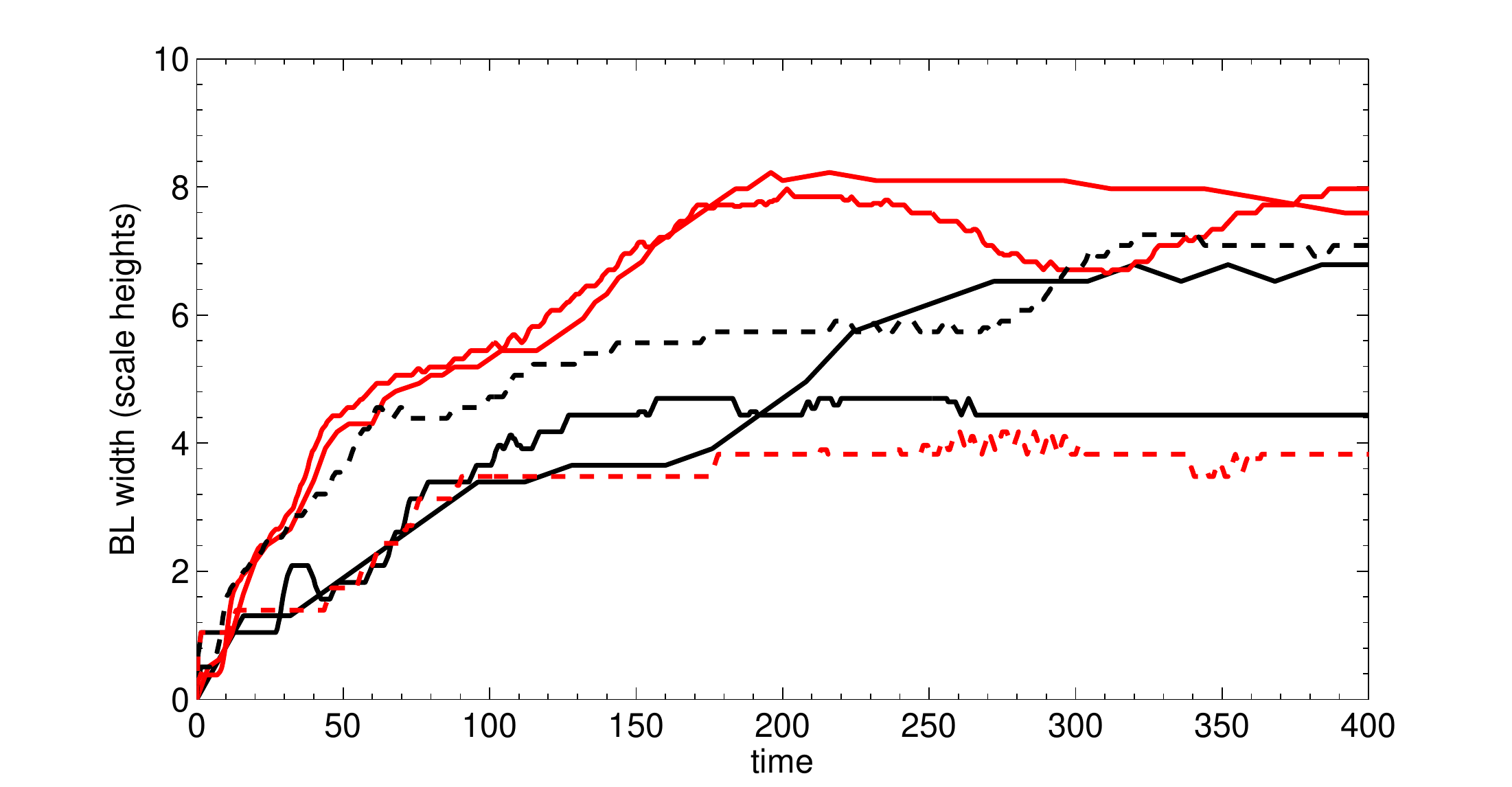}
\caption{Width of the BL in scale heights as a function of time. The
  dashed curves correspond to 3D stratified simulations and the solid
  lines to 3D unstratified simulations spanning the full $2\pi$ in
  azimuthal angle. The color differentiates Mach number with $M=9$ in
  black and $M=6$ in red.}
\label{blwidthfig}
\end{figure}

Fig. \ref{blwidthfig} shows the width of the BL (in scale
heights (equation[\ref{sceq}])) as a function of time
for both our 3D unstratified simulations that span the full $2 \pi$ in
azimuthal angle (solid lines) and the 3D stratified simulations
(dashed lines). The color denotes Mach number with black corresponding
to $M=9$ and red corresponding to $M=6$. 

We point out that there is a
sizeable degree of scatter among the simulations in
Fig. \ref{blwidthfig}, but that the mean is
around 6 scale heights, which is consistent with 2D simulations. One
possible reason for the larger amount of scatter in the 3D case is
that the 2D simulations were run for a longer time, and the measurement
of BL width was performed at a later time in the runs. Thus, the
BL in our 3D simulations may still be settling down to its steady state width.

We now
turn to the modal structure of the BL, since it is ultimately waves
which transport angular momentum in our simulations. Thus, understanding
the dispersion relation of the BL is instrumental to understanding
angular momentum transport by these waves.


\section{Boundary Layer Modes}
\label{disrel}


We find in our simulations that the shear in the boundary layer can
generate three distinct types of $k_z = 0$
acoustic modes. These three types of modes are directly related to the
upper, lower, and
middle branches discussed in \citet{BR} for a 2D plane parallel vortex
sheet in the supersonic regime. However, the dispersion relation for
these modes is modified in a simple way by the radial stratification in
the star and the Coriolis force in the disk. To begin our
discussion, we first review the dispersion relation of the 2D plane
parallel vortex sheet.


\subsection{Plane Parallel Vortex Sheet}
\label{vortsec}

In the absence of stratification or Coriolis force, the dispersion
relation for a 2D vortex sheet (with $x$ and $y$ coordinates locally 
representing $\cp$ and $\phi$ coordinates in the BL) with velocity 
profile
\ba
V_y(x) = \left\{
   \begin{array}{lr}
     V, &  x > 0 \\
     0, &  x < 0,
   \end{array}
   \right.
\ea
sound speed $s_+$ for $x > 0$, and sound speed $s_-$ for $x < 0$ in the limit
$V \gg s$ is given by \citep{BR}
\ba
\label{vortex_dis}
\omega/|k_y| \approx \left\{   
\begin{array}{lr}
     V-s_+, & \text{Upper branch}  \\
     V\left(\frac{s_+}{s_+ + s_-}\right), & \text{Middle branch} \\
     s_-, & \text{Lower branch}.
   \end{array}
   \right.
\ea
Note that by pressure balance $\rho_+/\rho_- = (s_-/s_+)^2$, assuming
$\gamma$ is everywhere constant. All three branches in equation
(\ref{vortex_dis}) are sonic in nature, with
outgoing sound waves emitted from the vortex sheet at $x = 0$.

The upper and lower branches have the property that the sound waves
emitted from the vortex sheet in the half-planes $x > 0$ and $x < 0$,
respectively, propagate
nearly parallel to the vortex sheet. In other words, $|k_y/k_x| \gg
1$ for the upper and lower branches in the half-planes $x > 0$ and $x
< 0$ respectively. The middle branch has the property
that $|k_y/k_x|$ is the same on both sides of the vortex
sheet. Thus, the wavefronts form the same angle with the vortex
sheet in both the $x > 0$ and $x < 0$ half-planes. This is true even if the
densities are different in the two half-planes. 
These sets of properties for the isothermal vortex sheet are instrumental in
determining how the dispersion relations for each of the three wave
branches are modified in going to the
disk-star system.

\begin{figure}[!h]
\centering
\subfigure[]{\includegraphics[width=0.29\textwidth]{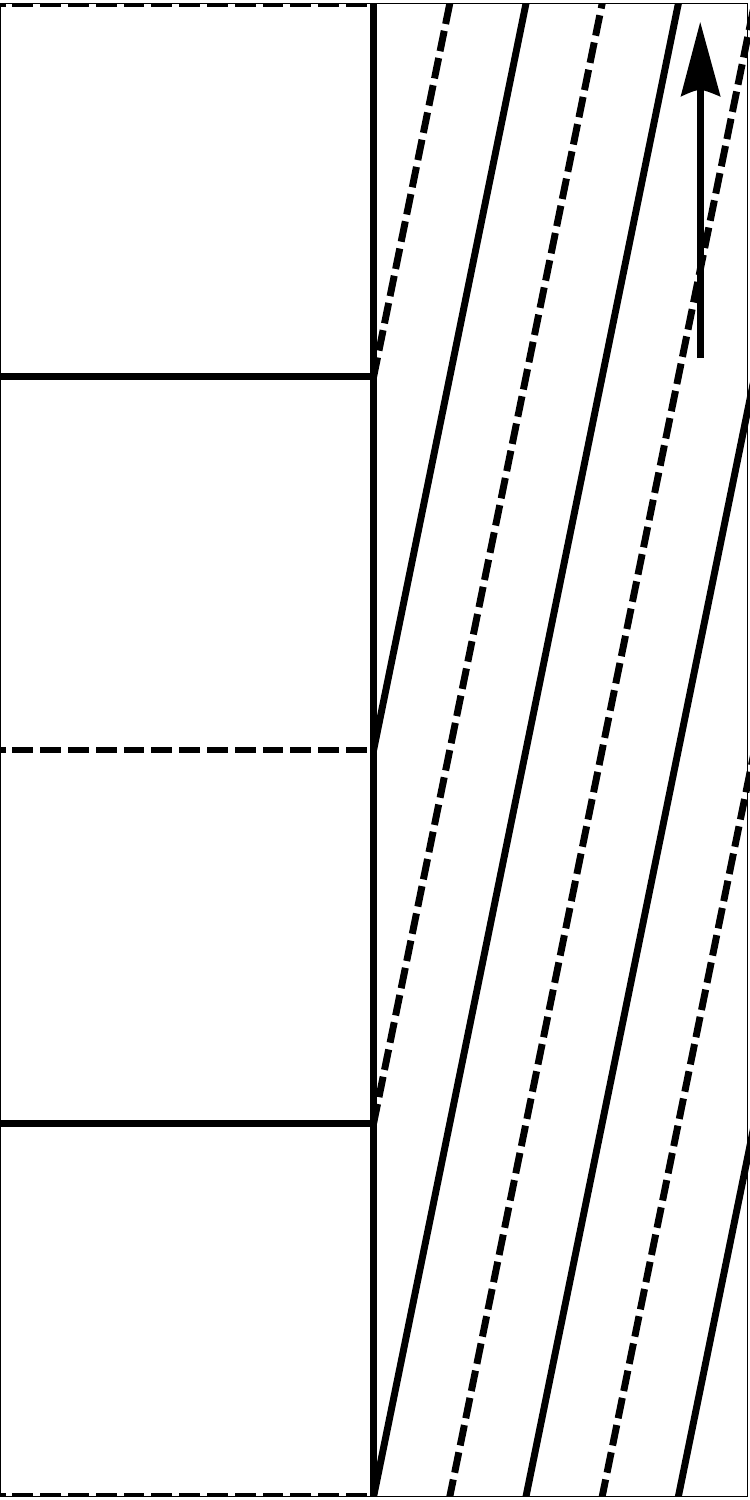}}
\qquad
\subfigure[]{\includegraphics[width=0.29\textwidth]{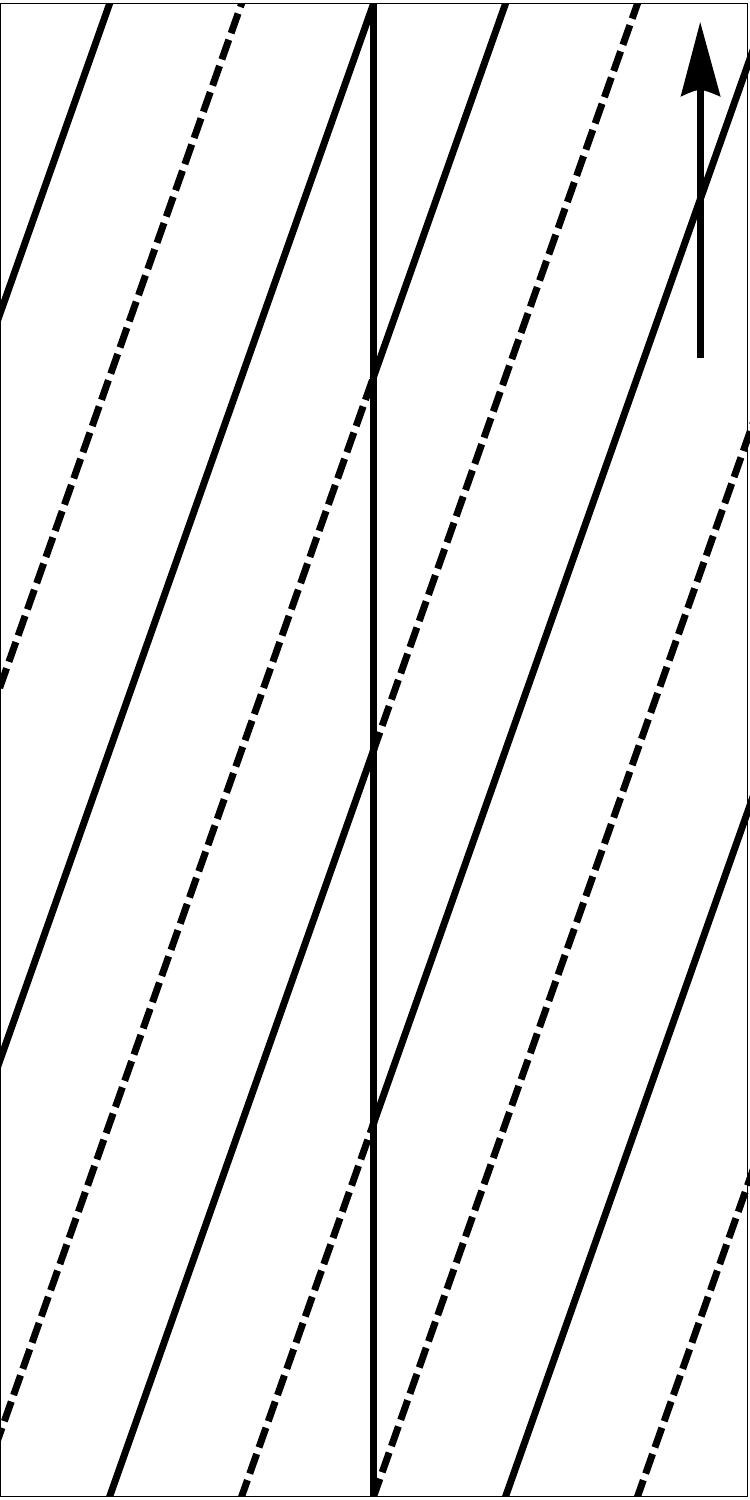}}
\qquad
\subfigure[]{\includegraphics[width=0.29\textwidth]{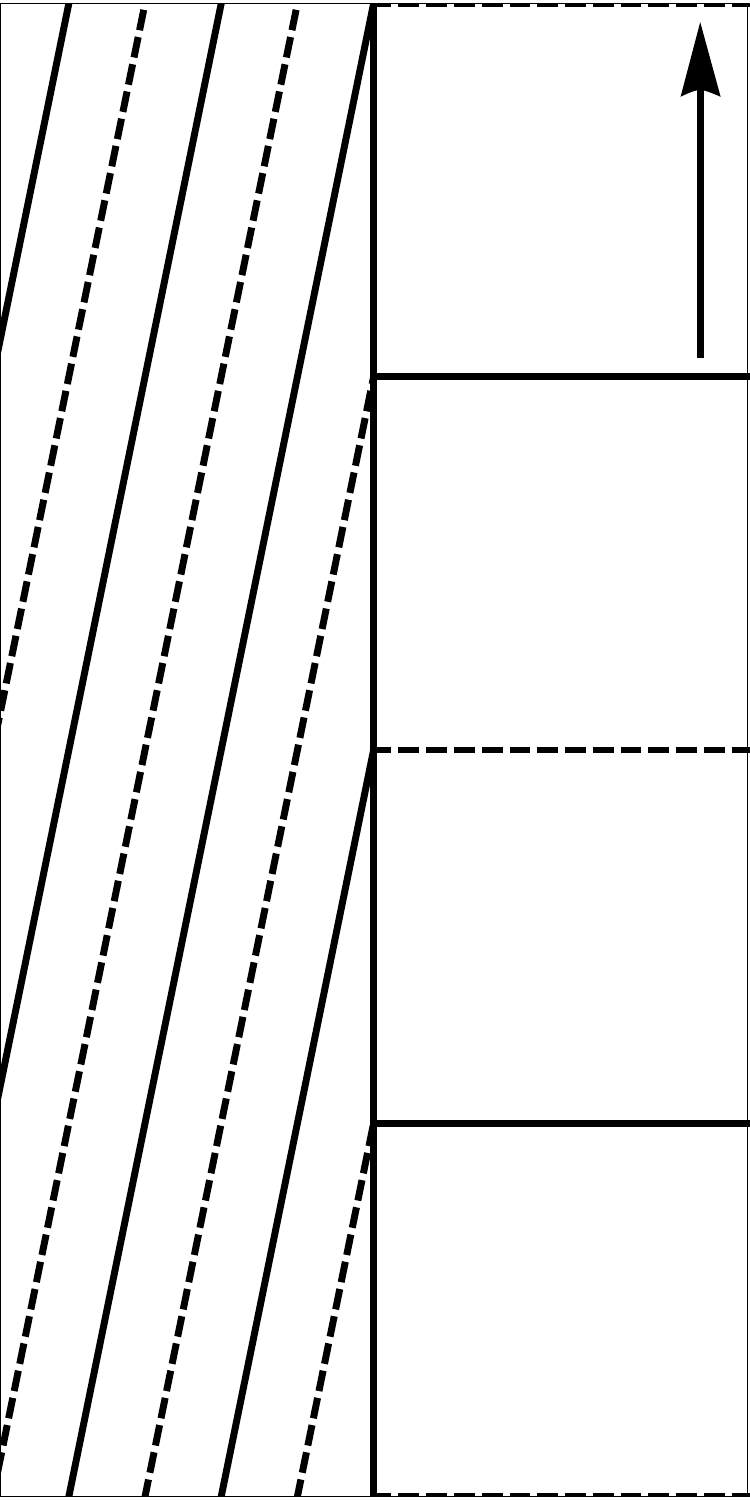}}
\caption{a) Wavefronts of $\delta v_x$ for the lower wave for
  $M=6$. The arrow
  depicts the direction of the flow in the $x > 0$ half-plane. The
  vortex sheet itself is depicted by the solid vertical line and is
  located at $x=0$. The
  solid lines depict wave crests, and the dashed lines depict
  troughs. Notice that the perturbation to $\delta v_x$ undergoes a
  change in sign
  (phase shift by $\pi$) at the vortex sheet. Panels b) and c) are the
  same as a), but for the middle and upper waves respectively. }
\label{modeschematic}
\end{figure}

Panels a-c of Fig. \ref{modeschematic} are schematics, which depict the geometry
of the wavefronts for $\delta v_x$, the perturbed
velocity perpendicular to the vortex sheet, for each of the three
branches. The arrows show the
direction of the flow, and the solid vertical line is the vortex
sheet at $x=0$. At the vortex sheet, $\delta v_x$ undergoes both a change in
amplitude and a change in sign (see equation (36) of
\citet{BR}). Thus, the wavefronts of
$\delta v_x$ appear shifted in phase by $\pi$ across the vortex sheet.

\subsection{Dispersion Relations for the Isothermal Boundary Layer}

We now discuss how the dispersion relations for each of the wave
branches in the isothermal vortex sheet given by equation
(\ref{vortex_dis}) are modified in going to the disk-star system. As a
result, we find
accurate, though, heuristic dispersion relations for the
generalization of these modes to the isothermal boundary layer studied
in our simulations.


\subsubsection{Upper Branch}
\label{uppersec}

As discussed in \S \ref{vortsec}, the upper branch for the vortex sheet
is characterized by a sound wave with phase speed $V-s_+$ in the
half-plane $x > 0$, which
propagates nearly parallel to the vortex sheet, i.e. $|k_y/k_x| \gg
1$. 

In order to generalize this to the disk-star system, we postulate
that in the disk-star system there is a branch characterized by a
density wave in the disk that has $|k_\phi/k_\cp| \gg 1$ at $\cp =
\cp_\star$. Here, and throughout the paper, we
shall define $k_\phi \equiv m/\cp$, where $m$ is the azimuthal pattern number.

Since, $|k_\phi/k_\cp| \gg 1$ in the disk near the BL, the wave fronts
are very loosely wound, and we cannot apply the
classical dispersion relation for a fluid disk in the tight-winding limit
\citep{GT78}. However, it
is straightforward to derive an approximate dispersion relation for a
loosely wound density wave, if we additionally assume $m \gg 1$
(Appendix \ref{diskr0}). The result is 
\ba
\label{upper_bl}
(\Omega - \Omega_P)^2 = \left(\frac{s}{\cp}\right)^2 +
\left(\frac{\kappa}{m}\right)^2.
\ea

Equation (\ref{upper_bl}) generally has two solutions with one having
$\Omega_P > \Omega$ and the other $\Omega_P < \Omega$. For the
isothermal vortex sheet, it turned out that the physical
solutions, corresponding to outgoing waves have a phase velocity $\omega/k < V$
\citep{BR}. By analogy, we take the solution to equation
(\ref{upper_bl}) having $\Omega_P < \Omega$ to be the physical one. In
that case, taking $\cp = \cp_\star$, $V = \cp_\star
\Omega(\cp_\star)$, and making the
associations $\omega = m \Omega_P$, and $k_y = m/\cp_\star$, we see that
equation (\ref{upper_bl}) is analogous to equation (\ref{vortex_dis}) for the
upper branch of the vortex sheet, except for the additional term
containing the epicyclic frequency, which is due to the Coriolis
force.

We can generalize equation (\ref{upper_bl}) to arbitrary values of
$k_\cp$ by writing
\ba
\label{masterdisk}
m^2(\Omega - \Omega_P)^2 =
  s^2\left(k_\phi^2 + k_\cp^2\right) + \kappa^2.
\ea
This approximate dispersion relation reduces to the one for
  tightly-wound waves in the limit $|k_\cp/k_\phi| \gg 1$
  \citep{GT78} and to equation (\ref{upper_bl}) in the limit $|k_\phi/k_\cp|
  \gg 1$. Thus, it is accurate in both the tight-winding and
  loose-winding limits. The essential modification to the
  tightly-wound dispersion relation is the replacement of the radial
  wavenumber with the
  total wavenumber inside the braces in equation (\ref{masterdisk})
  i.e. $k_\cp^2 \rightarrow k_\cp^2 + k_\phi^2$.

  An important feature of equation (\ref{masterdisk}) is that it predicts an
  evanescent region in the disk directly adjacent to the BL. This can
  be deduced by noting that the left hand side of equation (\ref{masterdisk})
  vanishes when $\Omega = \Omega_P$, implying that $k_\cp$ is
  imaginary at the corotation radius. 

  Table \ref{uppertable} compares the value of $\Omega_P$ observed in
  simulations when the upper wave is present to that predicted by
  the dispersion relation (\ref{upper_bl}). For simplicity, we set
  $\cp = \cp_\star$ and use the Keplerian values of $\Omega$ and $\kappa$
  when calculating the
  predicted values from equation (\ref{upper_bl}), i.e.
all variables are evaluated just outside the BL. Nevertheless, the
  agreement between the
  analytical predictions and the simulations is good to the level of
  several percent for both 2D and 3D simulations and irrespective of
  stratification. However, some of our theoretical estimates
  tend to be slightly high, especially at
  late times ($t \gtrsim 60$ orbits) when the boundary layer has thickened
  substantially. One possible cause for the slight deviation between
  the measured and analytically predicted pattern speeds is that the
  boundary layer is not
  infinitesimally thin, but we have assumed so in evaluating the
  dispersion relation at $\cp = \cp_\star$.

\begin{table}[!htbp]
\centering
\begin{tabular}{|c|c|c|c|c|}
\hline
label & time & $m$ & $ \Omega_P$ measured & $\Omega_P$
predicted \\
\hline
2D6a & 20 & $\approx$ 40 & .84  & .831  \\
2D6b & 20 & $\approx$ 40 & .835 & .831 \\
2D6c & 20 &  $\approx$ 40 & .835 & .831 \\
3D6c & 120 & 9 & .77 & .80 \\
3D9a & 100 & 10 & .83  & .850  \\
3D9d & 60 & 12 & .85 & .861 \\
2D9a & 30 & $\approx$ 29 & .87  & .884  \\
2D9a & 280 & 11 & .815 & .856  \\
2D9b & 30 & $\approx$ 24 & .87  & .881 \\
2D9c & 30 & $\approx$ 32 & .88 & .885 \\
\hline
\end{tabular}
\caption{Comparison between the analytical value of $\Omega_P$ and the
  value measured in simulations for the upper branch. The
  columns are from left to right: simulation label, time
  since the start of the simulation, $m$, pattern speed measured from
  the simulation, pattern speed predicted from equation (\ref{upper_bl}).}
\label{uppertable}
\end{table} 

Table \ref{uppertable} shows that the behavior of the phase speed
of the upper mode is in reasonable agreement with the simple vortex sheet
prediction (\ref{vortex_dis}). Indeed, the latter would result in
$\Omega_P=\omega/(k_y\cp_\star)\approx 0.83$ and $0.89$ 
for $M=6$ and $9$ correspondingly. 
This agrees with the trend seen in our simulations, where a higher
value of $M$ results in a somewhat larger value of $\Omega_P$. Also,
the magnitude of $\Omega_P$ is close to $\Omega_K(\cp_\star)$, in accord with
equation (\ref{vortex_dis}) for high Mach number.

\subsubsection{Lower Branch}
\label{lowsec}

For the case of the isothermal vortex sheet (\S \ref{vortsec}), the
lower branch is characterized by
a sound wave which propagates nearly parallel to the vortex sheet in
the half-plane $x < 0$ with phase velocity $s_-$. We now discuss the
modifications that need to be made to this dispersion relation in
order to apply it to the disk-star system.

We begin by postulating that the lower branch in the disk-star system
is characterized by a
sound wave which propagates azimuthally in the star. One immediate
difference as compared to the isothermal vortex sheet
case is that the star is stratified, so the phase velocity of the
lower branch is not simply $s$, but is modified by the effects of gravity and
stratification.

In a non-moving, plane-parallel
atmosphere stratified in the $x$-direction with constant sound speed, $s$,
and gravity, $g$, the dispersion relation for gravito-sonic waves (p and g
modes) is given by e.g. \citet{Vallis} as
\ba
\label{disVallis}
\omega^2 &=& \frac{s^2 l^2}{2}\left[1 \pm \left(1 - \frac{4
    (\gamma-1) k_y^2}{\gamma^2 h_s^2 l^4}\right)\right] \\
l^2 &\equiv& k_x^2 + k_y^2 + 1/4 h_s^2 \\
\label{sceq}
h_s &\equiv& s^2/\gamma g,
\ea
where $h_s$ is the pressure scale height. 
Note that the dispersion relation (\ref{disVallis}) applies to {\it all
wavelengths}, not just to those that are short compared to the scale
height of the atmosphere.

The upper sign inside the braces on the right hand side of equation
(\ref{disVallis}) corresponds to p-modes and the lower sign
to g-modes. An isothermal equation of state ($\gamma = 1$) is neutrally
buoyant, so g-modes have $\omega = 0$, and we will
not consider them further. On the other hand, the dispersion relation for
p-modes with $\gamma = 1$ reduces to
\ba
\label{disVallisiso}
\omega^2 &=& s^2\left(k_x^2 + k_y^2 + \frac{1}{4 h_s^2}\right).
\ea
Note that the effect of gravity and stratification enters into the
dispersion relation
through the $1/4 h_s^2$ term, and in the limit $h_s \rightarrow
\infty$ we recover the dispersion relation for a sound wave in a
uniform medium. However, since we typically find $\lambda \gtrsim h_s$ in our
simulations for the dominant modes, the stratification term in
equation (\ref{disVallisiso}) is very important for comparing with simulations.

We can use equation (\ref{disVallisiso}) to derive a dispersion relation
for the generalization of the lower branch to the disk-star
system. We start by setting
$k_y = m/\cp$, and by analogy with the vortex sheet, we assume that
$|k_y/k_x| \gg 1$ inside the star. Making
again the association, $\omega = m\Omega_P$, we have
\ba
\label{disstar}
\Omega_P &=& 
s\sqrt{\left(\frac{1}{\cp_{0}}\right)^2 + 
\left(\frac{1}{2 m h_s(\cp_{0})}\right)^2}
\\
&=&\Omega(\cp_\star)\frac{\cp_\star}{\cp_0}\sqrt{M^{-2} + 
\left(\frac{M}{2 m}\frac{\cp_\star}{\cp_0}\right)^2}.
\label{disstar1}
\ea
Since $\Omega_P$ must be independent of radius, we have chosen an
``effective radius'', $\cp_0 \lesssim \cp_\star$, to use in the expression
on the right-hand side of
equation (\ref{disstar}). Although the choice of an effective
radius may seem like an ``ad-hoc'' prescription, we provide
theoretical justification for it in Appendix \ref{discylindrical}. 
It is further justified by the agreement between the value of
$\Omega_P(m,M)$ measured in
simulations and that given by equation (\ref{disstar}).

\begin{figure}[!h]
\centering
\includegraphics[width=0.8\textwidth]{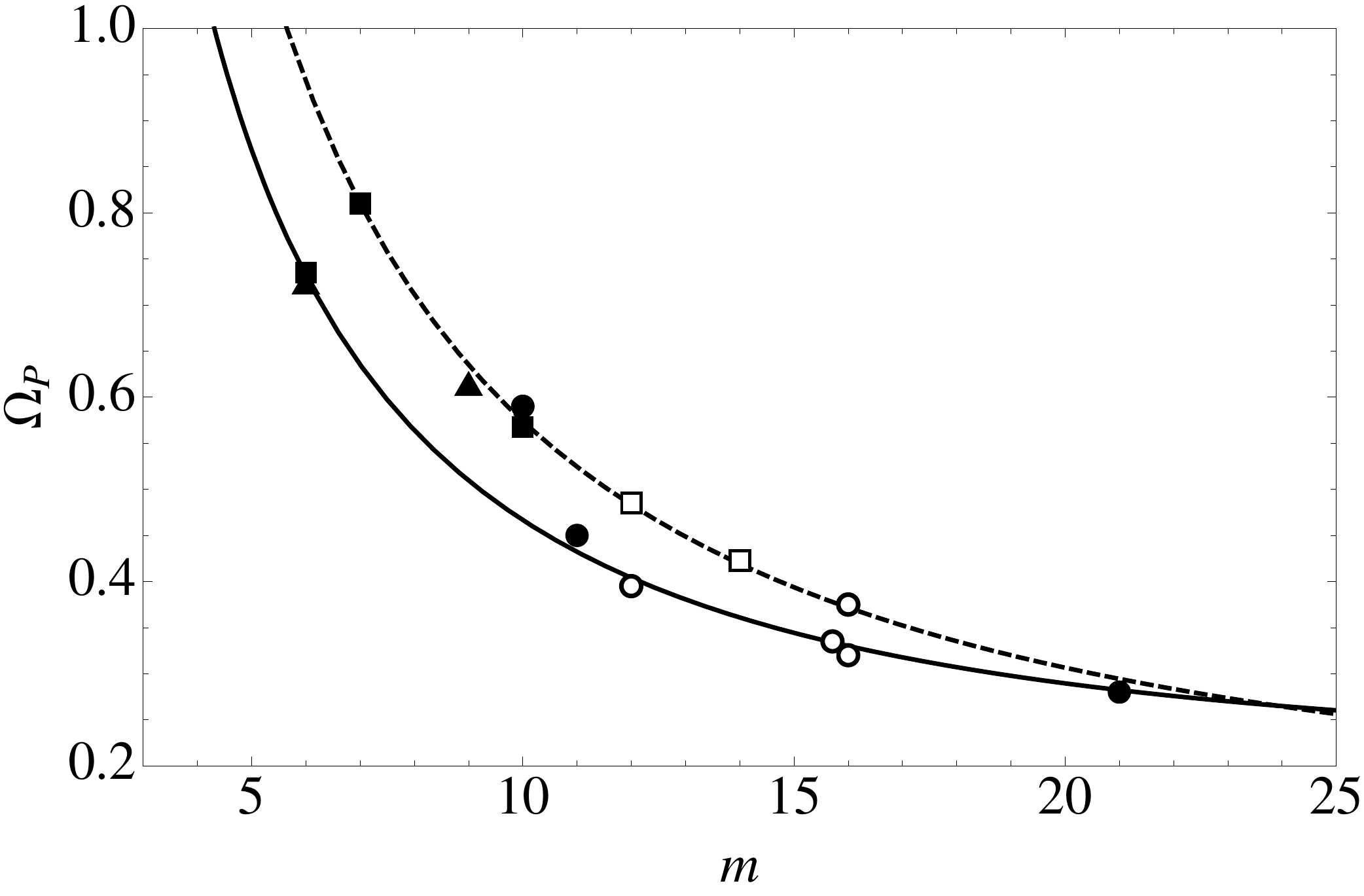}
\caption{Comparison between the analytical dispersion relation, $\Omega_P(m,M)$,
  and simulation results for the lower branch. The solid line is the
analytical prediction from equation (\ref{disstar}) for $M=6$ using an
  effective radius of $\cp_0
= .84$, and the dashed line is for $M=9$ using $\cp_0 = .90$. The
points show values of $\Omega_P$ and $m$ measured from simulations
  when the lower branch is dominant. The circles represent 2D
  simulations, the squares 3D unstratified
  simulations, and the triangles 3D stratified simulations. Filled points represent
  simulations spanning the full $2 \pi$ in azimuthal angle, and open points
  represent simulations for a wedge which spans an azimuthal angle $ <
  2\pi$. There is a solid triangle and a solid square that have merged near
  the solid curve at $M=6$.}
\label{lowerfigure}
\end{figure}

Figure \ref{lowerfigure} shows a plot of $\Omega_P$ vs. $m$ from our
simulations when the lower wave is observed. It is evident that the agreement
between equation (\ref{disstar}) and the data is quite good over a
wide range of values for $m$ and $\Omega_P$. Moreover, the 2D
simulations (circles), 3D unstratified simulations (squares), and 3D
stratified simulations (triangles) lie on the same
curve for both $M=6$ and $M=9$, implying that the behavior is
fundamentally the same in both 2D and 3D and with or without stratification.

It is of note that the lower mode in the BL has phase speed 
considerably higher than $\Omega_P=M^{-1}\Omega(\cp_\star)$ predicted 
by the simple vortex sheet model, see equation (\ref{vortex_dis}). 
This is because the results presented in \S \ref{vortsec} were obtained
assuming an {\it unstratified} star, in which case $h_s\to\infty$ and
our BL dispersion relation (\ref{disstar}) would reduce to its vortex 
sheet analog. However, in an isothermally stratified star, the density 
gradient strongly modifies the dispersion relation, as evidenced
by the second term in equation (\ref{disstar1}), and makes $\Omega_P$ 
not only a function of $M$ but also of  $m$, the azimuthal wavenumber.


\subsubsection{Middle Branch}

For the isothermal vortex sheet with a constant sound speed everywhere
throughout the domain ($s_+ = s_-$), the phase velocity of the middle
branch is simply $V/2$ (\S \ref{vortsec}). Moreover, the wavefronts
make the same angle with respect to the vortex sheet both above and
below it, which is true even for the case of
non-equal densities above and below the vortex sheet \citep{BR}. This
last fact can be expressed mathematically as $k_{x,+}/k_y =
k_{x,-}/k_y$, where the
plus and minus signs denote the value of $k_x$ in the $x > 0$ and $x <
0$ half-planes, respectively. 

We now postulate that for the disk-star system the wavefronts in the
star and in the disk form the same angle with respect to the azimuth
across the boundary layer. Using
equations (\ref{masterdisk}) and (\ref{disVallisiso}), we can write
\ba
\label{kxky+}
\left(\frac{k_{\cp,\text{disk}}}{k_\phi}\right)^2 &=& \frac{(\omega -
  m\Omega)^2 + \kappa^2}{(s k_\phi)^2} - 1 \\
\label{kxky-}
\left(\frac{k_{\cp,\text{star}}}{k_\phi}\right)^2 &=&
\frac{\omega^2}{(s k_\phi)^2} - \frac{1}{(2 h_s k_\phi)^2} - 1
\ea
Setting equations (\ref{kxky+}) and (\ref{kxky-}) equal to each other
  at $\cp = \cp_\star$ (the condition that the wavefronts in the disk
  and star form the same angle with respect to the azimuth across the BL),
the dispersion relation for the middle branch is
\ba
\label{dismiddle}
\Omega_P =
\frac{\Omega(\cp_\star)}{2}\left[1+\frac{1}{m^2\Omega^2(\cp_\star)}\left(\kappa^2(\cp_\star)
+ \frac{s^2}{4 h_{s,\star}^2}\right)\right].
\ea

Table \ref{middletable} lists the pattern speeds and $m$-numbers
measured in simulations when the middle branch is clearly dominant and
compares the pattern speed to that predicted by equation
(\ref{dismiddle}). For simplicity, we again assume a Keplerian profile
for $\kappa$ and $\Omega$ when computing the predicted pattern
speed. The agreement between the predicted and measured values is at
the several percent
level. Also, $\Omega_P$ is close to $\Omega(\cp_\star)/2$, in agreement with
the vortex sheet model, see equation (\ref{vortex_dis}).
The discrepancy between the measured and analytically-predicted pattern
speeds could arise because the condition $k_{\cp,+}/k_\phi =
k_{\cp,-}/k_\phi$ is not exactly satisfied for the disk-star system
as it is for the isothermal vortex sheet. Nevertheless, we find that equation
(\ref{dismiddle}) provides a good approximation to the pattern speed
observed in our simulations when the middle branch is dominant. We
also note that we have fewer measurements for the middle branch as
compared to the upper and lower branches, because the middle branch is
a transient phenomenon observed only during the linear growth regime
of the sonic instabilities, if at all.

\begin{table}[!htbp]
\centering
\begin{tabular}{|c|c|c|c|c|}
\hline
label & time & $m$ & $ \Omega_P$ measured & $\Omega_P$
predicted \\
\hline
3D6a & 10 & $\approx$ 42 & .52 & .502 \\
3D6d & 10 & $\approx$ 36 & .515 & .503 \\
3D6c & 10 & $\approx$ 44 & .52 & .502 \\
3D9e & 20 & $\approx$ 28 & .57 & .515 \\
\hline
\end{tabular}
\caption{Comparison of the analytically predicted value of $\Omega_P$,
  and the value measured from simulations for the
  middle branch. The columns of the table are from
  left to right: simulation label, time since the start of
  the simulation, $m$, pattern speed measured from the
  simulation, pattern speed predicted by equation (\ref{dismiddle}).}
\label{middletable}
\end{table} 


\subsection{Spatial Morphology of the Upper, Middle, and Lower Branches}
\label{morph}

We have identified three branches in our simulations that
are the generalizations of the upper, middle, and lower branches of the 2D
isothermal vortex sheet. We now discuss the structural morphology of
each of these branches.

\begin{figure}[!h]
\centering
\subfigure[]{\includegraphics[width=0.5\textwidth]{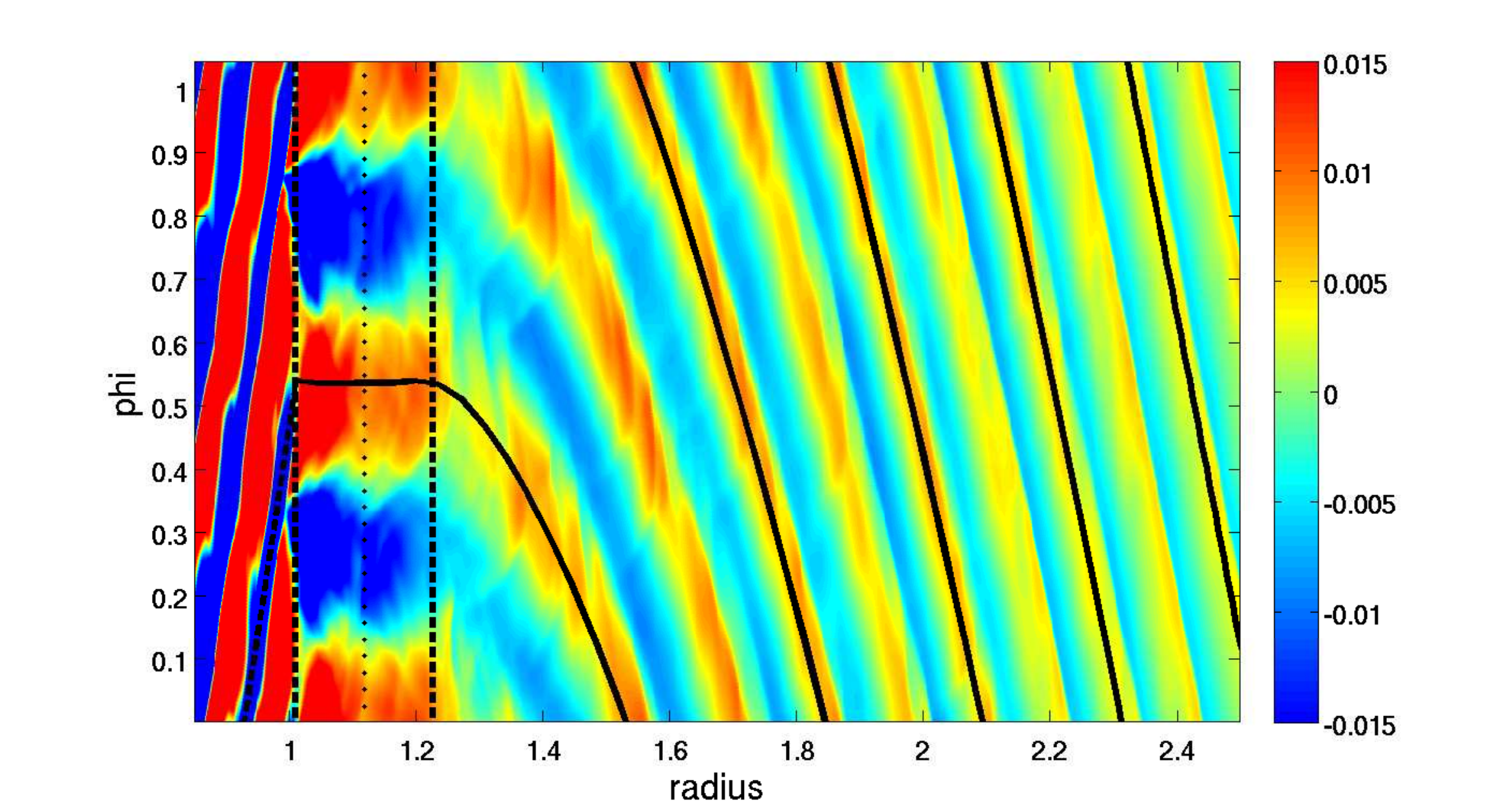}}
\subfigure[]{\includegraphics[width=0.5\textwidth]{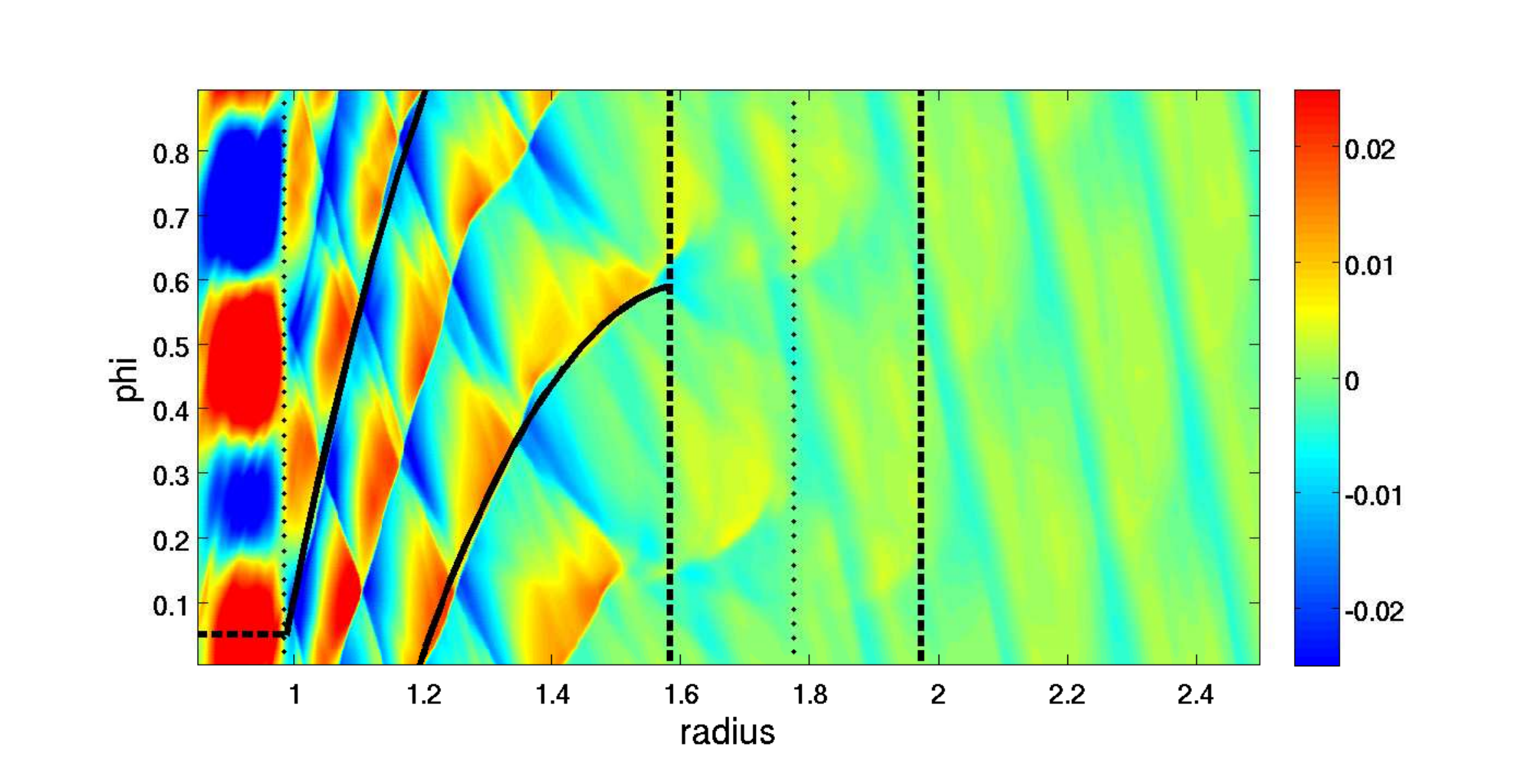}}
\subfigure[]{\includegraphics[width=0.5\textwidth]{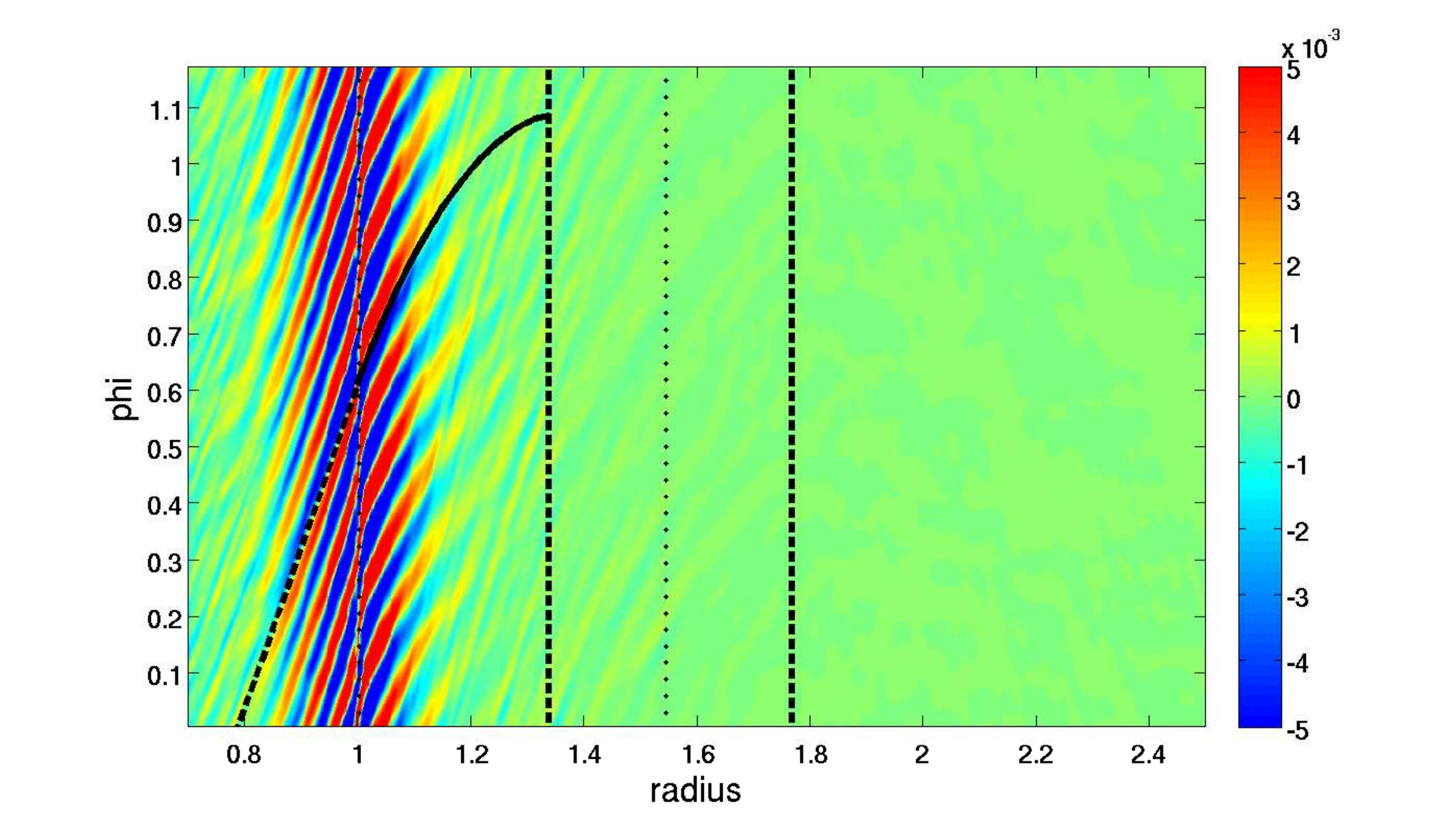}}
\caption{Panels a, b, c show snapshots of $\cp \sqrt{\Sigma} \La v_\cp
  \Ra_z$, when the upper, lower, and middle branches
  are dominant. Panel a corresponds to simulation 3D9d at $t=65$, b
  to simulation 3D9e at $t=320$, panel c to simulation 3D6a at
  $t=10$. Dashed vertical lines depict the boundaries of evanescent
  regions and dotted vertical lines depict corotation radii.}
\label{morphfig}
\end{figure}

Fig. \ref{morphfig} shows snapshots of $\cp \sqrt{\Sigma} \La v_\cp \Ra_z$
vertically-averaged in the $z$-direction for each of the three
wave-branches from 3D unstratified simulations. The quantity $ \cp \sqrt{\Sigma}
 \La v_\cp \Ra_z$ is approximately constant throughout the star and
disk, so it is especially useful for visualizing the radiation
pattern. This is because waves conserve energy and angular momentum
current, which are both of order $\cp^2 \Sigma \delta v^2$, where
$\delta v$ is the magnitude of the velocity
perturbation. The dashed and dotted vertical lines in
Fig. \ref{morphfig} depict
the edges of evanescent regions and corotation radii
respectively. Notice that there are always two corotation radii: one
inside the boundary layer at $\cp \approx 1$, and one
inside the evanescent region within the disk. In principle, there
is also an evanescent region around the corotation radius inside
the BL, but this region is narrow,
due to the steep radial gradient of angular velocity within the BL.

The sharp features seen in this Figure (especially in panel b) are
the shocks into which the propagating acoustic modes evolve as a
result of their nonlinearity. Even though the nonlinearity is relatively 
mild, with $v_\cp\sim 0.1 s$, implying density jump of $\sim 10\%$
across the shock fronts in this particular simulation (see also Fig.
\ref{shockprofiles}), the waves propagating into the disk break
and become weak shocks. In our simulations we typically find 
this to occur not too far from the BL, at separations of 
$(0.2-1)\cp_\star$ away from their launching point.

The solid and dashed curves in Figure \ref{morphfig} trace the shape
of shock fronts or wave fronts if the wave has not shocked. The
formula for the shape of a front is simply 
\ba
\label{shocktrace}
\frac{d \phi}{d \cp} = \pm \frac{k_\cp}{k_\phi},
\ea
where $k_\cp$ is given by formula
(\ref{masterdisk}) inside the disk or (\ref{disVallisiso}) inside the
star. The sign on the
right hand side of equation (\ref{shocktrace}) determines whether the
wave front is angled up or down, assuming we take $k_\cp$ and $k_\phi$
to be positive. We see from Figure \ref{morphfig} that the analytic
curves for the shock fronts are in excellent agreement with
snapshots from the simulations, which is due to the fact that waves
are only weakly nonlinear.

For the upper branch, which has high $\Omega_P$ close to unity, the
evanescent region extends all the way
down to the boundary layer, so density waves emitted from the BL that
propagate into the
disk have a pattern speed that is faster than the angular speed of the
disk material and wind up with increasing radius. For the lower and 
middle branches, however, the evanescent region in the disk is
well-separated from the boundary layer, so density waves emitted from
the BL have a slower pattern speed than the angular speed of the disk material.
These density waves unwind as they travel towards the forbidden
region, partially reflecting off it, with very little transmission
(see \S \ref{uldisk} for a lower bound on the reflection coefficient
measured from simulations). 

As a result of this
reflection, a standing shock pattern can be set up in the
disk between the BL and the forbidden region (Fig. \ref{morphfig}b)
if a resonance criterion is satisfied. This criterion states that in order to
achieve standing shocks, the total azimuthal angle subtended by a
shock in traveling from the BL to the forbidden region and back must
be an integer multiple of the angle subtended by one wavelength of the
mode. Mathematically, this criterion is
\ba
\label{rescrit}
\frac{2 \pi p}{m} &=& 2 \Delta \phi_{\text{shock}} \\
\Delta \phi_{\text{shock}} &=& \int_{\cp_\text{BL}}^{\cp_0} d\cp
\frac{d\phi}{d\cp},
\ea
see \citet{BRS} for more details.
Here, $\cp_0$ is the inner edge of the forbidden region in the disk, $\cp_{BL}
\approx 1$, and $d \phi/ d \cp$ is given by equation
(\ref{shocktrace}). The integer $p\ge 1$ gives the number of shock
crossings for a shock traveling from the BL to the forbidden region or
vice versa, counting the reflection point at the edge of the
evanescent region as a shock crossing. For instance, solving equation
(\ref{rescrit}) for the
$m=14$ mode depicted in Fig. \ref{morphfig}b we find $p = 6.03$. Thus, $p$
is very close to an integer, which allows the standing shock pattern
to form, and it is evident that each outgoing and incoming shock
undergoes six shock crossings (counting the point of reflection).

Although $p$ does not necessarily need to be an integer, it is likely
that modes with an azimuthal wavenumber, $m$, that is close
to an integer value of $p$ are reinforced relative to modes with a
similar value of $m$, but for which $p$ is not close to an integer by
the resonance condition.


\subsection{Modes with $k_z \ne 0$}
\label{kz0sec}

The upper, lower, and middle branch modes are all two-dimensional in
the $\cp - \phi$ plane in
the sense that they do not require the $z$-dimension to operate. In
principle, there are also Kelvin-Helmholtz (KH) modes for vertical
wavenumbers $k_z/k_\phi \gtrsim M$ \citep{BR}. 

To determine the relative importance of the $k_z \ne 0$ vs. the
$k_z = 0$ modes, we compute the power spectrum from the 
unstratified simulations along the $z$-dimension. We work in the
variable $\delta \rho/\sqrt{\Sigma_0}$, which by the
definition of $\delta \rho$ (equation [\ref{deltarhoeq}]) sums
to zero when integrated over the entire simulation domain. This is a
natural variable to use, since one expects the time-averaged value of $|\delta
\rho/\sqrt{\Sigma_0}|$ to be roughly constant for a wave
across the entire simulation domain up to factors related to the
geometry and rotation profile.

To obtain the power spectrum, we first perform a discrete Fourier
transform in the
$z$-direction. In order to carry out summations, we let $\delta
\rho/\sqrt{\Sigma_0}$ be a function of the grid points $(l,m,n)$ in
the $(\cp,\phi,z)$, directions respectively. The Fourier transform in
the $z$-direction then becomes
\ba
\mathcal{F}_n(l,m) \equiv \frac{1}{\sqrt{N_z}} \sum_{n=0}^{N_z-1}
e^{-i 2 \pi n z/N_z} \frac{\delta
    \rho(l,m,n)}{\sqrt{\Sigma_0(l)}},
\ea
where $N_z$ is the number of cells in the $z$-dimension. The power
spectrum is then defined as
\ba
P_n \equiv \frac{1}{N_\cp N_\phi N_z}\sum_{l=0}^{N_\cp-1}
\sum_{m=0}^{N_\phi-1}
(\mathcal{F}_n(l,m))^2.
\ea  
We note that by Parseval's theorem,
\ba
\sum_{n=0}^{N_z-1} P_n = \frac{1}{N_\cp N_\phi N_z}
\sum_{n=0}^{N_z-1} \sum_{m=0}^{N_\phi-1}
\sum_{l=0}^{N_\cp-1} \left(\frac{\delta
    \rho(l,m,n)}{\sqrt{\Sigma_0(l)}}\right)^2. 
\ea

\begin{figure}[!h]
\centering
\includegraphics[width=0.7\textwidth]{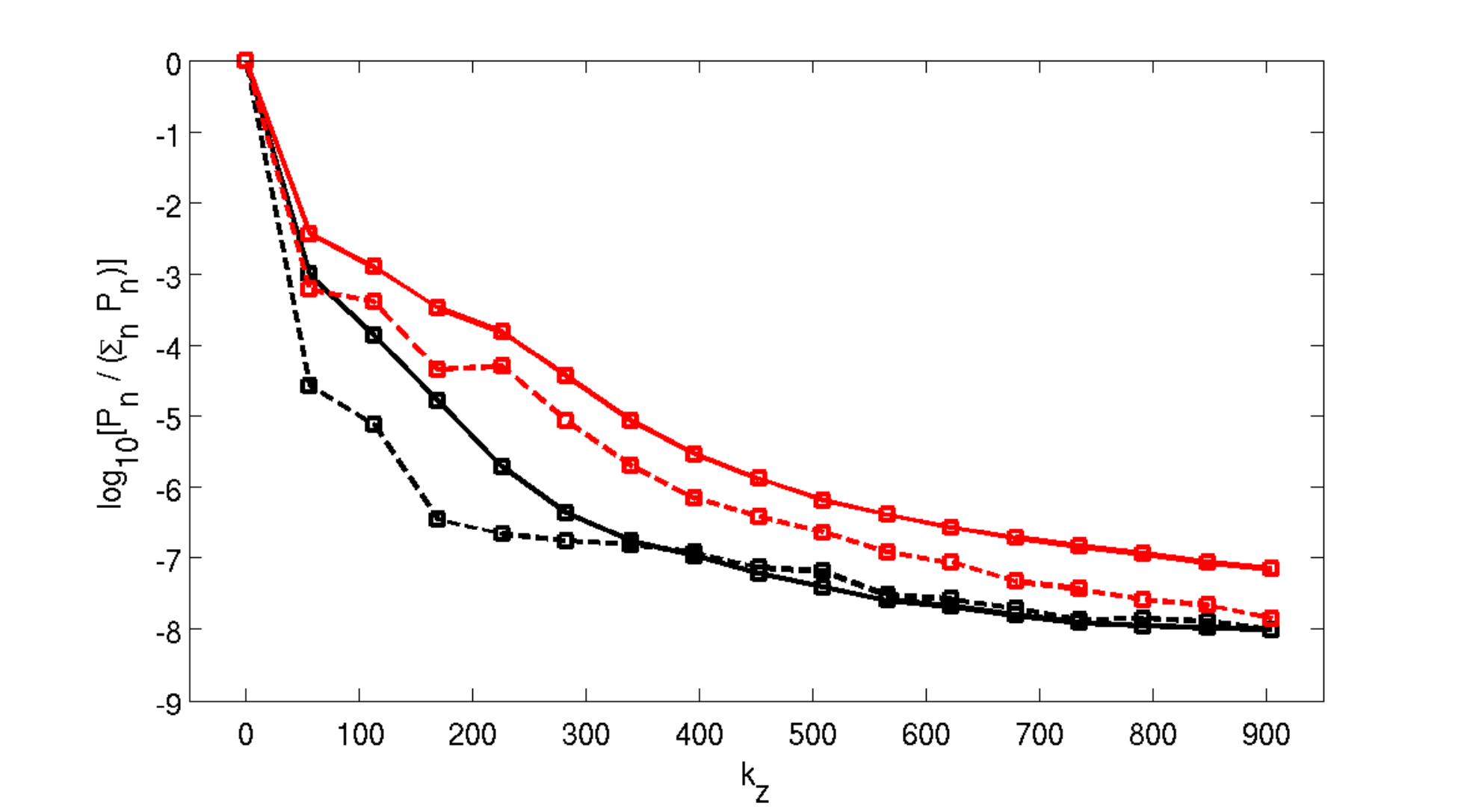}
\caption{The power spectrum, $\log_{10}(P_n/\sum_n P_n)$ as a function
  of $k_z$.
The solid lines are for the upper mode and the dashed lines for the
lower mode. Simulation 3D9e is shown in black and 3D9f in red. Note
that simulations 3D9e and 3D9f have the same vertical extent,
  $\Delta_z$, but a different resolution in the
$z$-direction. When comparing across simulations with different
  resolutions, one should compare $k_z$ rather than $n$, since modes
  with the same value of $k_z$ correspond to the same vertical
  wavelength, which is not true for modes with the same $n$ if the
  simulations have different resolutions.}
\label{powerspectrum}
\end{figure}

Fig. \ref{powerspectrum} shows power spectra computed from the
simulations. Plotted is $\log_{10}(P_n/\sum_n 
P_n)$ vs. $k_z$,
where $k_z \equiv 2 \pi n/ \Delta z$, and $\Delta z$ is the width of
the simulation domain in the $z$-direction. The black lines are from
simulation 3D9e, and the red lines are from simulation 3D9f which has
the same parameters but twice the $z$-resolution. The solid lines
correspond to a time when the upper mode is dominant, and the dashed
lines to a
time when the lower mode is dominant. From the figure, it is
clear that the $n = 0$ mode has at least several orders of magnitude
more power than
all of the $n > 0$ modes combined. Note that we only plot the modes
from $n=0$ to $n = N_z/2$, where $N_z/2 = 16$ for simulation
3D9e. This is because modes having  $n =
N_z/2 + 1$ to $N_z - 1$ (if $N_z$ is divisible by 2) are just the complex
conjugates of modes having $n = N_z/2-1$ to $n=1$, for a real signal
with no imaginary component, such as ours. Simulation 3D9f has $N_z/2=32$,
but we only show the modes up to
$n=16$, since there are no interesting features past $n > 16$, and the
power spectrum simply continues to decay with increasing $n$.

It is clear from Fig. \ref{powerspectrum} that there is more power in
the $k_z \ne 0$ modes relative to the $k_z = 0$ mode for the
higher resolution simulation as compared to the lower resolution
simulation. Thus, it is possible that our simulations are not fully
converged in the $z$-direction. Nevertheless, the fact that the power
in the $k_z \ne 0$
modes is still several orders of magnitude less than the power in the
$k_z = 0$ mode even for the higher resolution simulation suggests
that the $k_z = 0$ mode is indeed dominant irrespective of the resolution
in $z$. 

\begin{figure}[!h]
\centering
\subfigure[]{\includegraphics[width=0.8\textwidth]{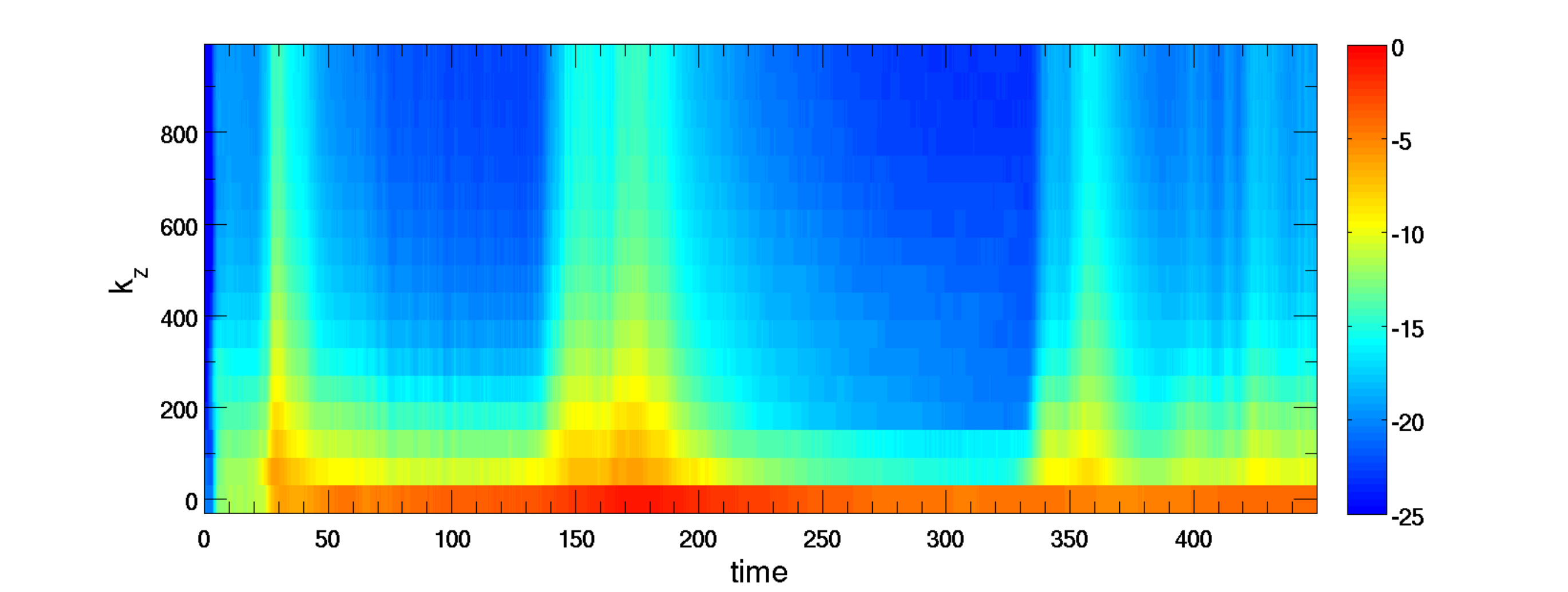}}
\subfigure[]{\includegraphics[width=0.8\textwidth]{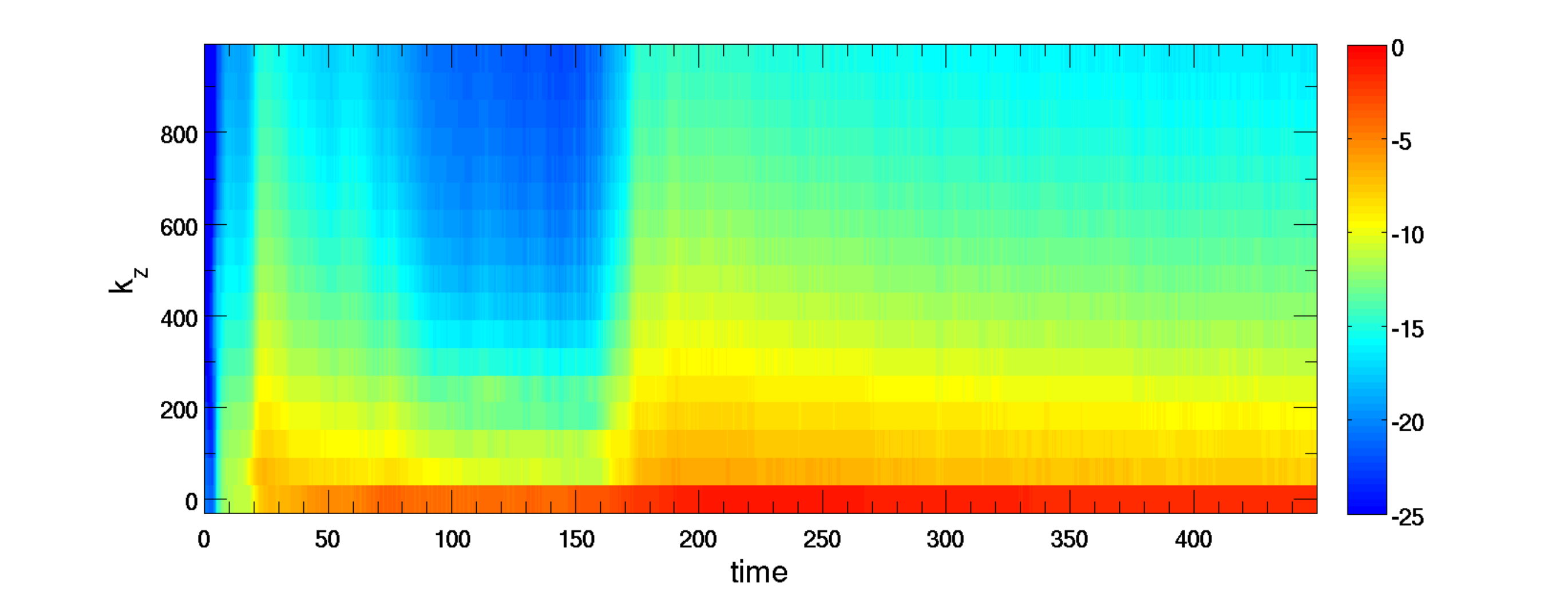}}
\caption{a) $k_z$-time plot of $\log_{10}(P_n)$ for simulation 3D9e. b)
 $k_z$-time plot of $\log_{10}(P_n)$ for simulation 3D9f, which has
 all the same parameters as simulation 3D9e, except that the
 $z$-resolution has been doubled. We again point out that comparing
 across simulations with different
  resolutions, one should compare $k_z$ rather than $n$, since modes
  with the same value of $k_z$ correspond to the same vertical
  wavelength.}
\label{powerspacetime}
\end{figure}

Fig. \ref{powerspacetime} shows a $k_z$-time plot of $\log_{10}(P_n)$ for
simulations 3D9e and 3D9f. One can see interesting temporal features, such as
rapid excitation of both $k_z \ne 0$ modes and $k_z = 0$ modes and
their gradual subsequent decay. One possible source for these periodic
outbursts are secondary KH instabilities \citet{BRS}. However, in the
3D case, unlike the 2D case, a kink-like inflection point in the
azimuthal velocity profile does not form, so this mechanism is not
certain. It is also evident in Fig. \ref{powerspacetime} that the higher
resolution simulation (3D9f) has a greater proportion of power in the $k_z
\ne 0$ modes, but again the $k_z = 0$ mode is clearly dominant for the
duration of the simulation even in the higher resolution case.  Thus, one would
expect the $k_z = 0$ modes to dominate the angular momentum transport
in the system, which we show explicitly in the next section.

\section{Angular Momentum Transport}
\label{angmomsec}


Here we discuss angular momentum transport due to the upper and lower
wave branches. We find good agreement between theory and simulation and
conclude that
the upper branch is more efficient at transporting angular momentum. We
do not consider the middle branch here, since it is dominant
only for short periods of time while the instabilities are still growing.

As we shall soon demonstrate, angular momentum transport in our
simulations is dominated not by turbulent stresses, but rather by
acoustic radiation of angular momentum away from the boundary layer both into
the star and into the disk. Moreover, the $k_z = 0$ modes dominate
the angular momentum transport, as hinted at in \S \ref{kz0sec}, which
allows us to use the theory of density waves in thin disks
to calculate the rate of angular momentum transport (e.g. Appendix J
of \citet{BinneyTremaine}).

The angular momentum current through radius $\cp$ is given by
equation (\ref{CLdef}).
By analogy with equation (\ref{FLdecompose}), it can be
decomposed into a stress term and an advective term. 
\ba
\label{CSgen}
C_S(\cp) \equiv 2 \pi \cp F_S \\
\label{CAgen}
C_A(\cp) \equiv 2 \pi \cp F_A.
\ea

Waves transport angular momentum exclusively through stresses, and we
can use equations (\ref{CSgen}) and (\ref{FSgen}) to calculate
$C_S$. Inside the star, $C_S$ for a given mode with
azimuthal wavelength $m$ is given in terms of the amplitude
of the density perturbation $\delta \Sigma_m(\cp)$, as
\ba
\label{CLmodestar}
C_{S,m}(\cp) = \pm\left(\frac{k_\cp \cp}{m}\right)\frac{\pi s^4
  \Sigma_0}{\Omega_P^2} \left(\frac{\delta
  \Sigma_m}{\Sigma_0}\right)^2. 
\ea
Here we have used the
relations in Appendix \ref{discylindrical} to express $\delta v_{\phi,m}$ and
$\delta v_{\cp,m}$ in terms of $\delta \Sigma_m$.
The total angular momentum current due to waves is the sum over
  all modes for both incoming and outgoing waves and is given by
\ba
\label{CSdiskwavtot}
C_S(\cp) = \sum_m C_{S,m}(\cp).
\ea
  
We next assume that all waves propagate outwards from the BL, and
  that $k_\cp \cp / m$ is relatively insensitive to $m$
  near the dominant $m$ value, which we denote as $\ol{m}$. This
  pair of assumptions will be verified later by comparison with simulation
  results. However, it fixes the sign in equation (\ref{CLmodestar})
  and allows us to use Parseval's theorem \citep{BRS} to write the total
  stress due to waves in the star as
\ba
\label{CLstar}
C_{S}(\cp) = -\left(\frac{k_\cp(\ol{m})\cp}{\ol{m}}\right)\frac{s^4
  \Sigma_0}{\Omega_P^2}\int_0^{2\pi} d\phi \left(\frac{\delta
  \Sigma}{\Sigma_0}\right)^2. 
\ea

The analog to equation (\ref{CLstar}) for the angular momentum current
due to density waves in the disk is \citep{BinneyTremaine,BRS}
\ba
\label{CLdisk}
C_S(\cp) = \left(\frac{k_\cp(\ol{m})\cp}{\ol{m}}\right)\frac{s^4
  \Sigma_0}{(\kappa/m)^2-(\Omega-\Omega_P)^2}\int_0^{2\pi} d\phi \left(\frac{\delta
  \Sigma}{\Sigma_0}\right)^2,
\ea
where we have again used Parseval's theorem and assumed that all waves
propagate away from the BL.

Waves propagating into the disk shock, resulting in $\partial
  C_S/\partial \cp \ne 0$ due to dissipation of the wave with
  increasing distance away from the BL. The shock-related 
change in angular momentum current carried by the wave 
as a function of radius in the disk is given by
  \citep{Larson, BRS}
\ba
\label{dEdmeq}
\frac{\partial C_S}{\partial \cp} = -m \cp \Sigma_0 \frac{dE}{dm}.
\ea
Here, $dE/dm$ is the energy converted into heat per unit mass of fluid
  crossing the shock, which comes at the expense of the energy
  contained in the shock and causes it to damp. For the isothermal
  case, the expression of \citet{Larson} for $dE/dm$ is
\ba
\label{dEdmapprox}
\frac{dE}{dm} = \frac{s^2}{6}\frac{\eps^3}{(1+\eps)^2}
\ea
in the limit $\eps \ll 1$, where $\eps \equiv \Delta \Sigma/\Sigma$ is
the jump in density across the shock divided by the preshock density.

It is also possible to derive an expression for $dE/dm$ which is valid for
  arbitrary values of $\eps$ in the isothermal limit
  $\gamma \rightarrow 1$ (Appendix \ref{appiso}) 
\ba
\label{dEdmiso}
\frac{dE}{dm} = s^2 \frac{\eps(2+\eps) - 2(1+\eps)\ln(1+\eps)}{2(1+\eps)}.
\ea
This converges to Larson's result (equation [\ref{dEdmapprox}]) to
leading (third) order in $\epsilon$.

In contrast to waves propagating into the disk, waves propagating away
from the BL and into the star do not shock, because their amplitude
is proportional to $\rho^{-1/2}$ by conservation of energy and angular
momentum. Thus, as waves propagate into the star in our simulations,
their amplitude is reduced, nonlinear effects are mitigated,
and $\partial C_S/\partial \cp = 0$. Inside a real star, this
reduction in amplitude could not continue indefinitely, and the waves
would eventually be absorbed, depositing their energy and angular
momentum to the stellar fluid. However this absorption could
potentially occur deep inside the star, making the BL appear cooler
than if the energy were dissipated locally.

We now apply these results to understand angular momentum transport
by the upper and lower acoustic modes, which appear in our simulations 
most often.


\subsection{Upper Branch}
\label{ubsec}

\subsubsection{Disk}
\label{ubsecdisk}

\begin{figure}[!h]
\centering
\includegraphics[width=0.7\textwidth]{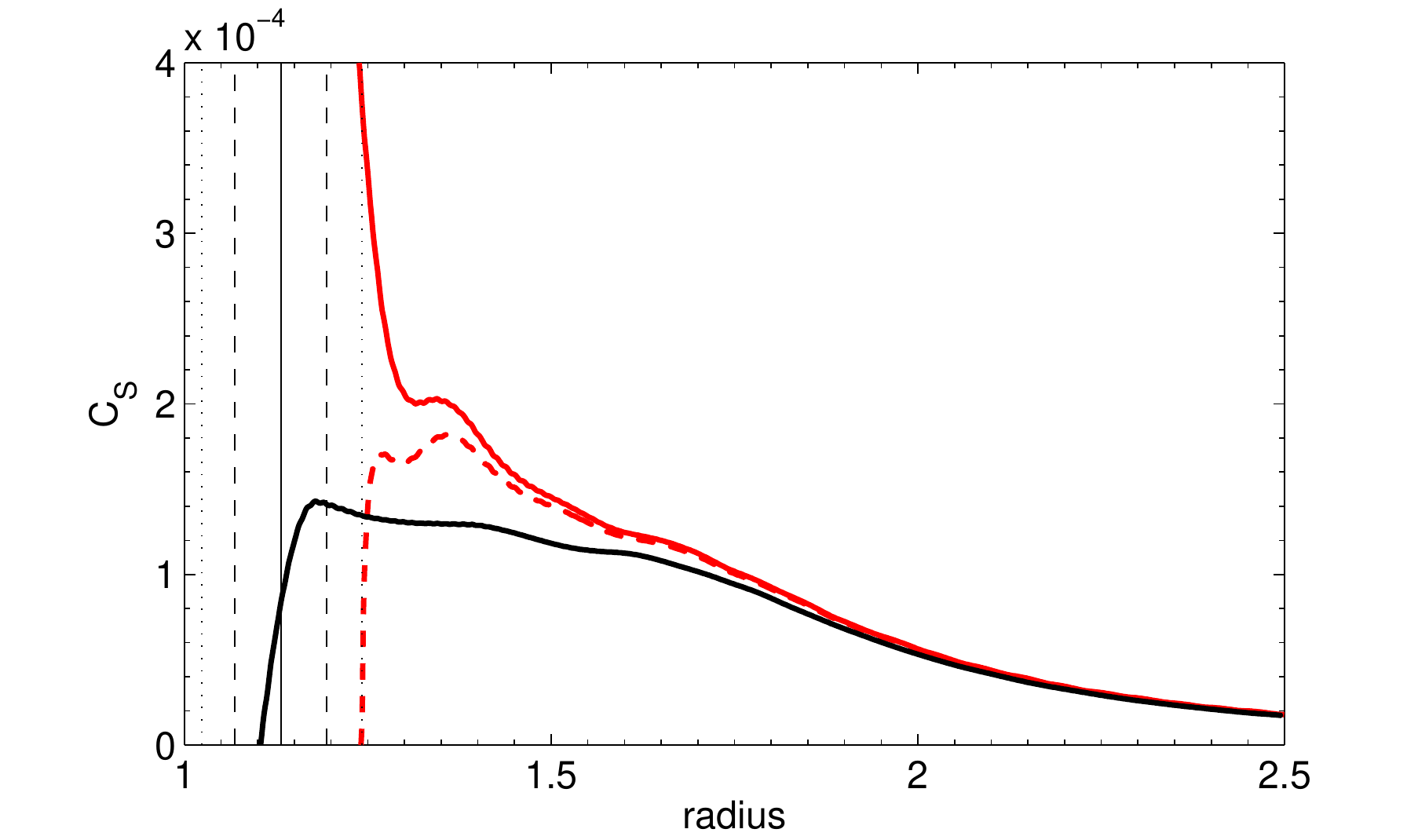}
\caption{Plot of $C_S$ for simulation 3D9d at time $t=75$ within the disk.
See text for details.}
\label{CLdiskfig}
\end{figure}

Fig. \ref{CLdiskfig} shows $C_S$ inside the disk for simulation 
3D9d at $t \sim 75$, which
corresponds to approximately the same time as Fig. \ref{morphfig}a,
when the upper mode is
clearly dominant. The solid black curve shows the measured
value of $C_S$, using equations (\ref{FSgen}) and (\ref{CSgen}), and
the red curves are the analytically-expected values of $C_S$,
calculated using equation (\ref{CLdisk}) together with the value of
$\Delta\Sigma/\Sigma$ across shock fronts measured from 
simulations. The solid red curve uses the
standard WKB expression for $k_\cp$ \citep{GT78}, and the
dashed red curve uses $k_\cp$ from equation (\ref{masterdisk}). The
standard WKB expression can be obtained from equation (\ref{masterdisk})
by eliminating the $k_\phi$ term in the brackets. For simplicity, we
use the Keplerian values of $\kappa$ and $\Omega$ in equation
(\ref{CLdisk}) when calculating the analytically-expected value of
$C_S$ (red curves). This is a good approximation, since the rotation
profile is very nearly Keplerian in the disk, away from the direct
vicinity of the BL.

Regardless of which
form for $k_\cp$ is employed, equation (\ref{CLdisk}) itself is only valid in
the tight-winding approximation. Thus, the red curves converge to the
black at large $\cp$, but deviate significantly as we approach the
evanescent region around corotation, where the upper mode is loosely
wound. For the solid red curve, the
evanescent region is located between the vertical dashed lines, 
which correspond to
the two Lindblad radii, and for the dashed red curve, the evanescent
region is located between the dotted vertical lines. The corotation
radius is marked by a solid vertical line. The fact that the red
curves converge to the black far away from the evanescent region
around corotation is decisive evidence that angular momentum is
radiated away from the BL in the form of density waves.

Something else to note is
that the black curve in Fig. \ref{CLdiskfig} switches
sign close to corotation. This is exactly what would be expected if
angular momentum were carried by waves, since the angular momentum
density of a wave undergoes a change in sign at the corotation
radius \citep{BinneyTremaine}.

\begin{figure}[!h]
\centering
\subfigure[]{\includegraphics[width=0.49\textwidth]{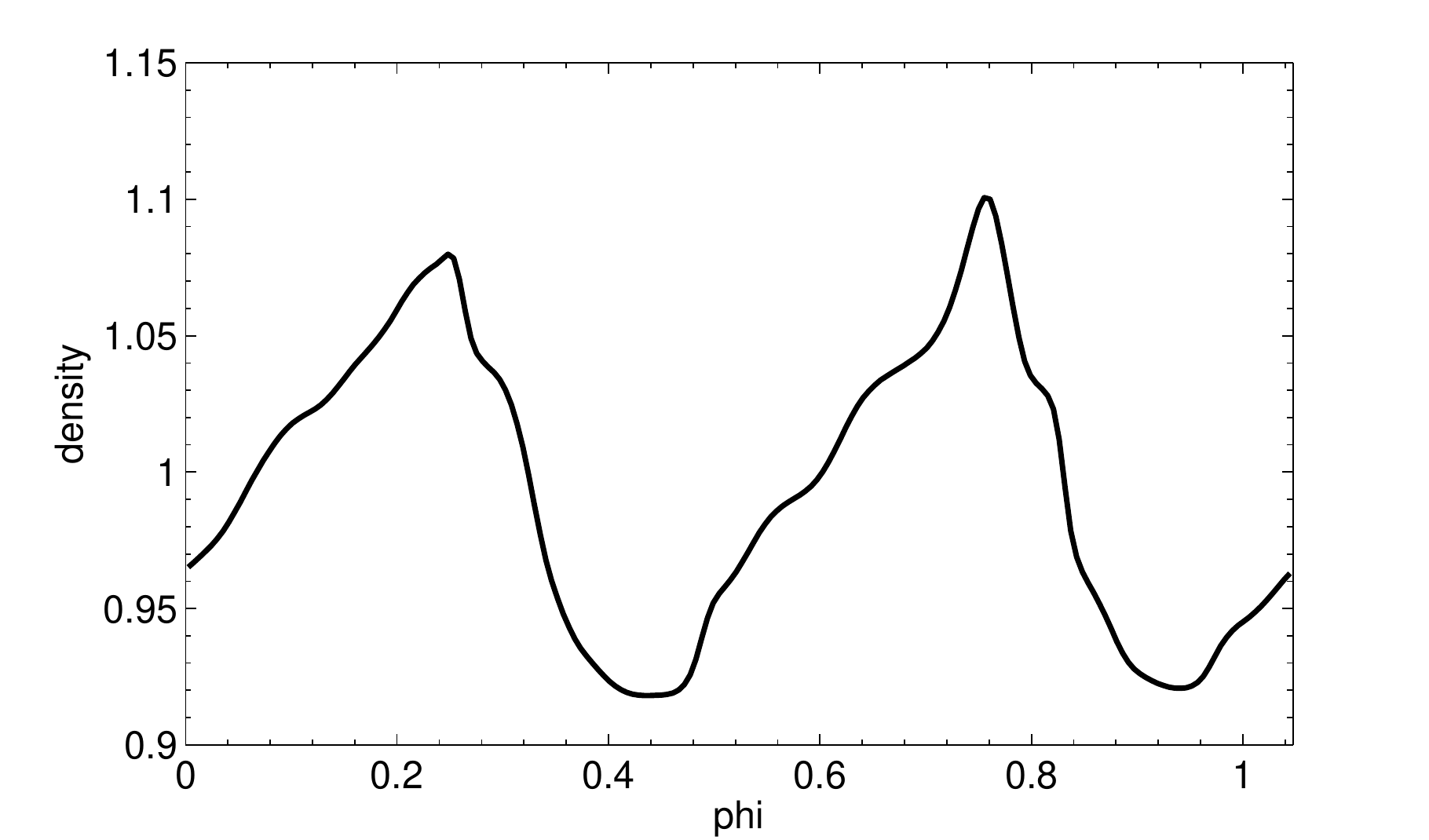}}
\subfigure[]{\includegraphics[width=0.49\textwidth]{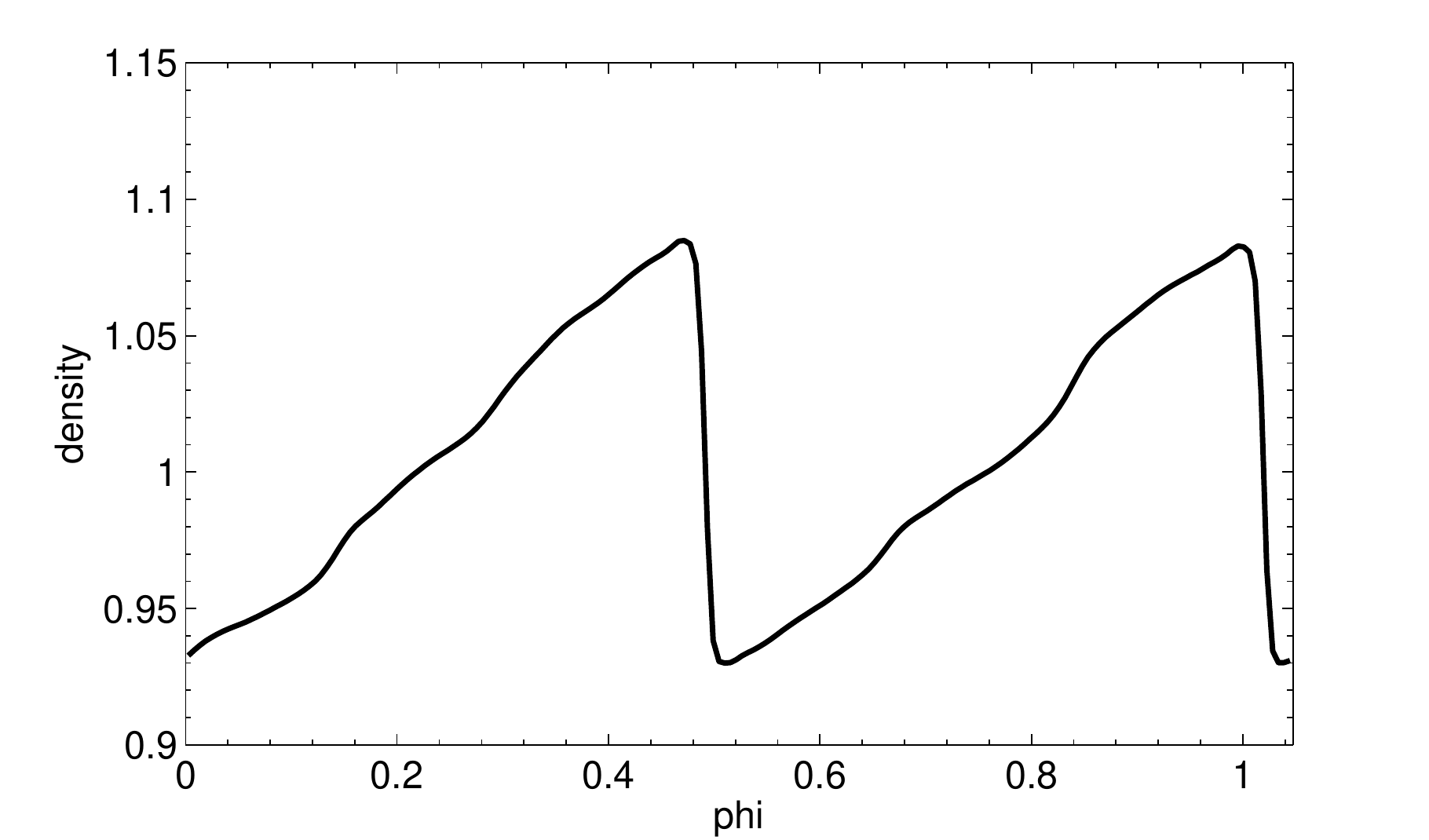}}
\caption{Azimuthal profiles of the wave before shocking (panel a) at
  $\cp = 1.5$ and after shocking (panel b) at $\cp = 1.9$.}
\label{shockprofiles}
\end{figure}

We also point out that in Fig. \ref{CLdiskfig}  the flux is
approximately constant between the edge of the evanescent region and
$\cp \approx 1.7$, but then starts to decrease more rapidly with
radius. This reduction in flux is caused by shocking of the wave, 
which is illustrated in Fig. \ref{shockprofiles}. In this Figure 
we show the profile of the density perturbation $\Delta\Sigma/\Sigma$ 
in azimuthal direction, i.e. at fixed $\cp$, before
(a, at $\cp= 1.5$) and after (b, at $\cp = 1.9$) breaking 
for simulation 3D9d at $t=75$. It is obvious that in 
the latter case the wave has shocked, with clear
discontinuity in the wave profile, which converges towards the 
``N-wave'' shape \citep{LL}. Note that the analytically-expected 
value of $C_S$ (equation [\ref{CLdisk}]), automatically takes into
account the dissipation of the wave due to shocking, through the
factor of $(\Delta \Sigma)^2$.

\begin{figure}[!h]
\centering
\includegraphics[width=0.7\textwidth]{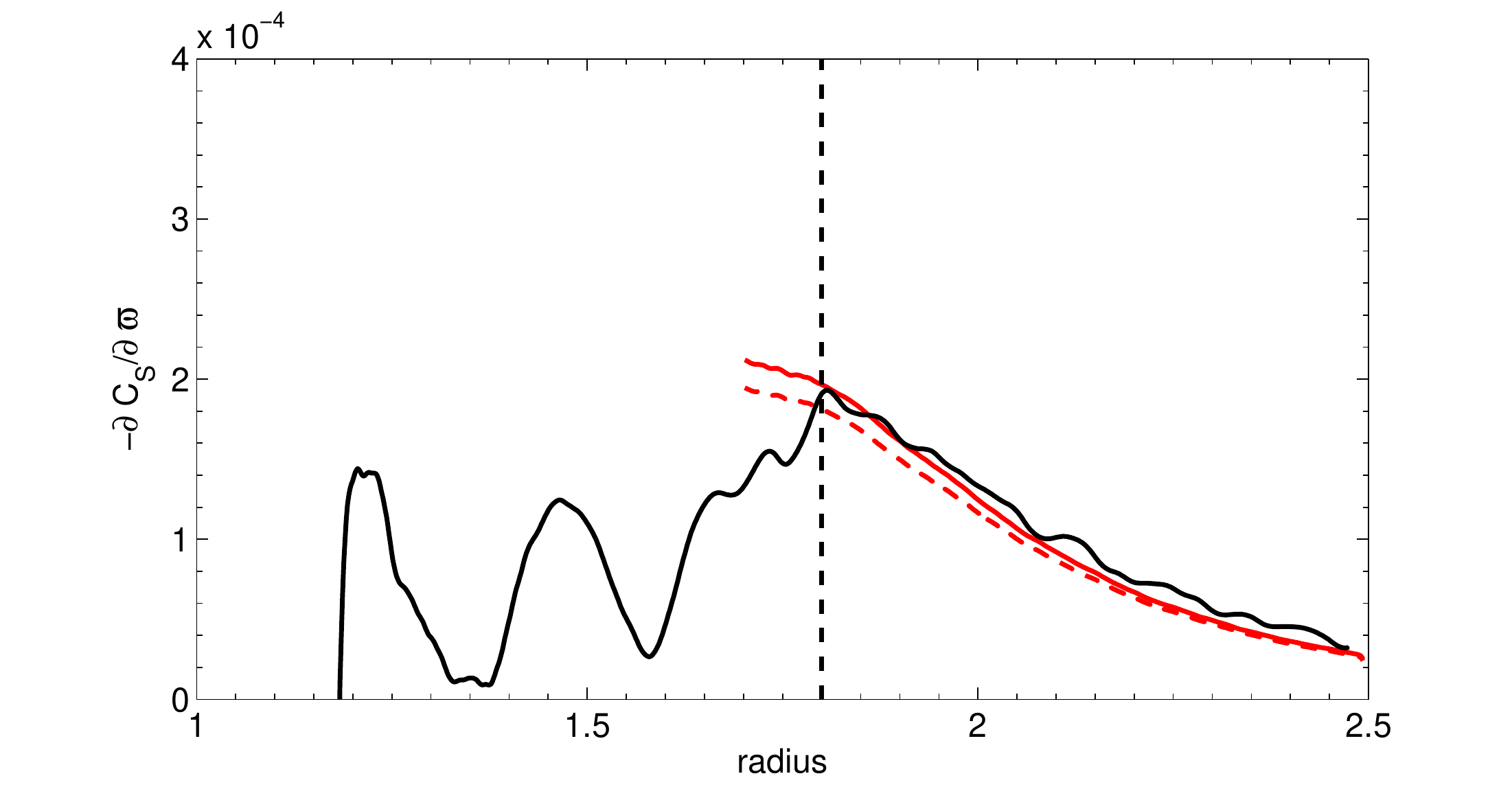}
\caption{Plot of $-\partial C_S/ \partial \cp$ for simulation 3D9d at time
  $t=75$. The dashed black line at $\cp  = 1.8$ denotes the radius
  at which the entire wave has shocked.}
\label{dCLdrfig}
\end{figure}

We now turn to the reduction of the wave flux with radius after the wave has
shocked. Fig. \ref{dCLdrfig} shows $-\partial C_S/\partial \cp$ as a function of radius
for simulation 3D9d at $t=75$. The black curve shows the value of
$-\partial C_S/ \partial \cp$ computed from the simulation data, where $C_S$ is
calculated using equation (\ref{CSgen}). The red curves are the
analytical expectations for $-\partial C_S/\partial \cp$ due to wave
dissipation by shocks calculated using equation (\ref{dEdmeq}). The
dashed red curve uses the value of $dE/dm$
from equation (\ref{dEdmapprox}), and the solid red curve uses the value of
$dE/dm$ from equation (\ref{dEdmiso}). 

In measuring $\Delta \Sigma/
\Sigma$ in order to calculate $dE/dm$, we find it necessary to correct
for the finite width of the
shock in the simulation. This is because the shock in our simulations
extends over several cells and is not a true discontinuity. Thus, we
extrapolate the shock density profile from a simulation and
estimate what the value of $\Delta \Sigma/
\Sigma$ would be for a true discontinuity. For our resolution, this correction
increases $\Delta \Sigma/\Sigma$ by $\approx 8 \%$. Given the stiff
dependence of $dE/dm$ on $\Delta \Sigma/\Sigma$ (leading order
dependence is $(\Delta \Sigma/ \Sigma)^3$, see equation 
(\ref{dEdmapprox})), this correction is
necessary to obtain good agreement between theory and simulations.

The agreement between the measured value of $\partial C_S/\partial \cp$ and the
theoretically-predicted one indicates that
angular momentum radiated away from the evanescent region into the disk is
redistributed by shock dissipation. This dissipation is most
vigorous after the wave has completely shocked around $\cp \approx
1.8$, but the steepest parts of the wave may shock even 
earlier at $\cp < 1.8$, so the wave angular momentum starts to 
be redistributed to the disk fluid even before the entire
wave has shocked. Since $\Omega_P > \Omega$
for the upper branch outside the evanescent region in the disk, the
wave carries {\it positive} angular momentum, and dissipation of the wave
{\it spins up} the disk fluid, pushing matter {\it outward}, see 
\S \ref{sect:Mdot}.


\subsubsection{Star}
\label{starcl}

Radiation from the BL is emitted not only into the disk, but into the
star as well. The amplitude of acoustic waves traveling into the star
is proportional to $\rho^{-1/2}$, so they don't shock in our
simulations and propagate out through the inner boundary of the
simulation domain.

The inner radial ``do nothing'' boundary condition is not perfectly
absorbing, so waves incident upon it are reflected with a reflection
coefficient 
\ba
\label{epseq}
\eps_R \equiv \frac{\delta \Sigma_R}{\delta \Sigma_\text{out}}.
\ea
Here, $\delta \Sigma_\text{out}$ is the amplitude of the outgoing
  wave (with respect to the BL), and $\delta \Sigma_R$ is the
  amplitude of the reflected
  wave. As long as $\eps_R$ is not close to 1,
the reflected waves will not fundamentally affect the dynamics of the
instabilities in the boundary layer, since the ``overreflection''
mechanism described in \citet{BR} only works if $\eps_R \approx
1$. However, the angular momentum current due to stresses measured by equation
(\ref{CSgen}) will underestimate the true current, since the
measured current contains components due to both the true outgoing
waves (with respect to the BL) and the artificial reflected waves from
the inner boundary. Nevertheless, the flux due to only the outgoing
component, $C_{S,\text{out}}$, can be estimated in terms of the measured
flux, $C_{S,\text{meas}}$, as 
\ba
\label{CLcorr}
C_{S,\text{out}} = \frac{C_{S,\text{meas}}}{1 - \eps_R^2}.
\ea
This equation can be derived by considering a linear superposition of
an outgoing wave and the reflected wave. The reflected wave has the
opposite sign of $C_S$ as the outgoing wave and an amplitude that is
second order in $\eps_R$, since angular momentum flux is a second
order quantity.

\begin{figure}[!h]
\centering
\subfigure[]{\includegraphics[width=0.7\textwidth]{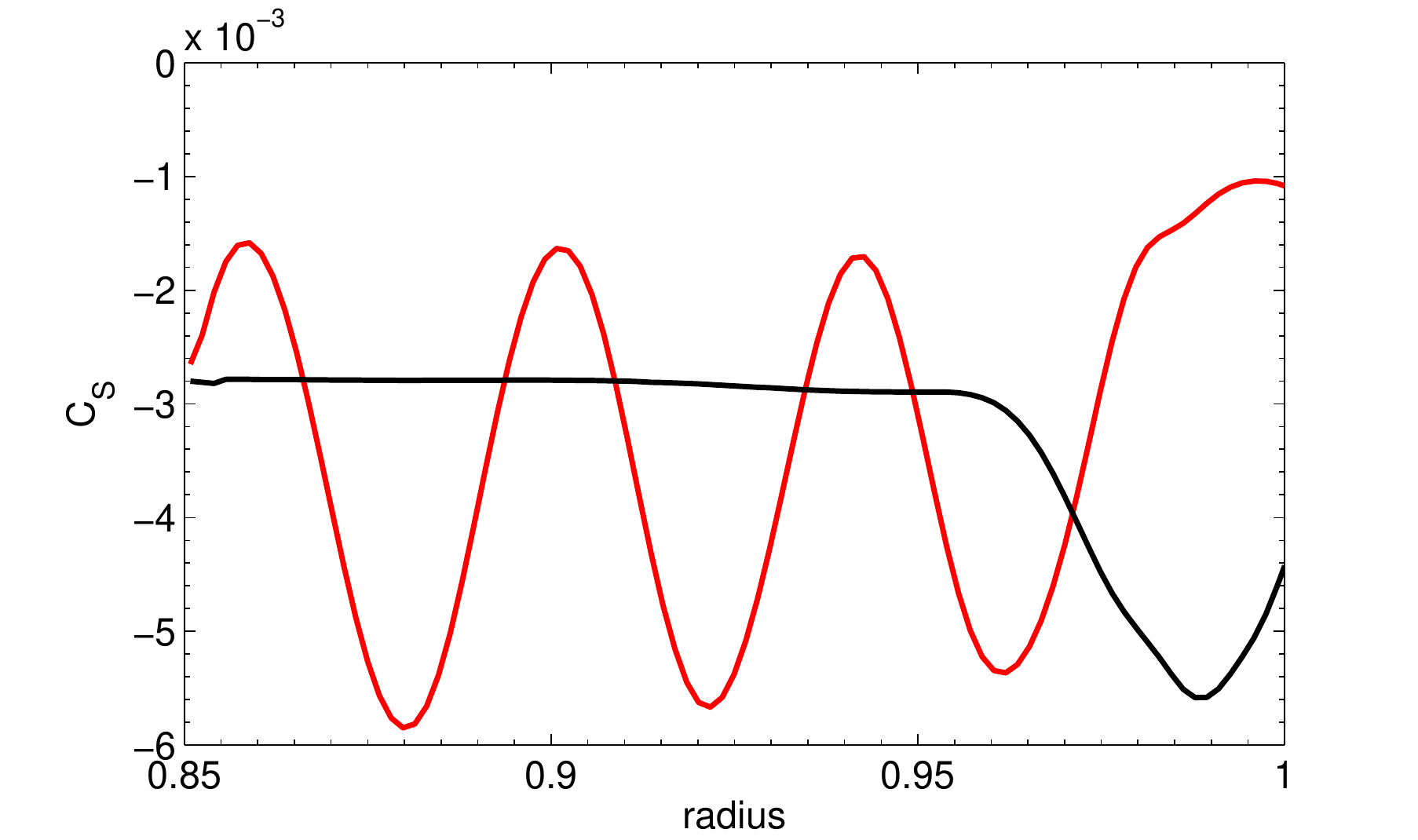}}
\subfigure[]{\includegraphics[width=0.7\textwidth]{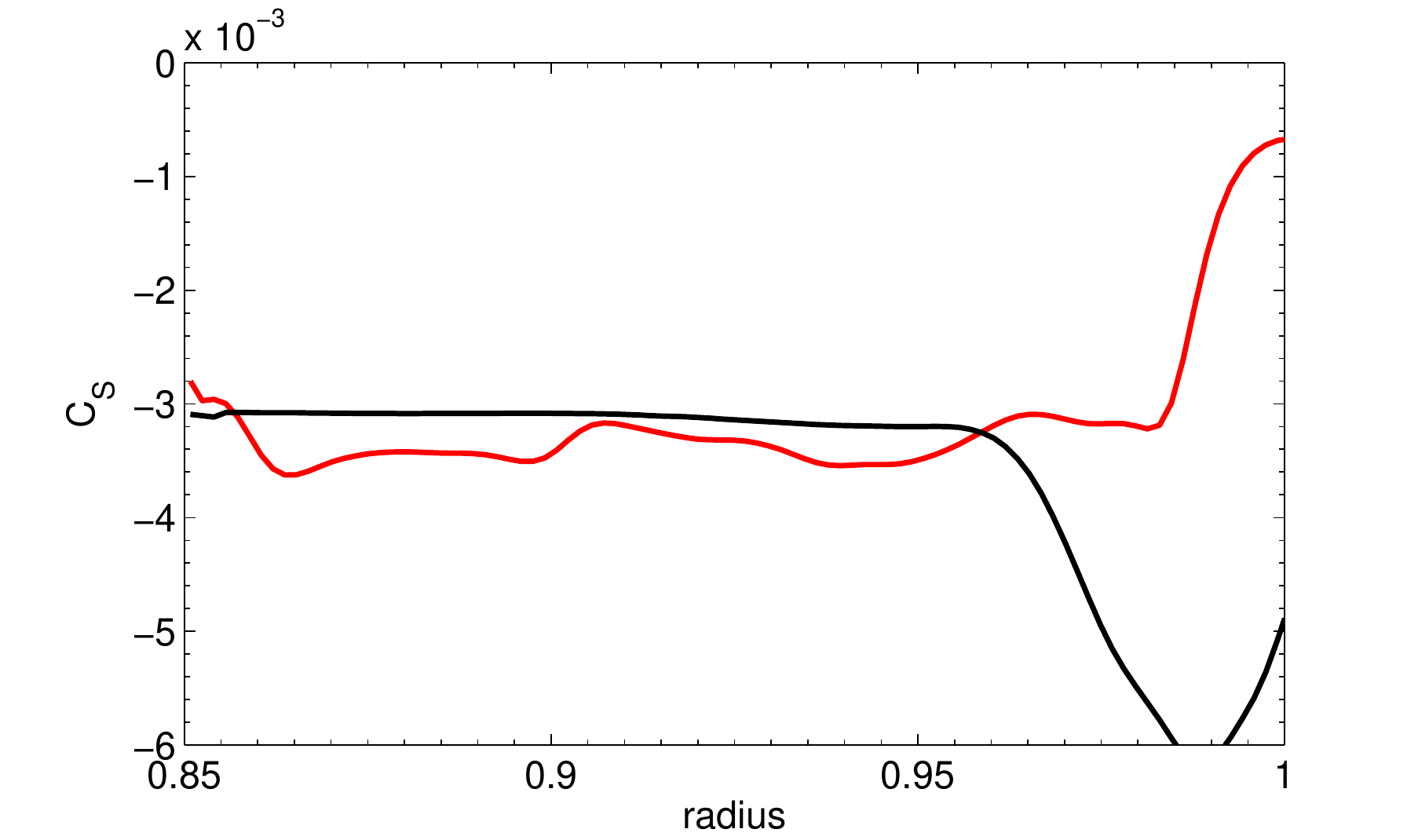}}
\caption{a) Uncorrected current due to stresses inside the star computed using
  equations (\ref{CSgen}) (black line) \& (\ref{CLstar}) (red
  line). b) Corrected current due to outgoing waves computed using
  equations (\ref{CLcorr}) (black line) \& (\ref{CLcorr1}) (red line).} 
\label{CLstarfig}
\end{figure}

Fig. \ref{CLstarfig}a shows the measured current (black line)
calculated using equation
(\ref{CSgen}), and the analytically-predicted current due to waves (red
line) calculated using equation (\ref{CLstar}). The
analytically-predicted current
oscillates around the measured current, which is roughly constant for
$\cp \lesssim .96$. These
oscillations can be explained by the presence of a reflected wave from
the inner boundary with amplitude $\eps_R \approx .3$. This amplitude can be
estimated from the straightforward-to-derive relation
\ba
\eps_R = \frac{C_{S,\text{max}}^{1/2} -
  C_{S,\text{min}}^{1/2}}{C_{S,\text{max}}^{1/2} +
  C_{S,\text{min}}^{1/2}},
\ea
where $C_{S,\text{max}}$ and $C_{S,\text{min}}$ are the maximum and
minimum values of the oscillations in the red, analytically predicted
flux curve in Figure \ref{CLstarfig}a. 

The reason for the oscillations in the red curve of
Fig. \ref{CLstarfig}a is that equation (\ref{CLstar}) is an accurate
estimator of the current due to waves if there are {\it only}
outgoing waves present; it does not account for reflected
waves. However, it is straightforward to modify equation (\ref{CLstar})
so that it estimates the current due to only the outgoing component. To
do this, we assume that there is a single outgoing mode present with a
constant value of $k_\cp$ that
is reflected at the inner boundary into an incoming mode with the
same value of $k_\cp$ and an amplitude which is a fraction
$\epsilon_R$ of the amplitude of the outgoing mode. Under these
assumptions, the current
due to only the outgoing component, $C_{S,\text{out}}$, can be
expressed in terms of the uncorrected current given by equation (\ref{CLstar}),
$C_{S,\text{unc}}$, as
\ba
\label{CLcorr1}
C_{S,\text{out}} = \frac{C_{S,\text{unc}}}{1 - 2\eps_R \cos(2 k_\cp
  \cp + \psi) + \eps_R^2},
\ea
where $\psi$ is a free phase parameter.

The red curve in Fig. \ref{CLstarfig}b shows $C_{S,\text{out}}$
as computed from equation (\ref{CLcorr1}), and the black curve shows
$C_{S,\text{out}}$ according to equation (\ref{CLcorr}). The
oscillations in the red curve have all but disappeared after applying the
correction to compute the flux from only outgoing waves. Moreover, the red
and black curves are seen to be roughly constant for $\cp \lesssim
.96$ and to have very similar amplitudes. This proves that in the
star, just as in the disk, angular momentum is carried away from
the BL by waves. However, unlike in the disk, the waves do not shock as
they penetrate into the star, but rather diminish in amplitude to
conserve energy and angular momentum.

Comparing the scales on Figs. \ref{CLdiskfig} and \ref{CLstarfig}b, we
see that the angular momentum flux into the star is larger than the
angular momentum flux into the disk by an order of magnitude. This means that
most of the angular momentum lost by the inner part of the disk goes
into spinning up the star and only a small fraction is transported
outward into the disk.


\subsection{Lower Branch}
\label{uldisk}

\begin{figure}[!h]
\centering
\subfigure[]{\includegraphics[width=0.7\textwidth]{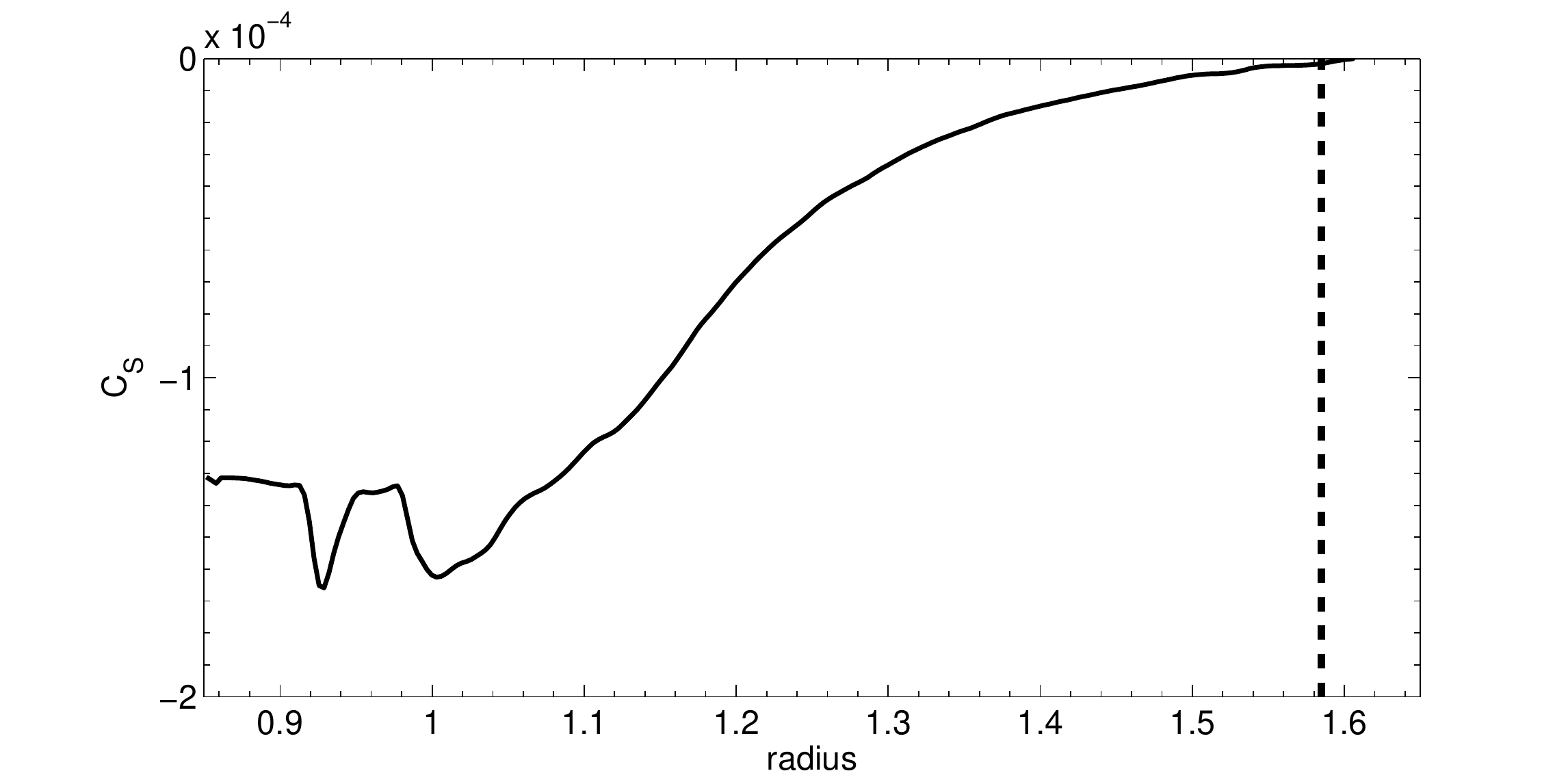}}
\subfigure[]{\includegraphics[width=0.7\textwidth]{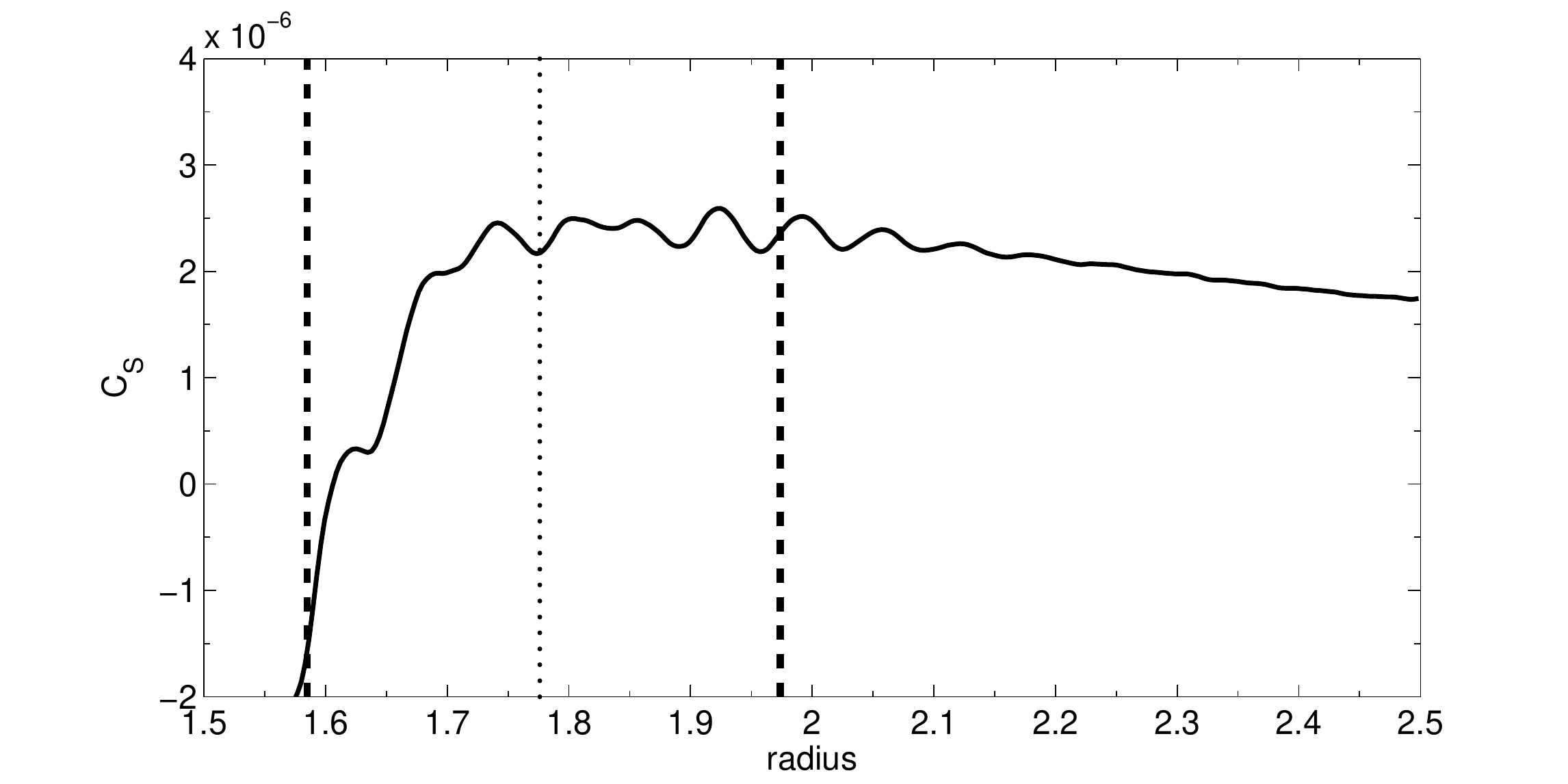}}
\caption{a) $C_S$ for simulation 3D9e at time $t=320$ inside the inner
  edge of the
  evanescent region in the disk. The dashed line marks the inner edge
  of the evanescent region in the disk b) Same as a, but $C_S$ is
  shown within the evanescent region and beyond it. The evanescent
  region itself is located within the dashed lines and the dotted line
  is the corotation radius.} 
\label{lowerbranchcl}
\end{figure}

We turn now to angular momentum transport for the lower
branch. Fig. \ref{lowerbranchcl} shows $C_S$ calculated using equation
(\ref{CSgen}) for simulation 3D9e at time $t=320$;
the same time as Fig. \ref{morphfig}b, when the lower branch is clearly
dominant. Panel a of Fig. \ref{lowerbranchcl} depicts $C_S$ from the
inner edge of the simulation domain up to
the inner edge of the evanescent region in the disk at $\cp
= 1.58$. Panel b, which has a different scale than panel a, depicts
$C_S$ from just inside the inner edge of the evanescent region to the
outer edge of the simulation domain. The dashed lines in panel b
depict the edges of the evanescent region, and the dotted line is the
corotation radius ($\Omega_P \approx .42$). We point out that although
the magnitude of $C_S$ for the lower mode in Fig. \ref{lowerbranchcl} 
is smaller
than the magnitude of $C_S$ for the upper mode in Fig. \ref{CLstarfig}, the
amplitude of either mode is not constant in time. In fact the
magnitude of $C_S$ inside the star
may be as high for the lower mode as for the upper mode at earlier times
(see Fig. \ref{stressspacetime} in \S \ref{transevol}).

In contrast to the upper branch, it is more difficult to estimate the
current due to waves using equations (\ref{CLdisk}) and (\ref{CLstar}) in
both the star and the disk for the lower branch. In the disk, the
difficulty comes about due
to the fact that there are both outgoing and reflected shocks between
the surface of the star and the Lindblad resonance in the disk
(Fig. \ref{morphfig}b). Equation (\ref{CLdisk}) only applies if there
are exclusively outgoing waves, and it is not straightforward to
modify it to account for interference effects from reflected
waves. Without modification, the plot of the analytically predicted
current, $C_S$, from equation (\ref{CLdisk}) would show large spikes at
shock crossings, and only in between shock crossings when
the outgoing and reflected waves are well-separated is there reasonable
agreement between the measured and analytically predicted values of
$C_S$ (see Fig. 5 of \citet{BRS}).

The complication with applying equation (\ref{CLstar}) to analytically
predict $C_S$ inside the star is
that $k_\cp \approx 0$ (to zeroth order in $M^{-1}$) for the lower mode
(\S \ref{lowsec}), which makes the
current estimated from equation (\ref{CLstar}) equal to zero. Thus, in
order to use equation (\ref{CLstar}) we would need to know $k_\cp$ to first
order in $1/M$, and we have chosen not to pursue such a calculation in 
this work.

One thing to notice in Fig. \ref{lowerbranchcl} is that $C_S$
vanishes close to the edge of the
evanescent region in the disk. This is to be expected if we have near
perfect reflection, since $C_S=0$ at the point of reflection if the
amplitudes of the
outgoing and reflected shocks are the same. $C_S \ne 0$
throughout the entire region between the BL and the inner edge of the
evanescent region due to shock dissipation, which causes the shock
amplitude, and consequently $|C_S|$, to diminish with increasing
path length traveled by the shock. This makes $C_S < 0$ between the BL and the
point of reflection, because the shock pattern has a negative angular
momentum density inside corotation, and the outgoing shocks, which
have a larger amplitude than the incoming ones, propagate in the
$+\cp$ direction.

We can estimate a lower bound on the reflection coefficient from the
evanescent region by comparing the scales of
Figs. \ref{lowerbranchcl}a and \ref{lowerbranchcl}b. The absolute
value of $C_S$ (for outgoing shocks only) is $\sim 100$ times lower
outside the evanescent region than inside it. Since $C_S$ is a second
order quantity in $\delta \Sigma$, the reflection coefficient as
defined in equation (\ref{epseq}) is
$\eps_R > .9$. This is only a lower bound on the reflection coefficient,
since outside the evanescent region $C_S$ could be dominated by an
upper branch component.  

We also point out that
$C_S$ changes sign within the evanescent region close to the
corotation radius in the disk ($\cp_\text{CR} = 1.78$), as expected
for angular momentum transport by waves. $C_S$ does not change sign at
the corotation radius located in the BL at $\cp \approx 1$, since
both the direction of wave propagation, and the sign of the
angular momentum density change there. In other words, inside
corotation at $\cp \approx 1$ the angular momentum density is
positive, but waves propagate in the $-\cp$ direction so $C_S <
0$. Just outside it, the angular momentum density is negative 
(since $\Omega_P<\Omega$ there), but waves propagate in the 
$+\cp$ direction so $C_S < 0$ as well.


\subsection{Relation of Angular Momentum Transport to Temporal Evolution}
\label{transevol}

We now discuss how angular momentum transport in our simulations
controls the temporal evolution of the system. One can use a
characteristic value of $|C_S| \sim 3 \times 10^{-3}$ taken from
Fig. \ref{CLstarfig} to
estimate a typical ``evolution time'', for the system when the
upper branch is dominant. We define $t_\text{evol}$ as the time it
takes for the evanescent
region (upper branch) in the disk to lose half its angular momentum. Taking the
the evanescent region for upper branch to extend from $1 < \cp <
1.25$, we estimate
\ba
t_{\text{evol}} \sim 300.
\label{tevol}
\ea  
 
\begin{figure}[!h]
\centering
\includegraphics[width=0.9\textwidth]{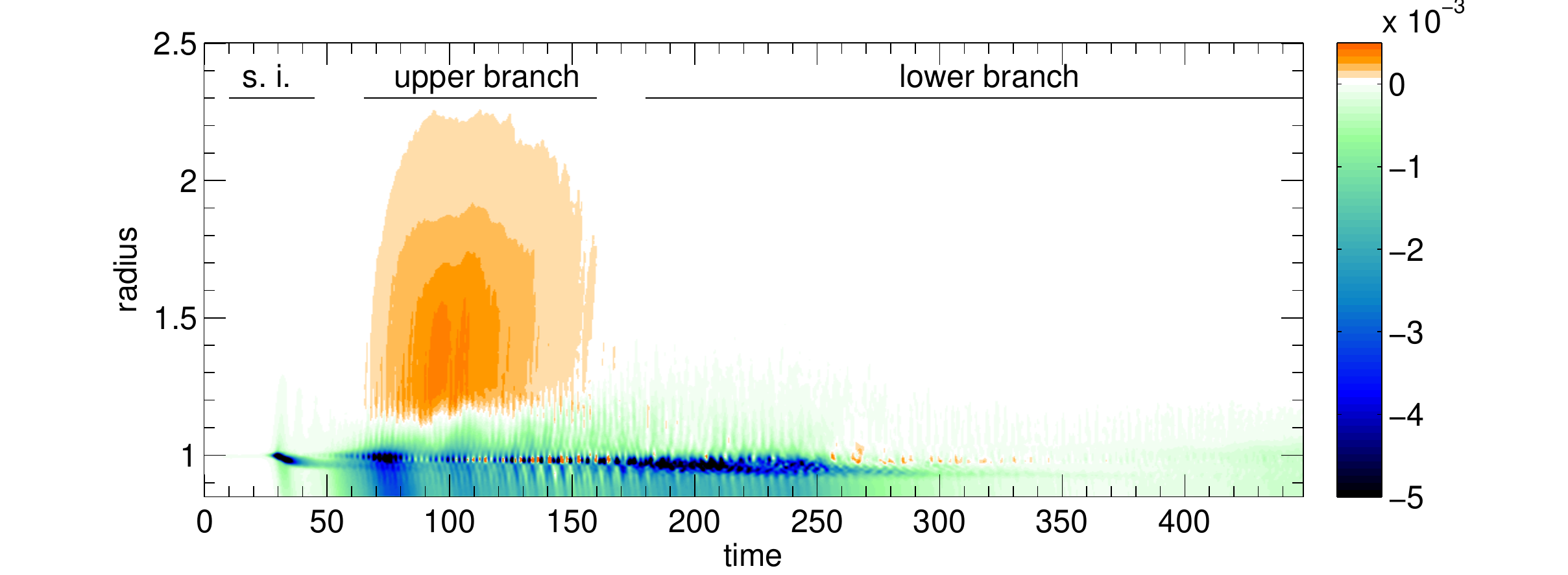}
\caption{Radius-time image of $C_S$ for simulation 3D9a. The
  horizontal bars show
the periods of time when the sonic instabilities are active, the upper branch
is dominant, and the lower branch is dominant.} 
\label{stressspacetime}
\end{figure}

Fig. \ref{stressspacetime} shows a radius-time plot of
$C_S$ for simulation 3D9a. From the figure, it is clear that
$t_{\text{evol}}$ is within a factor of several of the span of time
during which the upper branch is dominant. Moreover, we see as well
from Fig. \ref{stressspacetime} that $t_\text{evol}$ is also a good
estimator of the timespan during which there is a significant angular
momentum current radiated away from the BL and the accretion rate is
highest.

The fact that the evolution time is comparable to both the timespan
over which the upper branch is dominant and the timespan for
significant mass accretion in the simulations suggests the
following scenario for the temporal evolution of the system. The
amplitude of the upper branch is initially
high, but diminishes when it
has depleted the evanescent region in the inner part of the disk of
angular momentum. The system then switches to lower branch behavior
with continuing mass accretion. However, without a mechanism of
resupply for the inner disk,
which does not exist in our simulations, the rate at which the inner disk is
depleted of angular momentum must drop. Thus, the angular momentum
current and the mass accretion rate both plummet for $t
\gtrsim t_\text{evol}$ as seen in Figs. \ref{spacetimedensity}b and
\ref{stressspacetime}. This occurs in step with a reduction in the
amplitude of the acoustic modes, which are responsible for the angular momentum
transport.

In contrast to our simulations, there does exist a mechanism of
resupply for the inner disk in an astrophysical system,
namely accretion due to turbulent stresses. These stresses are not
present in our simulation, since we do not have magnetic fields, which
lead to the MRI instability in the disk. However, one may speculate
that if the accretion time due to turbulent stresses were comparable to
or shorter than $t_\text{evol}$, then the acoustic modes would
operate at a high amplitude indefinitely. Indeed, the amplitude of the
modes themselves could potentially be controlled by the time for MRI
to resupply the inner part of the disk.


\subsection{Relation of Angular Momentum Transport to Mass Accretion Rate}
\label{sect:Mdot}

Angular momentum carried by the waves launched in the BL is eventually
deposited into the disk fluid as a result of wave dissipation. In this 
section we relate the angular momentum transport rate to the
mass accretion rate through the disk, $\dot{M}$, defined in 
equation (\ref{Mdotdef}). 

We can use equations (\ref{1Dcont}) and (\ref{1Dangmom}) to write
$\dot{M}$ in terms of $C_S$ as 
\ba
\label{Mdoteq}
\dot{M} = \left[\frac{\partial}{\partial \cp}\left(\Omega
  \cp^2\right)\right]^{-1}\left[2 \pi \cp^3 \Sigma_0 \frac{\partial
    \Omega}{\partial t} + \frac{\partial C_S}{\partial \cp} \ \right].
\ea
Time evolution of $\Omega$ represented by the first term on 
the right hand side forces gas to move radially, since 
to conserve angular momentum fluid elements must readjust their 
radial distance, $\cp$, when $\Omega$ varies. The $\partial
  \Omega/\partial t$ term in equation (\ref{Mdoteq})
  vanishes for a steady state disk, but turns out to be important 
for our simulations, since the rotation profile evolves in time 
(Fig. \ref{densomega}). 

\begin{figure}[!h]
\centering
\includegraphics[width=0.7\textwidth]{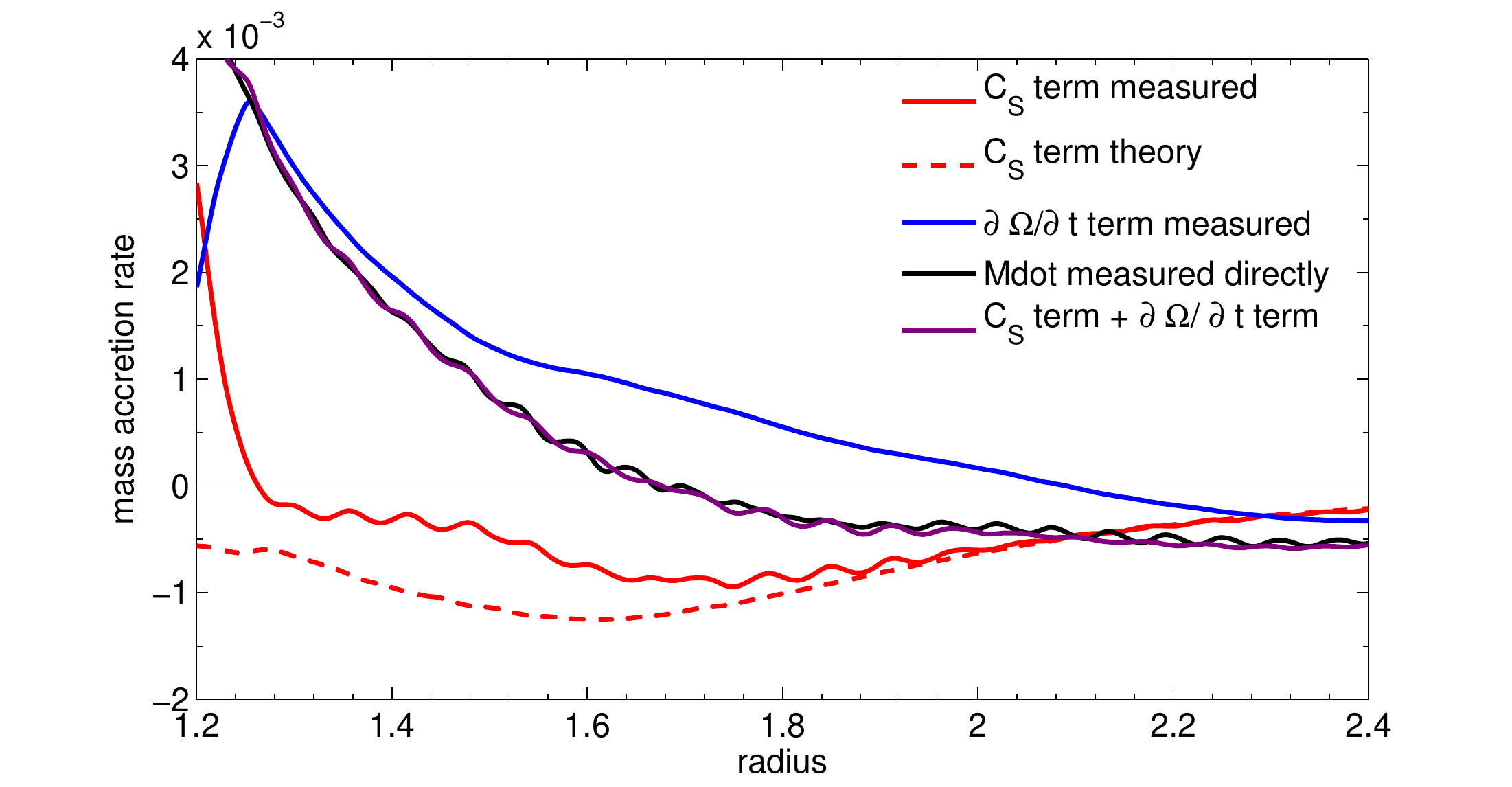}
\caption{Plot of $\dot{M}$ measured from the simulations (black line)
and the contributions to $\dot{M}$ from the $\partial C_S/\partial \cp$
(solid red line) and $\partial \Omega/\partial t$ (blue line) terms in equation
(\ref{Mdoteq}). The dashed red line shows the theoretical estimate for
the $\partial C_S/\partial \cp$ term assuming wave dissipation in shocks.} 
\label{Mdotfig}
\end{figure}

We plot in Fig. (\ref{Mdotfig}) the mass accretion rate in the BL 
as a function of radius for simulation 3D9a averaged between 
times $t=70-100$, when the upper mode with high $\Omega_P$ 
clearly dominates. The solid red
and blue curves show the contributions of the $\partial C_S/ \partial
\cp$ and $\partial \Omega/\partial t$ terms, respectively, to the mass
accretion rate
(equation [\ref{Mdoteq}]), as measured from the simulation, and the black
curve shows the total mass accretion rate measured directly from the
simulation using equation (\ref{Mdotdef}). It is clear that both the
$\partial \Omega/ \partial t$ and $\partial C_S/ \partial \cp$ terms
contribute significantly to the mass accretion rate, even far out in
the disk. As a consistency check, the sum of the solid red and blue
curves is shown in purple and is seen to equal the black curve up to small
wiggles that are artifacts of the measurement procedure.

Fig. \ref{Mdotfig} clearly shows that the $\partial\Omega/\partial t$
term is important for the calculation of $\dot M$. In our case, it is
significant because the inner disk is depleted of mass as a result of
angular momentum transport by acoustic modes. This causes 
time-varying density and pressure gradients in the disk, 
which is significant even quite far from the BL, at separations 
$\sim\cp_\star$ from it. Time evolution of this pressure support 
ultimately causes explicit time dependence of $\Omega$ and drives 
accretion even in the region where the wave has not shocked yet
(interior to $\cp\approx 1.8$ in Fig. \ref{Mdotfig}).

Prior to its dissipation the wave propagating through the disk 
cannot affect its state \citep{GN89}. Only after it shocks, 
transfer of its angular momentum to the disk drives additional 
mass accretion.
The dashed red curve in Fig. \ref{Mdotfig} shows the theoretically
predicted value for the component of mass accretion due to the
$\partial C_S/ \partial \cp$ term in equation (\ref{Mdoteq}), assuming
the $\partial C_S/ \partial \cp$ term is entirely due to shock 
dissipation. In other words, $\partial
C_S/\partial \cp$ is computed using equation (\ref{dEdmeq}), where we use
equation (\ref{dEdmiso}) to calculate $dE/dm$, and measure $\Delta
\Sigma/\Sigma$ in the same way described in \S \ref{ubsecdisk}. We
point out that the
formula for the $\partial C_S/\partial \cp$ term due to shocks for the
lower mode was given in Appendix B of \citet{BRS}, and the derivation
for the upper mode is nearly identical, see \S \ref{angmomsec}. 

The good agreement between
the solid and dashed red curves for $\cp \gtrsim 1.8$ in
Fig. \ref{Mdotfig} shows that the
$\partial C_S/\partial \cp$ component of the mass accretion rate is
well-described by dissipation in shocks. For smaller values of $\cp$,
the waves emitted from the BL have not yet fully shocked (\S
\ref{ubsecdisk}), so the agreement is not as good.

It is interesting to note that the mass accretion rate due to wave
dissipation is negative in Fig. \ref{Mdotfig}, i.e. the disk material is 
{pushed away} from the BL beyond the point where the wave has shocked. 
This is due to the fact 
that Fig. \ref{Mdotfig} corresponds to a time when the upper branch is
dominant. As discussed in \S \ref{uppersec}, the upper branch has a fast
pattern speed with a corotation radius in the disk at $\cp_{\text{cr}} \approx
1.12$ for this simulation. Outside of corotation ($\cp >
\cp_{\text{cr}}$), $\Omega_P >
\Omega$ so the wave has a positive angular momentum density. Thus,
dissipation of the wave {\it increases} the angular momentum of the fluid
elements in the disk, causing them to move {\it outwards} and resulting in
a {\it negative} mass accretion rate. 

The situation is different for the
lower and middle branches. Unlike the upper branch which has an evanescent
region directly adjacent to the boundary layer, the middle and lower
branches have their evanescent regions in the disk separated from the
boundary layer. Waves can propagate between the BL and the evanescent
region in the disk. Since $\Omega_P < \Omega$ in this region, this
implies that the wave has {\it negative} 
angular momentum density and dissipation of
the wave results in a {\it positive} mass accretion rate. Beyond the
evanescent region in the disk, $\Omega_P > \Omega$ for both the lower
and middle branches so the mass accretion rate due to wave dissipation
is in principle negative in this region, just as for the upper
branch. However, very little flux penetrates the
evanescent region (\S \ref{uldisk}), so the contribution to mass accretion
from wave dissipation beyond the evanescent region in the disk is very
small for the lower and middle branches.


\section{Effect of Stratification on Acoustic Modes}
\label{stratsec}


Up to now we have discussed the results of unstratified
simulations, and we find that {\it density} stratification does not produce
fundamentally different behavior. In particular, the upper, lower, and middle
branches are all observed in stratified simulations, as well, and facilitate
angular momentum transport.  

\begin{figure}[!h]
\centering
\subfigure[]{\includegraphics[width=0.49\textwidth]{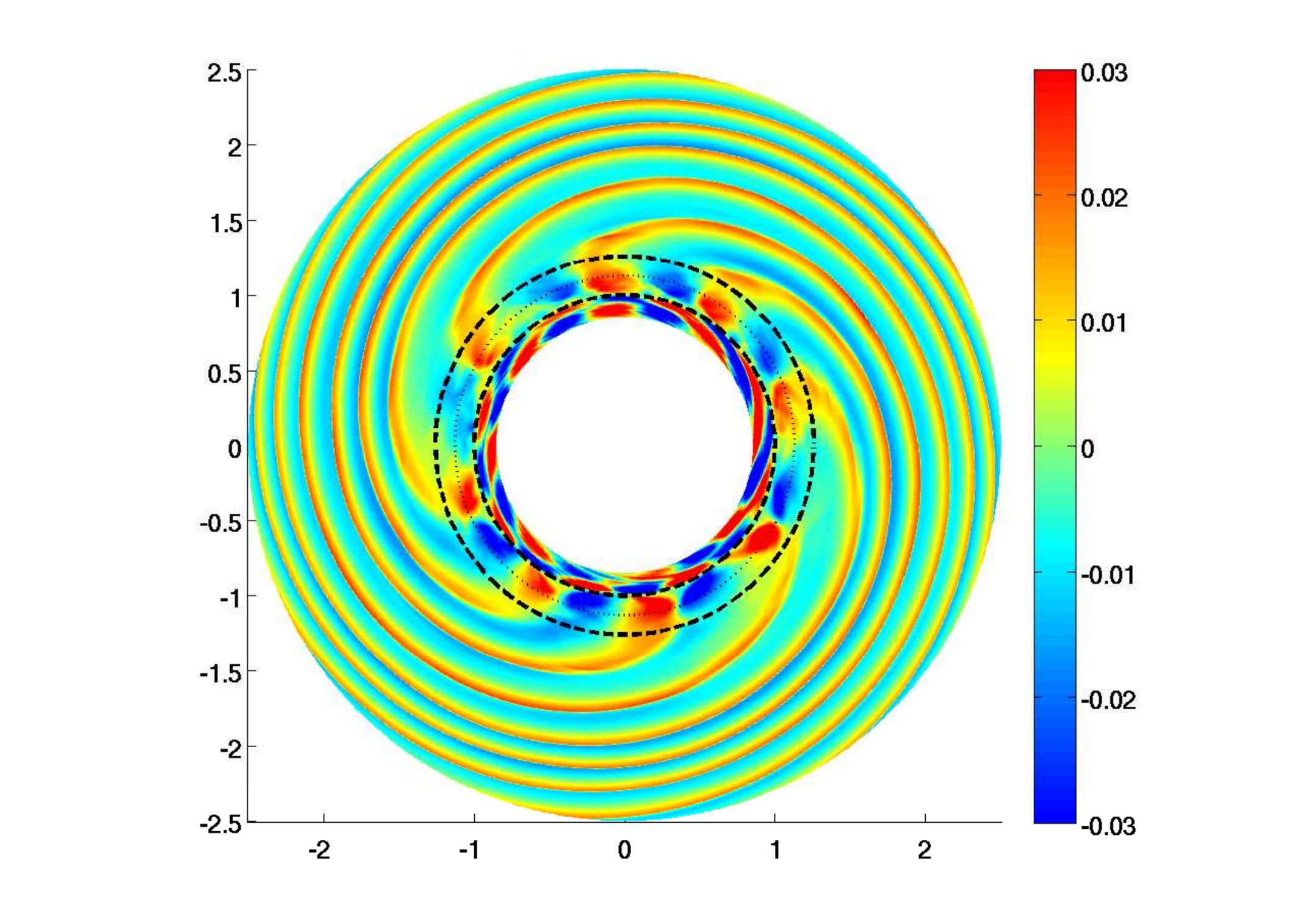}}
\subfigure[]{\includegraphics[width=0.49\textwidth]{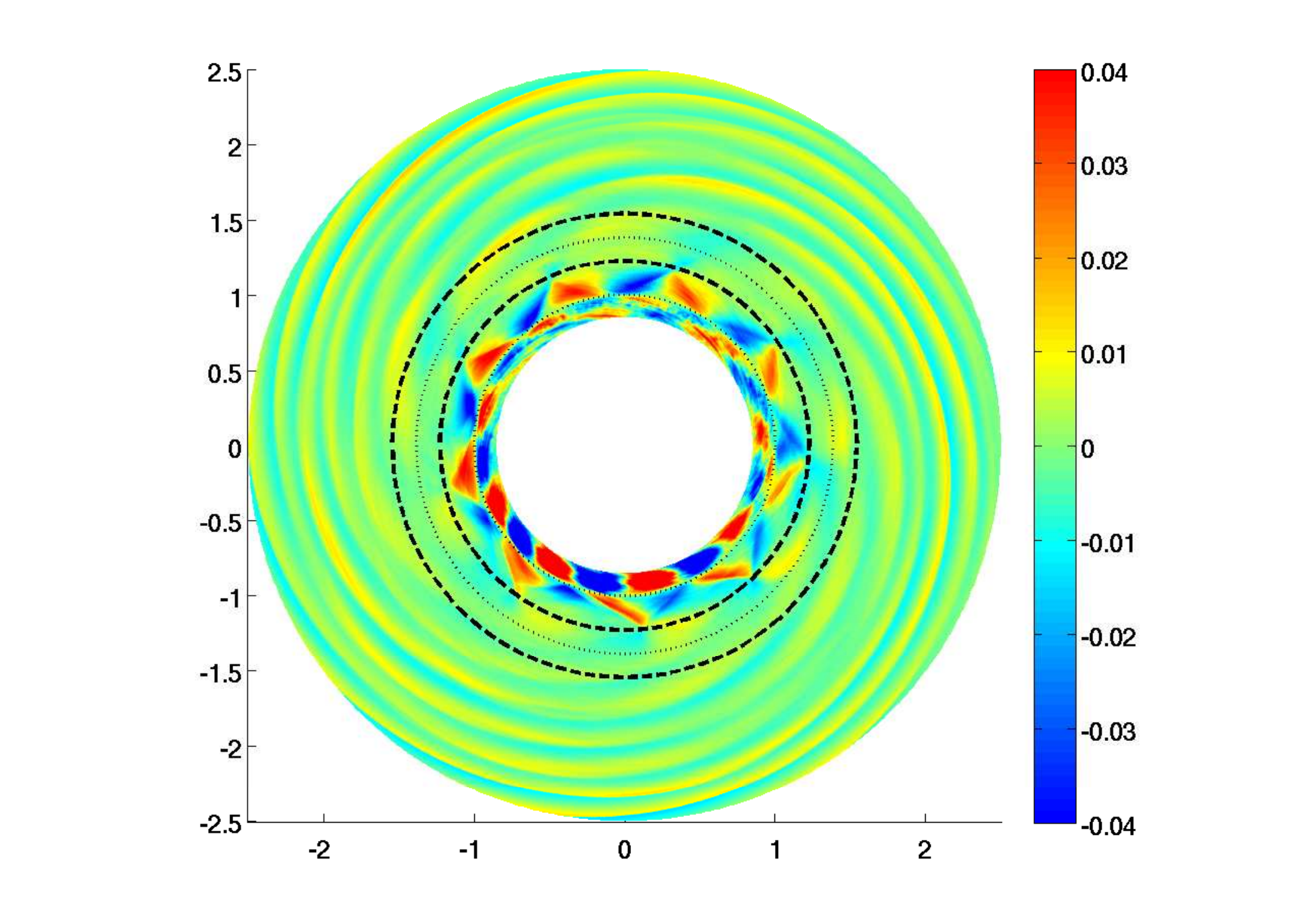}}
\caption{Panels a and b show snapshots of $\cp \sqrt{\Sigma} \La
  v_\cp \Ra_z$
  when the upper and lower waves are dominant, respectively, although
  there appears to
  be a superposition of both types of waves in the panels. The dashed and
  dotted circles depict the boundaries of evanescent regions
  and corotation radii, respectively. In panel a, the evanescent
  region in the disk extends all the way to the BL.}
\label{stratfig}
\end{figure}

Panels a and b of Fig. \ref{stratfig} show snapshots of the upper and
lower branches for simulation 3D9c at times $t=160$ and $t=320$ in the
color variable $\cp \sqrt{\Sigma} \La v_\cp \Ra_z$. Comparing
Fig. \ref{stratfig} with Fig. \ref{morphfig}, we see that the spatial
morphology of the wave branches is the same in the stratified and
unstratified cases. In addition,
we showed in \S \ref{disrel} that the dispersion relations (\ref{upper_bl})
(upper branch),
(\ref{disstar}) (lower branch), and (\ref{dismiddle}) (middle branch)
provided a good
estimate of the pattern speed in simulations irrespective of
stratification. Thus, we conclude that stratification has little
effect on the dynamics and morphology of the modes for an isothermal
equation of state.

\begin{figure}[!h]
\centering
\subfigure[]{\includegraphics[width=0.7\textwidth]{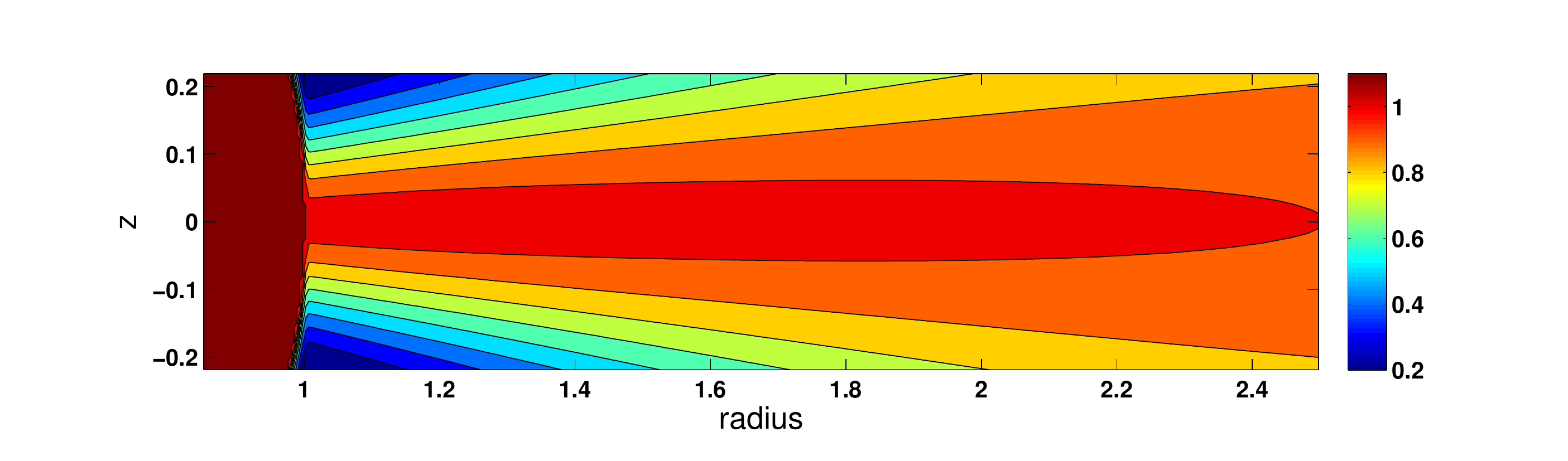}}
\subfigure[]{\includegraphics[width=0.7\textwidth]{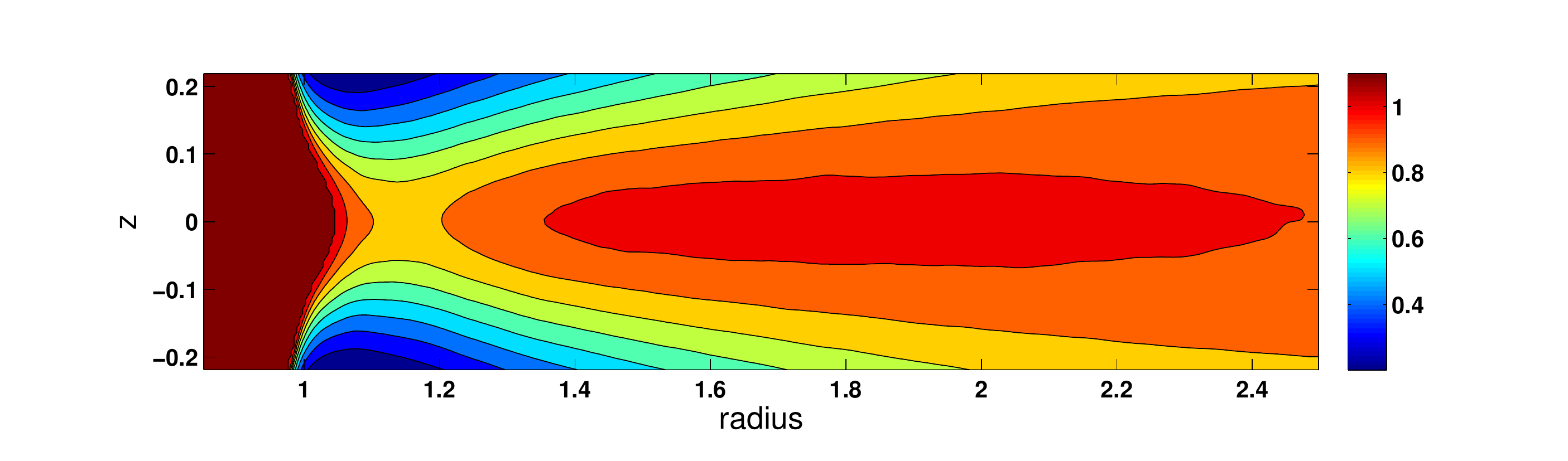}}
\caption{Panels a and b show contours of the azimuthally-averaged
  density profile at $t =
  0$ and $t=150$, respectively, for simulation 3D9c. Formation of a neck
  in the density distribution is clearly visible in panel b.}
\label{dcontfig}
\end{figure}

Although stratified and unstratified simulations exhibit similar
behavior, there are some obvious differences. For instance, in stratified
simulations we observe the formation of a neck in the density profile,
which is shown in Fig. \ref{dcontfig}. The formation of this neck
implies that sonic instabilities operate preferentially near the 
equator and spin up an equatorial belt on the star.

\begin{figure}[!h]
\centering
\subfigure[]{\includegraphics[width=0.45\textwidth]{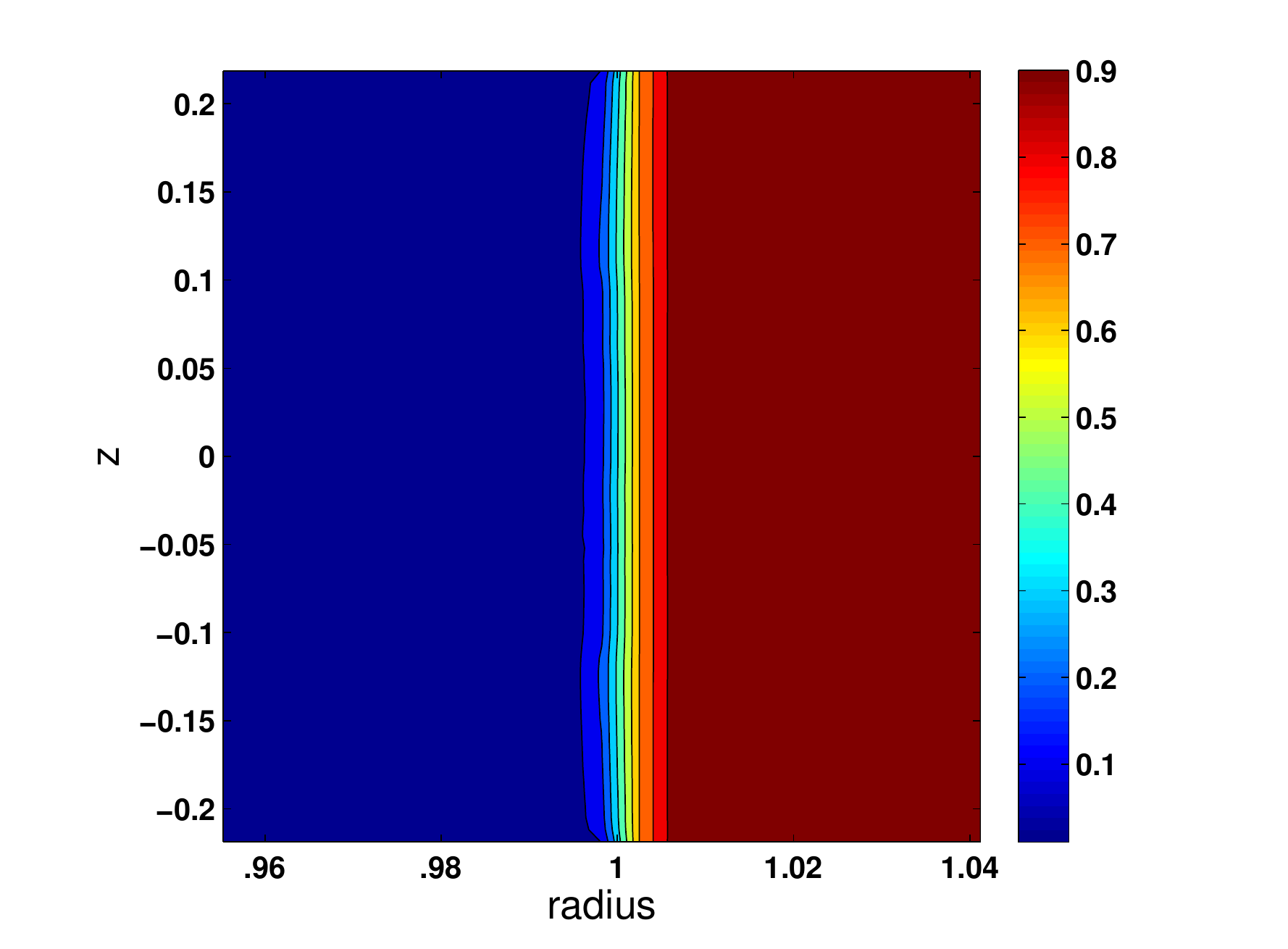}}
\subfigure[]{\includegraphics[width=0.45\textwidth]{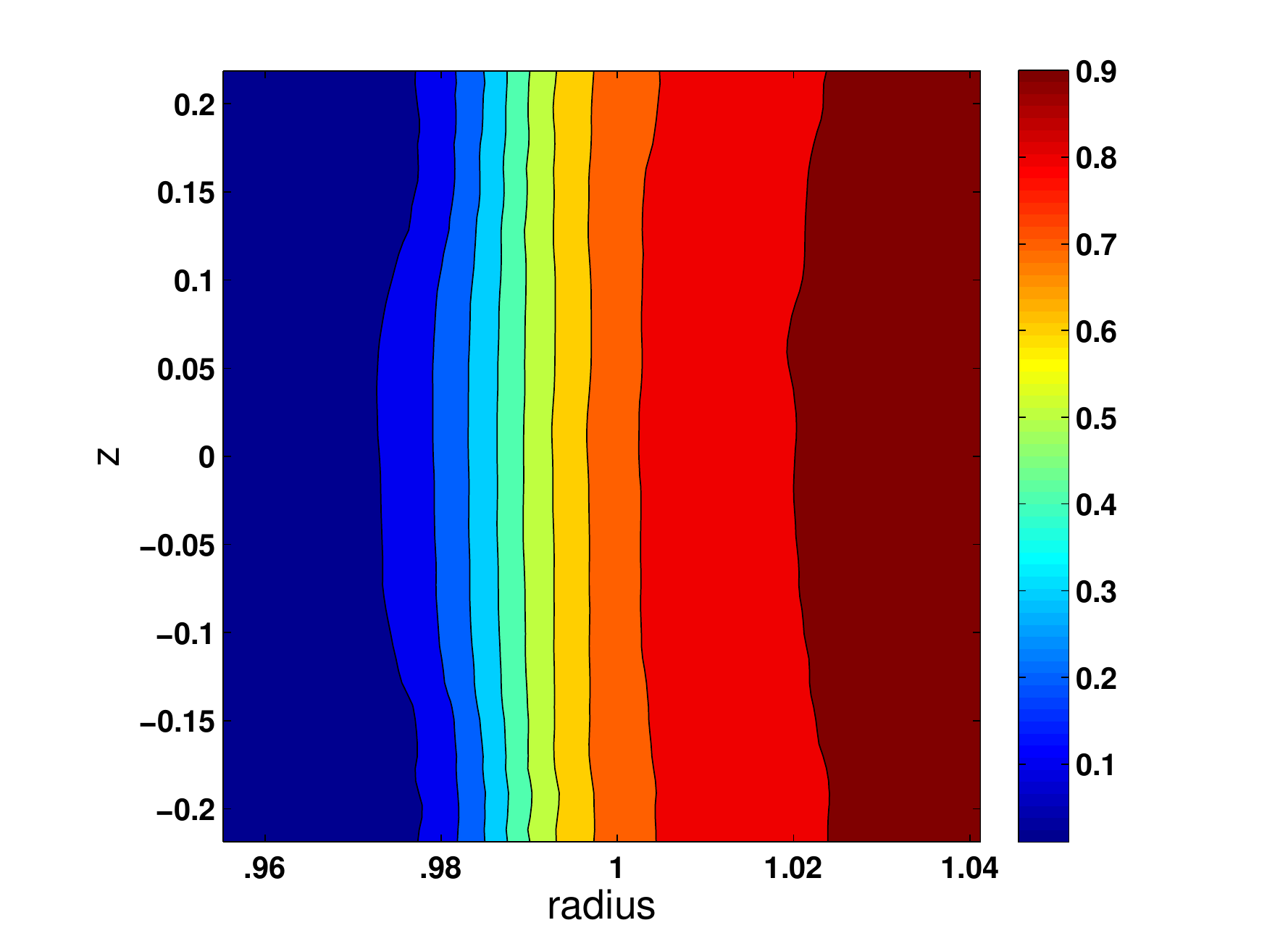}}
\caption{Panels a and b show contours of the density-weighted azimuthal
  average of $v_\phi$ at $t=0$ and $t=150$, respectively, for
  simulation 3D9c. There is a slight deviation from rotation on
  cylinders in panel b with the equatorial plane rotating faster than
  higher latitudes.}
\label{ocontfig}
\end{figure}

Further evidence in favor of a spun up
equatorial belt can be seen in Fig. \ref{ocontfig}, which shows
contours of the density-weighted azimuthal average of $v_\phi$ at
$t=0$ (panel a) and $t=150$ (panel b). It is clear from panel b
that the equatorial region of the star is spinning faster than higher
latitudes. For an isothermal equation of state or more generally for
any equation of state that is a function of the pressure alone,
deviation from rotation on cylinders as see in Fig. \ref{ocontfig}b
implies the existence of meridional circulation. Thus, although we do
not see evidence of significant azimuthal spreading of disk material
over the surface of the star, we do observe the sonic instabilities to
induce meridional flows. 

In principle, one may also interpret the vertical extent of 
the equatorial bulge of the spun-up fluid in our stratified simulations 
as evidence for the emergence of a spreading layer 
\citep{InogamovSunyaev,PiroBildsten} on the stellar surface. 
However, given the simplified nature of our runs (e.g. isothermal EOS), 
it difficult to unambiguously claim observation of this phenomenon in
the simulations.

Another difference between stratified and unstratified simulations
is that $t_\text{evol}$ as defined in equation (\ref{tevol}) is a
factor of two or so higher in the stratified case. This implies that
the acoustic modes have lower amplitudes and take a longer time to
transport angular momentum away from the inner disk in the stratified case.
Thus, mass accretion and angular momentum transport can occur at an
appreciable rate for a longer time before shutting off when the
inner disk has been depleted of material (\S \ref{transevol}).

\begin{figure}[!h]
\centering
\subfigure[]{\includegraphics[width=0.9\textwidth]{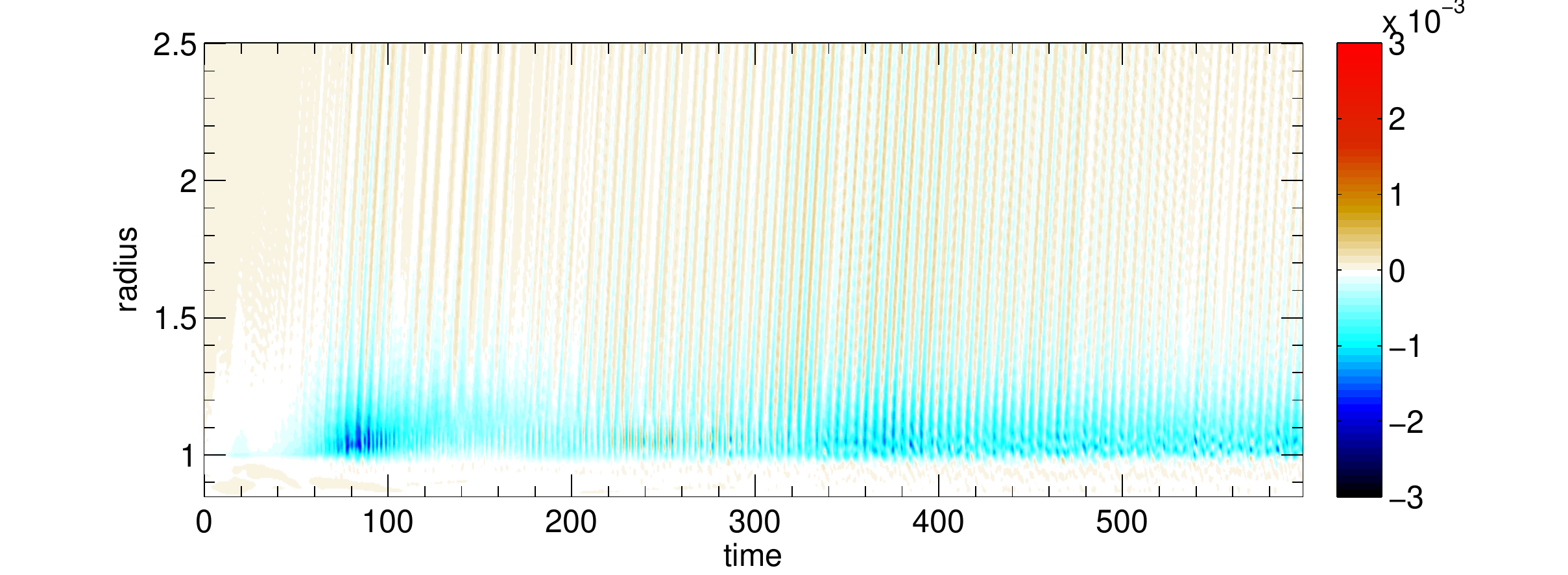}}
\subfigure[]{\includegraphics[width=0.9\textwidth]{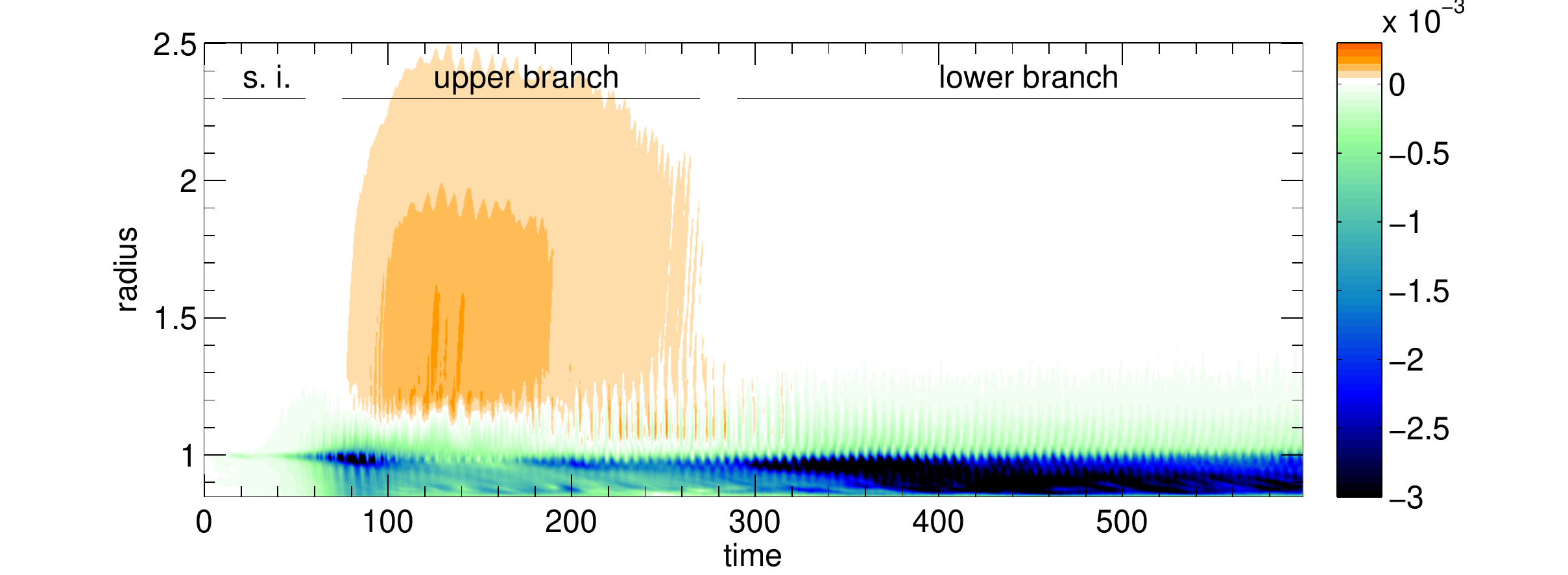}}
\caption{Panels a and b show radius-time plots of $\La v_\cp \Ra$ and
  $C_S$ for simulation 3D9c.}
\label{ststratfig}
\end{figure}

Panels a and b of Fig. \ref{ststratfig} show radius-time
plots of $\La v_\cp \Ra$ and $C_S$, respectively, for the stratified
simulation 3D9c. These radius-time plots can be compared with those
for the unstratified case shown in Figs. \ref{spacetimedensity}b and
\ref{stressspacetime}. Both the unstratified and stratified
radius-time plots look fundamentally similar. However the features in the
stratified simulation appear to be ``stretched'' in time as compared
to the unstratified one, providing evidence that $t_\text{evol}$ is
longer in the stratified case.

\section{The Incompatibility of $\alpha$ Models and Angular
  Momentum Transport by Waves}
\label{sect:alpha}


As we have shown, angular momentum transport in our simulations is
facilitated by quasi-2D acoustic waves, rather than turbulent
stresses. This assertion follows from
the fact that $C_S$ is well-described by
theoretical formulas applicable for waves \S \ref{ubsec}, and that most
of the power in the fluctuations is in $k_z = 0$ modes \S
\ref{kz0sec}. Because $\alpha$-models are based on the notion of a local
effective viscosity due to turbulent stresses, whereas waves are
inherently nonlocal, transporting angular momentum over long distances
before they are absorbed or dissipate, it is clear that any
conventional model of $\alpha$ viscosity is inapplicable to the BL. 

We can make this point more concrete with three specific reasons why an
$\alpha$ model is not applicable to transport by waves. First, in
conventional $\alpha$ models the stress vanishes (and $C_S$ changes
sign) when $\partial \Omega/ \partial\cp = 0$, and this is not 
observed in our simulations. Instead, in a wave model, the
stress vanishes (and $C_S$ changes sign) when $\Omega = \Omega_P$,
and we do see evidence of this (see \S \ref{ubsecdisk} and 
\ref{uldisk}). Although there is a corotation resonance in the BL, 
which could in principle be located close to the point where 
$d \Omega/d \cp = 0$, this is a fundamentally
different criterion. Moreover, for waves there is another corotation resonance
in the disk, where $C_S$ changes sign, which is an effect not captured by
$\alpha$ models. 

Second, equations (\ref{CLstar}) and (\ref{CLdisk}), which describe
transport by waves are fundamentally different in structure from
equations (\ref{alphadef1})-(\ref{alphadef3}) and depend on
quantities such as $\Delta \Sigma/
\Sigma$ and $\Omega_P$. The former is a function of time that depends
in a complicated way on the details of the sonic instability, and
the latter depends on the particular wave branch. Thus, although
possible, it would be contrived to boil down an expression containing
these quantities into an effective viscosity parameter; any such
parameter would be a complicated function
of the wave branch (upper, lower, or middle), azimuthal wavenumber, wave
amplitude, Mach number, radius, etc.

Third, Navier Stokes viscosity with an enhanced value of the
viscosity coefficient, $\nu$, due to turbulence assumes
{\it local} transport of angular momentum, which is
the reason why the stress is proportional to $\partial \Omega/\partial
\cp$. When angular momentum is carried by waves over long distances, the
transport is {\it nonlocal} --- different parts of the disk can 
exchange angular momentum only when waves dissipate, which often 
happens at large separation from their launching site. 
Thus any model, such as $\alpha$ viscosity, which is based on 
Navier-Stokes viscosity no longer applies. The possibility 
of non-local momentum transport has been hinted at by
\citet{InogamovSunyaev2}, who postulated that angular momentum
  transport within the spreading layer (a BL with substantial 
extent in the $\theta$ direction) of a neutron star occurs by 
wave damping, in order to avoid excessive heating at the base 
of the layer.

For these reasons, we believe it is better to use
equations (\ref{CLstar}) and (\ref{CLdisk}) as they are for
estimating angular momentum
transport in the BL. We do not provide an effective $\alpha$ for our
simulations, since it is incorrect to adopt an $\alpha$ viscosity model
when angular momentum is transported by waves.


\section{Discussion}


The major result of this work is a new picture of the 
momentum, energy, and mass transfer in astrophysical BLs. 
We have found that the isothermal boundary layer naturally 
supports three $k_z = 0$ branches of acoustic modes when there is a
supersonic velocity drop across it. These three types of acoustic modes
are directly related to the three types of waves found for the 
isothermal vortex sheet in \citet{BR}. Each of the three 
acoustic mode wave branches is associated with a different spatial
morphology. However, two features common to all of the wave branches
are emission of sound waves from the BL into both the star and the
disk and the presence of two corotation radii -- one within the disk and
the other within the BL. The corotation radius within the disk lies
inside a broad evanescent region (see Fig. \ref{morphfig}).  

The behavior of a particular
mode is affected by the density stratification 
inside the star and the differential rotation inside the disk. The
latter can potentially lead to the formation of a pattern of trapped
shocks inside a resonant cavity in the disk near 
the BL. For this to occur, the inner edge of the evanescent region
inside the disk must be well-separated from the BL.

An important point is that the dispersion relations for all three
acoustic mode wave branches remain the same in 2D, 3D 
unstratified, and 3D stratified simulations. This can 
be seen from Tables \ref{uppertable} \& \ref{middletable} and
Figure \ref{lowerfigure}. Thus, the proposed picture of angular
momentum transport in the BL is robust with respect to the 
dimensionality of the problem, strengthening the results
previously obtained by \citet{BRS}. 

An acoustic mode with pattern speed $\Omega_P$ couples to fluid at
the corotation radius within the BL ($\Omega(\cp_\text{c}) = \Omega_P$),
sapping it of angular momentum and driving
accretion onto the central object. Thus, the fluid in the BL loses
angular momentum to waves, which propagate away from the BL. 

A fraction of the angular momentum radiated away from the BL
is redeposited in the disk by wave dissipation
in shocks, which drive the evolution of the disk surface density 
(\S \ref{sect:Mdot}). An additional factor in the time-evolution of the
disk surface density is the time-dependence of $\Omega$,
which is connected with the temporal evolution of the radial pressure
gradient in the disk. The time-dependence of $\Omega$ is an important
effect in our naturally-evolving simulations.
However, in a truly steady-state model of the BL, this 
contribution to the temporal evolution of the disk surface density
would be absent.

The remainder of angular momentum lost by fluid in the BL is radiated 
into the star. Waves propagating away from the BL and into the star
weaken in amplitude as a result 
of angular momentum conservation in a medium with a rising density.
For this reason, waves inside the star do not shock --- their
nonlinear evolution is too weak and in our 
simulations they simply propagate out through the inner 
boundary of the simulation domain. In a real system, they would 
presumably be damped by some {\it linear} dissipation process 
such as conduction or radiation, ultimately transferring their
angular momentum and energy to the star. 

The radiation of angular momentum by waves away from the 
BL and its subsequent redistribution by damping of these 
waves are manifestly nonlocal processes, see \S \ref{sect:alpha}. 
This casts serious doubt on the utility of semi-analytic models 
employing an (intrinsically local) $\alpha$-prescription for 
viscosity in predicting BL structure and observational 
manifestations.

The general picture of acoustic wave-mediated BL 
dynamics advanced in this work may potentially have 
several important consequences that we discuss next. 
We warn, however, that until simulations with a more 
realistic equation of state are performed, some of the 
following predictions may be somewhat speculative.


\subsection{Implications: Energy Transport in the BL}
\label{sect:energy}

Waves propagating away from the BL carry not only angular 
momentum but also energy, which is released by
accreting material as it is spun down. Wave damping 
in the disk or star results in the {\it nonlocal} release of
this accretion energy --- energy is deposited into the
disk or star {\it at the location where the wave damps}.
For example, the rate of thermal energy release in the disk per 
unit radial distance due to dissipation of waves with 
pattern speed $\Omega_P$ is  
\ba
\frac{\partial \dot{E}_d}{\partial\cp}=
\left[\Omega_P-\Omega(\cp)\right]\frac{\partial C_S}{\partial\cp},
\label{eq:Ediss}
\ea
where $\partial \dot{E}_d/\partial \cp$ denotes the energy dissipation
rate per unit radius
(i.e. $(\partial \dot{E}_d/\partial \cp) d \cp$ is the energy
dissipated in an annulus between $\cp$ and $\cp + d\cp$). By our
definition, which is consistent with the definition in \S 6.4.2 of
\citet{BinneyTremaine}, $\dot{E}_d \le 0$
and is non-zero only where $\partial C_S/\partial\cp \ne 0$,
(i.e. after the wave has shocked in the disk). This non-local energy 
transport may affect the structure and observational 
manifestations of the BL in the following way. 

Conventional wisdom based on local $\alpha$-models (e.g. 
\citet{Balsara}) states that the intense energy release 
has to take place in the BL itself, heating the gas there 
and considerably modifying the structure of the BL. This does 
not need to be true in our picture of wave-mediated 
transport: kinetic energy of material accreting through 
the BL can be transported away from the layer by waves and deposited
in both the disk and the star. This can easily prevent 
the layer from overheating, meaning that its structure may 
be affected by accretional energy dissipation {\it weaker} than was 
thought before.  This is a very important consequence of 
wave-driven transport.

Unfortunately, we cannot explore the effect of nonlocal
energy transport on BL structure in our simulations directly, 
since they 
employ an isothermal equation of state and may not accurately capture 
the details of wave dissipation, especially inside the star.
However, the removal of accretion energy from the BL in the form of
waves that is a prediction
of our model adds some support to the results of our purely 
isothermal simulations, which assume the fluid temperature does 
not change at all.

As we have seen in \S \ref{ubsec},
at least during some periods of time $|C_s|$ going
into star is much larger than that going into the disk. Since the
energy density of a wave is directly proportional to its angular
momentum density, $E_\text{w} = \Omega_P L_\text{w}$
\citep{BinneyTremaine}, it is natural
to expect that most of the accretional energy released in the BL is
ultimately deposited in the star, rather than in the disk. 
As a result, heating of disk material by wave dissipation, 
which is distributed over a radial range $\Delta\cp\sim \cp_\star$
is unlikely to exceed the energy release in the inner 
disk due to conventional anomalous viscosity, such as MRI. The latter does
not operate in our simulations since we have focused on the pure
hydro case. Thus, if acoustic waves dominate transport in
the BL one might expect the effective temperature of the 
inner disk and its overall spectrum {\it not to change much} 
compared to models not accounting for the presence of 
a BL, which seems to be precisely what is observed in some systems
\citep{Ferland, Polidan}.

Depending on the nature of the accreting object, deep heating 
occurring inside the star as a result of wave dissipation may 
affect its internal structure. This effect 
is likely to be minimal for white dwarfs and neutron stars, 
which are supported by degeneracy pressure, but may be 
important for accreting pre-main sequence stars. Indeed, 
a typical T Tauri star accreting at the rate 
$\dot M=10^{-6}M_\odot$ yr$^{-1}$ (typical during the 
so-called FU Orioni outbursts) should have a rate of accretional
energy release in the BL comparable to its own luminosity
$\sim L_\odot$. Depending on the depth at which the waves 
excited in the BL dissipate, the nonlocal energy transport
into the star may be able to affect its structure
\citep{Prialnik,Hartmann97,Popham97}. In fact, \citet{Kenyon88}
and \citet{Popham96} have previously shown that
during FU Orioni outbursts, T Tauri stars may increase their 
radii by a factor $\sim 2$, which might be explained in our 
model by accretional energy release inside 
the star mediated by acoustic waves. 

Observational signatures of the BL should also be affected 
by nonlocal transport of energy in the form of waves. Accretional energy 
carried away by waves is either buried in the star or transferred to
the disk, both of which
affect the spectrum of the BL. Given that wave 
dissipation is expected to only mildly heat the inner disk
above its standard temperature set by the visco-turbulent heating rate
\citep{SS}, it may even be natural that the spectrum of the system
will have a rather weak contribution from the BL as most of
the liberated energy is buried in the star. In the case of protostellar 
accretion, the sequestration of accretional energy released 
during FU Orioni outbursts inside the star may help explain 
the so-called ``luminosity problem'' --- an apparent 
discrepancy between the expected accretion luminosities 
of protostars and their observed luminosities 
\citep{Kenyon90,Evans}. BL formation is expected during 
FU Orioni outbursts with their high value of
$\dot M$, which allows the disk to overcome magnetic truncation. 
A sizeable fraction of the energy released during these intense 
bursts of accretion can then be deposited deep inside the star 
by dissipation of acoustic modes. This energy would later leak out of the 
star, on a timescale which is of the order of the thermal 
timescale at the depth where the energy carried by waves 
is deposited.

The same process could potentially explain why BL emission 
is missing from some observations of accreting 
systems \citep{Ferland, Polidan}.
In this framework, existing observations of high-energy 
spectral excesses from accreting systems, such as cataclysmic 
variables \citep{Pandel}, conventionally ascribed to BL emission, 
may in fact be due to some sort of coronal emission produced 
above the inner disk. Hard X-ray excesses found in accreting 
neutron star systems can be due to emission from a 
spreading layer \citep{Gilfanov,Revnivtsev}, which has 
a different morphology from the BL.


\subsection{Implications: stellar spinup}
\label{sect:spinup}

Deposition of angular momentum carried by waves deep 
into the star will cause stellar spinup to occur in a 
non-trivial fashion --- angular momentum will be added to 
the star where waves dissipate, spinning up the inner 
parts of the star, which are hidden from direct view. As a result, 
the surface layers of the star may rotate {\it slower} than 
would be expected if the angular momentum of accreted material  
were deposited locally, near the BL. Unfortunately, we
cannot directly explore this issue in detail in our 
simulations due to their simplified nature. 

Previously \citet{Barker} have proposed a similar 
picture of inside-out stellar spinup by dissipation of 
tidally excited internal gravity waves near the stellar 
center. Clearly, this mode of stellar spinup is quite 
different from the predictions of standard BL models 
based on the notion of local $\alpha$-viscosity 
\citep{KT}.


\subsection{Implications: mixing in the BL}
\label{sect:mixing}

The standard paradigm of local viscosity in the BL generally 
assumes that angular momentum transport in the BL 
is facilitated by turbulence resulting from nonlinear saturation 
of some local instability such as Kelvin-Helmholtz 
\citep{KT,InogamovSunyaev2}, baroclinic \citep{PiroBildsten2}, 
etc. Quite naturally, turbulence in the layer would also 
result in efficient mixing of fluid elements, leading 
to elemental homogeneity if different parts of the layer 
have different chemical compositions.

The situation is different if transport in the layer is 
facilitated by acoustic modes as we find in this work. 
Wave-driven transport in the supersonic case does not need
an exchange of fluid elements to occur and thus does not 
necessarily result in efficient mixing inside the layer. This
agrees with what we generally find in our simulations, which 
typically exhibit large-scale regular structures like the 
pattern of trapped shocks and show very little evidence for
small-scale turbulence that would efficiently mix the layer.
Thus, one may expect elemental mixing to be less prominent
in our picture of the BL, compared to conventional models.

Having said that we point out that previously \citet{BRS} 
found using 2D simulations that the regular pattern
of fluid motion in the vicinity of the layer can be 
occasionally interrupted by the short-term bursts of very 
irregular, rather turbulent behavior, which might result 
in efficient mixing of the layer. However, we do not observe such
bursts in our 3D simulations, possibly because in 3D as opposed to 2D,
our simulations do
not develop an inflection point in the rotation profile within the BL.
This important issue clearly needs further study. 


\subsection{Implications: variability of emission}
\label{sect:variability}

Another implication of transport by waves is that they 
naturally produce temporal variability, since
a pattern of acoustic modes rotating at a well-defined pattern speed
invariably leads to periodic emission from the BL. Note that this
mechanism of variability is completely different from previous 
suggestions such as a non-axisymmetric bulge in the disk 
near the BL \citep{Popham99} or shallow surface waves on 
a spreading layer \citep{PiroBildsten1}, which can provide 
additional sources of variability.

Our simulations
do not implement radiative transfer and thus cannot directly link
periodicity of fluid patterns to spectral variability of the BL.
Nevertheless, temporal behavior of the BL has the 
potential to provide the most direct link between our work and 
observations, since from our dispersion relations for the upper, 
middle, and lower branches it is possible to work out the {\it 
frequency spectrum} of the BL. It is clear that it should 
feature a peak at frequency $\omega=m\Omega_P$, which corresponds to
the apparent periodicity of the wave pattern to an external
observer. 

In our simulations, we typically find quite high
values of $\omega = m\Omega_P\sim (4-6)\Omega(\cp_\star)$ for the
most commonly
observed pattern of trapped lower branch modes. Moreover, from
equation (\ref{disstar1}), we have the following lower bound for the
frequency of the lower branch $m\Omega_P >
M\Omega(\cp_\star)/2$. This lower bound makes it unlikely that acoustic
modes (or at the very least the lower mode, which is prevalent at late
times in our simulations) can explain the origin of dwarf novae
oscillations (DNOs)
in cataclysmic variables \citep{Warner}, which typically have $M \sim
100$ and $m\Omega_P \lesssim \Omega(\cp_\star)$ \citep{Patt}.

On the other hand, the nature of our simulations may prevent
us from seeing all possible sources of variability in the BL. In 
particular, a well known peculiarity of the isothermal equation 
of state used in this work is that the Brunt-V\"{a}is\"{a}l\"{a} 
frequency 
\ba
N^2 = (\gamma -1)\frac{g^2}{s^2}
\ea
is zero if $\gamma = 1$. This means that gravity waves ($g$-modes) 
in the star are neutrally buoyant ($\omega = 0$) and thus do not 
show up in our simulations. 

A more realistic equation of state having $\gamma > 1$
would be able to support both acoustic modes ($p$-modes) and 
gravity waves ($g$-modes) inside the star. The latter correspond
to the lower (``$-$'') sign in the dispersion relation 
(\ref{disVallis}) and have lower frequency than acoustic 
modes, potentially making them relevant for explaining DNOs. 
In principle, density waves in the disk could couple 
to both $g$-modes and $p$-modes in the star, whereas only the 
latter coupling is possible for the isothermal equation of state. 
Thus, a real BL would likely support more than just the three acoustic 
wave branches described in this work and \citet{BR}, and the 
dispersion relation for gravity modes in the star coupled to 
density waves in the disk remains to be worked out.


\subsection{Future prospects}
\label{sect:future}

Our study employs two crucial simplifying assumptions: 
an isothermal EOS and lack of magnetic fields. Relaxing the 
first assumption requires using a more realistic EOS and properly 
characterizing radiation transfer in the system. In addition to
previously mentioned consequences --- e.g. appearance of 
new modes, see \S \ref{sect:energy} \& \ref{sect:variability} 
--- a better treatment of thermodynamics will have other 
implications. 

For example, damping of modes excited in
the BL depends on thermodynamics: it was previously shown by 
e.g. \citet{Lubow}, \citet{Bate} 
that {\it thermal} stratification results in wave action 
channeling towards the low density regions, which may 
accelerate the nonlinear evolution of modes and cause 
them to damp faster. This would affect the dissipation 
pattern of waves, bringing disk regions where wave 
energy and momentum are deposited closer to the BL and 
making the BL more pronounced from an observational 
point of view. Additionally, 
compressional heating of accreted material in the BL can 
promote the formation of a spreading layer 
\citep{InogamovSunyaev,PiroBildsten} on the surface of the accreting object.

Inclusion of magnetic field in our calculations should
also have interesting consequences. First, it would
result in the presence of MRI within the disk, which will 
provide a means of mass delivery from the outer disk 
to the BL region. At the moment, during the periods of 
lower mode dominance in our simulations, the fluid 
outside of the evanescent region in the disk remains unaffected by 
acoustic modes. As a result, mass depleted from the 
innermost disk does not get replenished by external 
accretion. Second, MRI would introduce turbulent motions in 
the disk, which may interact with the BL modes in a 
non-trivial fashion and enhance the importance of $k_z \ne 0$ modes
for angular momentum transport within the disk. Third, the number of
possible wave
branches the system could support would be even higher 
in the MHD case, since the disk would be able to support 
slow, fast, and Alfv\'{e}n waves, which could couple to
$p$ and $g$-modes in the star. Finally, field 
amplification by intense shear in the BL would affect 
the internal dynamics of the layer. We leave the detailed 
exploration of these issues to future study. 


\acknowledgements

Resources supporting this work were provided by the NASA High-End 
Computing (HEC) Program through the NASA Advanced Supercomputing 
(NAS) Division at Ames Research Center. We thank Jeremy Goodman and 
Anatoly Spitkovsky for useful discussions. The financial support 
for this work is provided by the Sloan Foundation and NASA grant 
NNX08AH87G.


\appendix
\section{Dispersion Relation for Loosely Wound Modes Having $m \gg 1$ in the Disk}
\label{diskr0}
We derive here the dispersion relation for loosely wound modes having
$m \gg 1$ in
an isothermal disk (equation [\ref{upper_bl}]). We assume a two
dimensional setup in the $\cp$-$\phi$ plane and ignore
stratification. Thus, we replace the density $\rho$ by the surface
density $\Sigma$.

We start by explicitly writing the continuity and momentum equations
in cylindrical coordinates (e.g. \citet{BinneyTremaine}), along with
the isothermal equation of state. We
denote the radial and azimuthal velocities by $u$ and $v$ respectively
\ba
\label{cyl:cont}
\frac{\partial \Sigma}{\partial t} +
\frac{1}{\cp}\frac{\partial}{\partial \cp} (\cp \Sigma u) +
\frac{1}{\cp} \frac{\partial}{\partial \phi}(\Sigma v) &=& 0 \\
\label{cyl:vr}
\frac{\partial u}{\partial t} + u \frac{\partial
  u}{\partial \cp} + \frac{v}{\cp} \frac{\partial u}{\partial \phi} -
\frac{v^2}{\cp} &=& -\frac{\partial \Phi}{\partial \cp} -
\frac{1}{\Sigma}\frac{\partial P}{\partial \cp} \\ 
\label{cyl:vphi}
\frac{\partial v}{\partial t} + u \frac{\partial
  v}{\partial \cp} + \frac{v}{\cp} \frac{\partial v}{\partial \phi}
+ \frac{v u}{\cp} &=& -\frac{1}{\cp}\frac{\partial \Phi}{\partial \phi} -
  \frac{1}{\Sigma \cp}\frac{\partial P}{\partial \phi} \\
\label{cyl:state}
P &=& \Sigma s^2
\ea

Next, we assume a constant equilibrium disk density, which through the
equation of state implies a constant equilibrium disk pressure. We also ignore
self-gravity so there is no perturbation to the potential. Specifying
first order quantities with a preceding $\delta$, the
linearized momentum and continuity equations then become
\ba
\frac{\partial \delta \Sigma}{\partial t} + \Omega \frac{\partial
  \delta \Sigma}{\partial \phi} + \frac{\Sigma}{\cp} \frac{\partial
  \delta v}{\partial \phi} + \frac{\Sigma}{\cp} \delta u + \Sigma
\frac{\partial \delta u}{\partial \cp}&=& 0 \\
\frac{\partial \delta u}{\partial t} + \Omega \frac{\partial \delta
  u}{\partial \phi} - 2 \Omega \delta v +
\frac{1}{\Sigma}\frac{\partial \delta P}{\partial \cp} &=& 0 \\
\frac{\partial \delta v}{\partial t} + \Omega \frac{\partial \delta
  v}{\partial \phi} + \frac{\partial (\cp
  \Omega)}{\partial \cp} \delta u  + \Omega \delta u  + \frac{1}{\Sigma \cp}
\frac{\partial \delta P}{\partial \phi} &=& 0 \\
\delta P - s^2 \delta \Sigma &=& 0
\ea

Taking perturbations in the form $\exp[i\int^\cp d\cp' k_\cp(\cp') + im(\phi -
\Omega_P t)]$ and eliminating $\delta P$ in favor of $\delta \Sigma$, the
linearized set of equations becomes
\ba
\label{cyl1}
i m(\Omega - \Omega_P) \frac{\delta \Sigma}{\Sigma} +
\frac{1}{\cp}(1 + i n) \delta u + \frac{i m}{\cp} \delta v  &=& 0 \\
\label{cyl2}
i m (\Omega - \Omega_P) \delta u - 2\Omega \delta v + \frac{i
  s^2 n}{\cp} \frac{\delta \Sigma}{\Sigma} &=& 0 \\
\label{cyl3}
i m (\Omega - \Omega_P) \delta v + \frac{\kappa^2}{2 \Omega} \delta u
+ \frac{i s^2 m}{\cp} \frac{\delta \Sigma}{\Sigma} &=& 0,
\ea
where we have defined
\ba
n(\cp) \equiv \cp k_\cp,
\ea
and used the epicyclic frequency
\ba
\kappa^2 = 2\Omega \left(2\Omega + \cp \frac{d \Omega}{d \cp} \right).
\ea

We can eliminate $\delta v$ in equations (\ref{cyl1}) and (\ref{cyl2}) by using
(\ref{cyl3}) to arrive at
\ba
\label{cyl4}
i\left[m^2(\Omega - \Omega_P)^2 - \frac{s^2 m^2}{\cp^2}\right]
  \frac{\delta \Sigma}{\Sigma} - \frac{
  m}{\cp} \left[\frac{\kappa^2}{2 \Omega} -
  (1+in) (\Omega-\Omega_P) \right] \delta u &=& 0 \\
\label{cyl5}
\left[m^2(\Omega - \Omega_P)^2 - \kappa^2\right] \delta u - \frac{s^2
  m}{\cp} \left[2 \Omega i - n (\Omega - \Omega_P) \right] \frac{\delta
  \Sigma}{\Sigma} &=& 0.
\ea
After some algebra, the above system of equations reduces to
\ba
\label{toteq}
\left[m^2(\Omega-\Omega_P)^2 - \frac{s^2 m^2}{\cp^2}\left( 1 +
\frac{n^2}{m^2} - i \frac{n}{m^2}\right) -
\kappa^2 \right] + \frac{i s^2}{\cp^2(\Omega - \Omega_P)}\left[-2
  \Omega (i - n) - \frac{n \kappa^2}{2 \Omega} \right] = 0.
\ea

We now make the approximations $m \gg 1$ and $n/m \ll 1$, and we
shall refer to the latter as
the ``loose winding'' approximation. In that case,
equation (\ref{toteq}) can immediately be simplified to
\ba
\label{toteq1}
\left[m^2(\Omega-\Omega_P)^2 - \frac{s^2 m^2}{\cp^2} -
\kappa^2 \right] + \frac{i s^2}{\cp^2(\Omega - \Omega_P)}\left(-2
  \Omega (i - n) - \frac{n \kappa^2}{2 \Omega} \right) = 0.
\ea
We see that the term in the square brackets is the dispersion relation
(\ref{upper_bl}). Therefore, it is left to show that the other terms
are negligible. To start, note that if we take in order
of magnitude $\kappa \sim \Omega$,
then the ``other'' terms are in order of magnitude
\ba
\label{toteq2}
\frac{1}{m(\Omega-\Omega_P)} \left[\kappa \frac{s^2 m^2}{\cp^2}
\max \left(\frac{1}{m},\frac{n}{m}\right)\right].
\ea
It is straightforward to show that regardless of whether $\kappa \gg sm/\cp$,
$\kappa \ll sm/\cp$, or $\kappa \sim sm/\cp$ the term in equation
(\ref{toteq2}) is smaller than $m^2(\Omega-\Omega_P)^2$ (and hence the
dominant terms in the square brackets in equation (\ref{toteq1}))
by a factor of {\it at least}
\ba
\max\left(\frac{1}{m},\frac{n}{m}\right) \ll 1.
\ea

Thus, as long as $m \gg 1$ and $n/m
\ll 1$, the dispersion relation is accurately given by equation
(\ref{upper_bl}). In our simulations $10 < m < 40$, so the first of
these conditions is typically well-satisfied. It is more difficult to get
a handle on how well the ``loose winding'' condition is satisfied.
However, the fact that Table \ref{uppertable} shows that there is
agreement between theory and simulations at the several percent level
is good justification for the use of equation (\ref{upper_bl}).

\section{Dispersion Relation for an Isothermal Stratified Atmosphere
  in Cylindrical Geometry}
\label{discylindrical}
We consider the dispersion relation for an unrotating stratified
atmosphere in cylindrical geometry. Our aim is to justify our use of
equation (\ref{disstar}), which is valid for plane-parallel
geometry. For simplicity, we
assume two-dimensional perturbations in the $\cp - \phi$ plane.
However, we do not initially specify the form of the
gravitational potential $\Phi(\cp)$, only specifying it when necessary
to make the calculations tractable. 

As in Appendix \ref{diskr0}, we start with equations
(\ref{cyl:cont})-(\ref{cyl:state}). We take small perturbations on top
of the equilibrium state, and again denote
small first order quantities with a preceding $\delta$. The linearized
forms of the continuity and momentum equations and the equation of
state are then given by
\ba
\label{lin:cont}
\frac{1}{\Sigma}\frac{\partial \delta \Sigma}{\partial t} -
h_s^{-1} \delta
u  + \frac{1}{\cp}\frac{\partial}{\partial \cp}(\cp \delta u)
  + \frac{1}{\cp}\frac{\partial \delta v}{\partial \phi} &=& 0 \\
\label{lin:vr}
\frac{\partial \delta u}{\partial t} + \frac{1}{\Sigma}\frac{\partial \delta
  P}{\partial \cp} + s^2 h_s^{-1} \frac{\delta \Sigma}{\Sigma} &=& 0
  \\
\label{lin:vphi}
\frac{\partial \delta v}{\partial t} + \frac{1}{\Sigma
  \cp}\frac{\partial \delta P}{\delta \phi} &=& 0 \\
\label{lin:state}
\delta P  - s^2 \delta \Sigma &=& 0,
\ea
where we have defined 
\ba
h_s(\cp) \equiv -\left(\frac{1}{\Sigma}\frac{d \Sigma}{d \cp}\right)^{-1}
\ea
to be the local scale height of the atmosphere.

We now assume perturbations in the form $
f(\cp)\exp[i(m\phi - \omega t)]$. Eliminating $\delta \Sigma$ and
$\delta v$ from equations (\ref{lin:cont})-(\ref{lin:state}) we have
\ba
-i \omega \frac{\delta P}{P} +
\left[\frac{1}{\cp}\left(1-\frac{\cp}{h_s}\right) + \frac{\partial}{\partial
    \cp}\right] \delta u + i \left(\frac{m^2}{\cp^2}\right)
\frac{s^2}{\omega} \frac{\delta P}{P} &=& 0 \\
-i \omega \delta u + \frac{s^2}{P}\frac{\partial \delta P}{\partial
  \cp} + \frac{s^2}{h_s} \frac{\delta P}{P} &=& 0. 
\ea 
Finally, eliminating $\delta u$ in terms of $\delta P$ we arrive at
the second order differential equation in terms of $\delta P/P$
\ba
\label{2ndordereq}
\left(\frac{\omega^2}{s^2} - \frac{m^2}{r^2}\right) \frac{\delta P}{P}
+ \left(1 -
\frac{\cp}{h_s}\right)\frac{1}{\cp}\frac{\partial}{\partial
  \cp}\left(\frac{\delta P}{P}\right) + \frac{\partial^2}{\partial
  \cp^2} \left(\frac{\delta P}{P}\right) = 0.
\ea

We can gain insight by comparing the differential equation
(\ref{2ndordereq}) to the second order equation for $\delta P/ P$ in
the plane-parallel case for constant gravity with stratification in the
$x$-direction. The latter is given by \citet{Vallis} as
\ba
\label{2ndorderVallis}
\left(\frac{\omega^2}{s^2} - k_y^2\right) \frac{\delta P}{P} -
\frac{1}{h_s} \frac{\partial}{\partial x}\left(\frac{\delta
  P}{P}\right) + \frac{\partial^2}{\partial x^2}\left(\frac{\delta
  P}{P}\right) = 0.
\ea
Equation (\ref{2ndordereq}) reduces {\it locally} to equation
(\ref{2ndorderVallis}) at a given radius $\overline{\cp}$ in the limit
\ba
\label{limitplane}
\frac{h_s(\overline{\cp})}{\overline{\cp}} \ll 1,
\ea
if we make the association $k_y \rightarrow m/\overline{\cp}$. In our
simulations, $h_s/\cp \sim .02$, so the condition (\ref{limitplane})
is well-satisfied. The local equivalence of equations (\ref{2ndorderVallis}) and
(\ref{2ndordereq}) thus provides a justification for our use of an
effective radius in the dispersion relation (\ref{disstar}). 

\section{Dissipation Rate in the Isothermal Limit}
\label{appiso}
We derive here a formula for the energy dissipation per unit mass due
to a shock in the isothermal limit, $\gamma \rightarrow 1$. This is
important in the context of our simulations, since the entropy is not
well-defined if $\gamma = 1$ This can be seen from the expression for
the entropy per unit mass for an ideal gas 
\ba
\label{entropypermass}
\frac{dS}{dm} = \frac{s^2}{T \gamma(\gamma-1)} \ln \left(P \rho^{-\gamma} \right)
\ea
is ill-defined if $\gamma = 1$. 

The energy
dissipation rate of the wave used in equation (\ref{dEdmeq}) for an
isothermal shock\footnote{An isothermal shock means that the temperature is
  the same on either side of the shock but not necessarily that
  $\gamma = 1$.} is given by
\ba
\label{dEdmentropy}
\frac{dE}{dm} &=& T\Delta \frac{dS}{dm}
\ea
where the $\Delta$ denotes the difference in a quantity across the shock. It is
important to show that $\Delta dS/dm$ and hence $dE/dm$ converges to a
well-defined value when
$\gamma \rightarrow 1$. We perform this below, and derive an exact result for
the dissipation rate $dE/dm$ in the isothermal limit. \citet{Larson}
has given an approximate expression for $dE/dm$ valid for arbitrary $\gamma$ in the
limit of weak shocks. We note
that our exact expression (equation [\ref{dEdmiso}])  matches the
expression of \citet{Larson}
when $\gamma = 1$ (equation [\ref{dEdmapprox}]) to
leading order (third) in $\Delta \Sigma/\Sigma$.

To start, we combine equations (\ref{entropypermass}) and
(\ref{dEdmentropy}) to write
\ba
\label{dEdmgen}
\frac{dE}{dm} = \frac{s^2}{\gamma(\gamma-1)}\ln \left[\frac{P_1}{P_0}
  \left(\frac{\rho_1}{\rho_0}\right)^{-\gamma} \right],
\ea
where the subscripts ``0'' and ``1'' denote pre and postshock
quantities respectively. Using the Rankine-Hugoniot relations and
  defining  $\gamma = 1 + \delta$ and $\eps \equiv
\rho_1/\rho_0 - 1$, equation (\ref{dEdmgen}) becomes 
\ba
\label{dEdmgammadelta}
\frac{dE}{dm} = \frac{s^2}{(1+\delta)\delta}\ln \left[\frac{2 +
  \epsilon(2+\delta)}{2 - \epsilon\delta} (1 + \epsilon)^{-(1
  + \delta)}\right].
\ea
Next, we take the isothermal limit $\delta \rightarrow 0$ and expand
  the expression inside the logarithm in powers of $\delta$. Equation
  (\ref{dEdmgammadelta}) then becomes 
\ba
\label{dEdmgd1}
\frac{dE}{dm} = \frac{s^2}{(1+\delta)\delta}\ln \left[1+
  \left(\frac{\eps(2+\eps) -
  (1+\eps)\ln(1+\eps)}{2(1+\eps)}\right)\delta
  +\mathcal{O}\left(\delta^2\right) \right].
\ea
Expanding the logarithm in powers of $\delta$ and keeping the highest
  order term we arrive at equation (\ref{dEdmiso})
\ba
\frac{dE}{dm} =
  s^2\frac{\epsilon(2+\epsilon)-2(1+\epsilon)\ln(1+\epsilon)}{2(1+\epsilon)}.
\ea
Note that this expression is valid for all values of $\epsilon$.
\end{document}